\documentclass[twocolumn,floats,floatfix,
    showpacs,prd,superscriptaddress,
    nofootinbib]{revtex4-2}

\usepackage{graphicx,epsfig}
\usepackage{amssymb,amsmath,amsthm,amsfonts,mathtools}
\usepackage{bm}
\usepackage[inline]{enumitem}
\usepackage{tensor}
\usepackage[usenames,dvipsnames]{xcolor}
\usepackage{url}
\usepackage{multirow}
\usepackage{array}
\usepackage{xspace}
\usepackage{comment}
\usepackage[caption=false]{subfig}
\usepackage{booktabs}
\usepackage{soul}

\usepackage{float}

\usepackage{accents}
\usepackage[linktocpage,breaklinks]{hyperref}
\hypersetup{colorlinks=true,
            citecolor=NavyBlue,
            linkcolor=NavyBlue,
            urlcolor=NavyBlue}

\definecolor{lightred}{RGB}{255, 120, 120}

\renewcommand{\vec}[1]{\boldsymbol{#1}}

\def\newacronym#1#2#3{\gdef#1{\gdef#1{#2\xspace}#3 (#2)\xspace}}
\def\bh#1{black hole#1 (BH#1)\gdef\bh{BH}}
\newacronym{\GW}{GW}{gravitational waves}
\newacronym{\qbs}{QBS}{quasi-bound state}
\newacronym{\qnm}{QNM}{quasi-normal mode}
\newacronym{\vsh}{VSH}{vector spherical harmonics}
\newacronym{\fkks}{FKKS}{Frolov-Krtou\v{s}-Kubiz\v{n}\'{a}k-Santos}
\newacronym{\ode}{ODE}{ordinary differential equation}
\newacronym{\pde}{PDE}{partial differential equation}

\def\dif{\textrm{d}}

\def\rS{r_{\rm{s}}}

\def\pc{\text{ pc}}
\def\ev{\text{ eV}}
\def\yr{\text{yr}}

\def\Lie{\mathcal{L}}

\begin{document}

\title{Black-hole hair from vector dark matter accretion}

\author{Fredric~Hancock}
\email{foh3@illinois.edu}
\affiliation{The Grainger College of Engineering,
Department of Physics \& Illinois Center for Advanced Studies of the Universe, University of Illinois Urbana-Champaign, Urbana, Illinois 61801, USA}

\author{Helvi Witek}\email{hwitek@illinois.edu}
\affiliation{The Grainger College of Engineering,
Department of Physics \& Illinois Center for Advanced Studies of the Universe, University of Illinois Urbana-Champaign, Urbana, Illinois 61801, USA}
\affiliation{Center for AstroPhysical Surveys, National Center for Supercomputing Applications, University of Illinois Urbana-Champaign, Urbana, IL, 61801, USA}

\begin{abstract}
    We model a single black hole in equilibrium with a dark photon-cold dark matter environment.
    Representing the dark photon as a Proca field,
    we show that a Schwarzschild black hole grows vector-field ``hair'' when allowed to accrete from an infinite homogeneous bath of particles far from the horizon.
    We solve the Proca equation in linear perturbation theory, separating it using the vector spherical harmonics and Frolov-Krtou\v{s}-Kubiz\v{n}\'{a}k-Santos approaches for the odd-parity and even-parity sectors, respectively. In the ``particle" dark matter regime, the field is purely infalling and exhibits a sharply peaked density profile, in concordance with the particle dark matter ``spikes" studied in the literature. In the ``wave" regime, the field exhibits standing waves, and the profile is smeared. We find a dark-matter density amplification upward of $10^7$ near the horizon. Though small for most black holes, we find the mass enclosed in the cloud can reach $\sim 1 \%$ of the black hole mass for large supermassive black holes. These black holes are also most susceptible to vector dark matter accretion, with mass accretion rates as large as $10 M_\odot/\yr$.
\end{abstract}

\maketitle
\tableofcontents

\section{Introduction}

The Proca field has received increasing interest in recent years. Taking its name from Romanian physicist Alexandru Proca, it is a spin-1 field obeying an extension of Maxwell's equations with an added mass term. While the model was originally introduced in the development of a theory of nuclear interactions \cite{1936JPhyR...7..347P},
Proca fields are, today, most often used in extensions to the standard model. There are many fundamental and phenomenological motivations for the extensions, but perhaps the most compelling come from the search for dark matter.

Similar to the (pseudo-)scalar axion, a Proca field behaves as cold dark matter if it is populated as a ``condensate", i.e., as a homogeneous, coherently oscillating configuration~\cite{Antypas:2022asj}.
This is known as ``dark photon'' dark matter.
For a Proca field $A_\mu$ of mass $\mu$, such a condensate has time dependence $A_\mu \sim e^{-i \mu t}$ in natural units.

Creating these condensates requires a nonthermal mechanism in the early universe, of which several have been proposed.
One is a misalignment mechanism, similar to the axion model, in which the rapid expansion of the early universe freezes the field at a random value, which gives way to a coherent oscillation at late times when $\mu$ passes below the Hubble scale \cite{Arias:2012az, Nelson:2011sf}. Another is production from inflation itself, where initial quantum Proca-field fluctuations are blown up into a condensate of particles, which then persists as a thermally decoupled spectator during the hot big bang \cite{Graham:2015rva}. These scenarios favor a field mass in the ultralight (``wave dark matter") regime, corresponding to $\mu < 1$ eV.
Various experimental results, including direct detection, astrophysical consistency, and cosmological measurement, place considerable constraints on $\mu$, particularly when its mixing with standard-model matter is large.
However, the ranges $10^{-10}\ev < \mu < 1 \ev$ and especially $\mu < 10^{-14} \ev$ are relatively open \cite{Fabbrichesi:2020wbt, Caputo:2021eaa}. The dark photon is thus a compelling dark matter candidate, and, moreover, the portions of its ultralight regime which are relatively unconstrained are compelling for \bh{} physics.

Several models in the UV
lend additional motivation for a hidden sector containing massive vector fields.
Many string compactification scenarios
such as standard compactifications of both $\mathrm{E}_8 \times \mathrm{E}_8$ heterotic and Type II theories
yield one or more hidden $\mathrm{U}(1)$ gauge groups \cite{Jaeckel:2010ni, Goodsell:2009xc}.
They can
acquire a mass via the Stueckelberg or
the Higgs mechanism, and they interact with the standard model kinetically through hypercharge mixing \cite{Holdom:1985ag, Curtin:2014cca, Adshead:2022ovo}. The fields' masses and mixing parameters vary by model, but this landscape of proposals motivates the existence of a hidden massive vector with a weak standard model coupling and a large range of possible masses.

We now turn to \bh{} physics.
When a bosonic field scatters off a rapidly spinning \bh{}, it may extract angular momentum from the horizon, an effect known as \bh{} superradiance. For a massive field with a Compton wavelength comparable to the \bh{} horizon radius, this effect may drive an instability, pushing the field into an exponentially growing regime \cite{Detweiler:1980uk, Dolan:2007mj,Shlapentokh-Rothman:2013ysa, Brito:2015oca}.
This occurs at $\mu \sim 10^{-10} \ev$ for the smallest stellar-mass ($\gtrsim 5 M_\odot$) \bh{s} and at $\mu \sim 10^{-20} \ev$ for the largest supermassive ($\sim10^{10} M_\odot$) \bh{s}---both viable under current experimental constraints. While less studied than the massive scalar instability, Proca field superradiance has received considerable attention. Aided by an increasingly sophisticated set of tools from \bh{} perturbation theory, multiple authors have characterized Proca-field superradiance in the frequency domain by finding unstable \qbs{} modes in Kerr spacetime~\cite{Pani:2012bp, Pani:2012vp,Dolan:2018dqv,Siemonsen:2019ebd, Cardoso:2018tly, Baumann:2018vus}. This has inspired multiple numerical studies showing the field's dynamical growth, both in the decoupling limit~\cite{Witek:2012tr, East:2017mrj} and with full backreaction~\cite{East:2017ovw, East:2018glu}.
Further work extended these models, e.g., by considering higher-order (self-interaction) terms in the field's potential~\cite{Clough:2022ygm}.

A primary theme running through these studies is a focus on the superradiant instability's end state; unstable superradiant growth, upon extracting enough \bh{} angular momentum to leave the superradiant regime, generically results in a diffuse cloud of particles surrounding the \bh{}. We may broadly separate these states into two categories. For a real field, these clouds reach a maximum mass, then begin to decay after spinning down the \bh{} until it can no longer support superradiant scattering. The decay timescale can be long in some astrophysical contexts, but it is not infinite. In the case of a complex Proca field, however, there exist nonperturbative stationary solutions with a sustained Proca buildup which branch off from regular Kerr spacetime just at the onset of the superradiant instability \cite{Herdeiro:2016tmi, Santos:2020pmh}. As true stationary spacetimes, these solutions are frequently discussed as violations of the no-hair conjecture---the famous statement that the endpoint of gravitational collapse is a stationary \bh{} described only by its mass, spin, and electromagnetic charge \cite{Ruffini:1971bza}. The presence of a superradiant cloud complicates the spacetime, introducing its mass and angular momentum as additional necessary configuration parameters. Here, we introduce another astrophysically viable mechanism to generate ``hair"---that of accretion from a particle bath. We discuss the no-hair conjecture and its violations in more detail later in this paper.

To understand this mechanism, we must now briefly turn away from vector fields. Minimally coupled massless scalar fields have traditionally been thought not to support any long-lived configurations in stationary \bh{} spacetimes. This was shaken, however, by a perturbative result from Jacobson, showing that such a field settles into a nontrivial configuration if allowed to grow linearly with time at spatial infinity~\cite{Jacobson:1999vr}. With the popularity of ultralight axion dark matter models, several authors have extended this result to the massive case, replacing the linearly growing boundary condition with an oscillating one in the same manner as an axion cold dark matter condensate~\cite{Hui:2019aqm, Clough:2019jpm, Bamber:2020bpu, Hui:2022sri}. Like superradiance, this mechanism creates a sustained cloud of field around the \bh{}, exciting interest both as a potentially detectable astrophysical effect and as a means for violation of the no-hair conjecture. Unlike superradiance, however, this mechanism applies regardless of the \bh{} spin, relying only on direct accretion, meaning it would exist in virtually any astrophysical context. Accretion-driven hair formation from an ultralight dark matter halo is thus an exciting novel effect, and its prominence should be understood properly for astrophysical \bh{s}.

\begin{figure}
    \centering
    \includegraphics[width=0.95\linewidth]{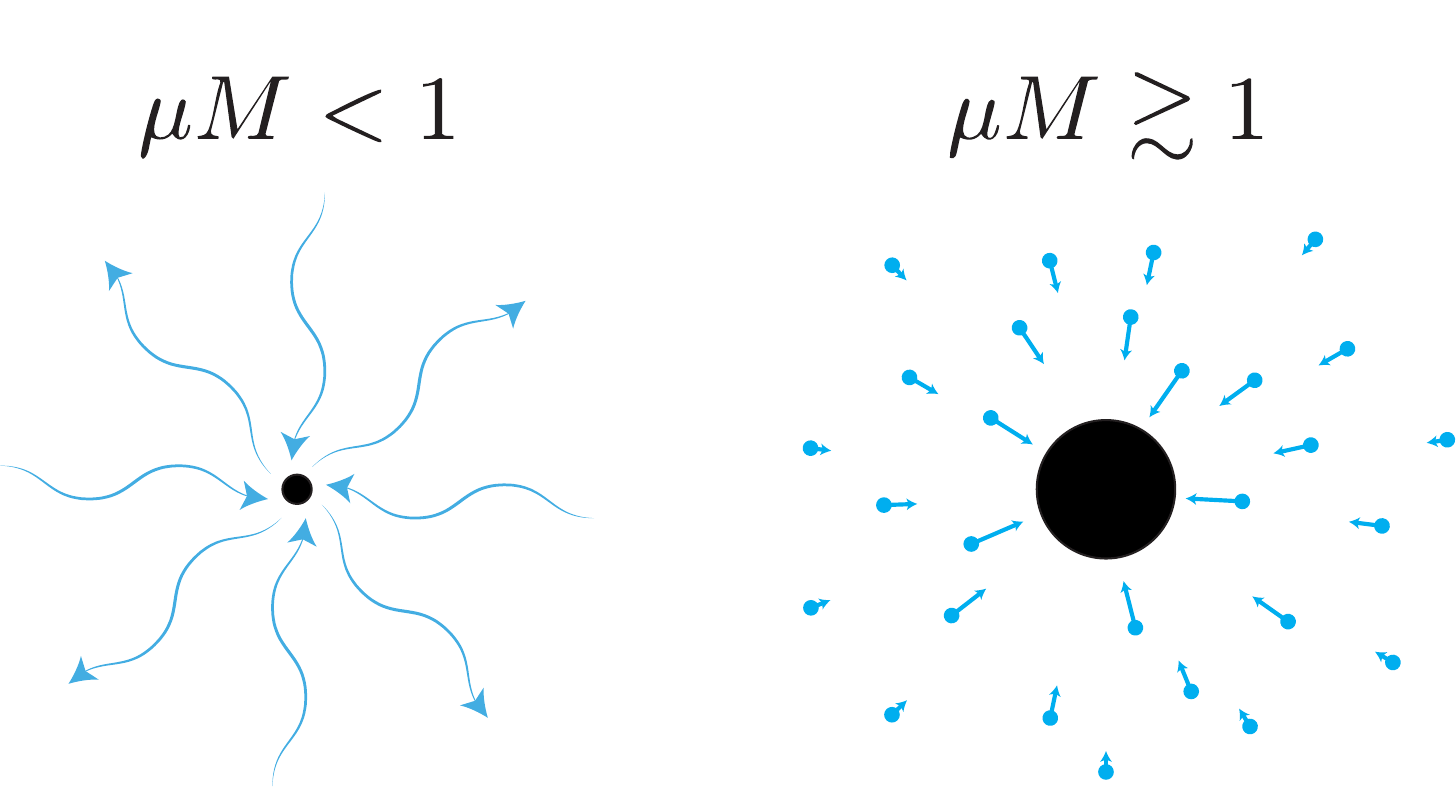}
    \caption{Cartoon of the Proca field's behavior in the wave ($\mu M < 1$) and particle ($\mu M \gtrsim 1$) regimes, shown in the left and right panels, respectively. Diffraction effects dominate in the former, suppressing accretion, while in the latter, the field acts as a collection of freefalling particles, and the problem loses all sensitivity to the field mass $\mu$.}
    \label{fig:accretion_cartoon}
\end{figure}

While studied in some detail for a massive scalar in spherical symmetry \cite{Hui:2019aqm, Clough:2019jpm}, this effect has not been studied for a massive vector field. In this work, we do just that; \textit{we find, for the first time, the profile a Proca field forms around a Schwarzschild \bh{} when allowed to accrete from an infinite, nonrelativistic, homogenous particle bath.} We do this using a perturbative scheme, assuming the field has negligible backreaction onto the spacetime. We characterize the field's radial profile using its density and mass function, and we estimate the rate at which it is accreted across the horizon, contributing to the \bh{'s} growth. We find that the field's phenomenology is characterized by two different regimes: a ``wave" and a ``particle" regime. These are controlled by the dimensionless parameter $\mu M$ (in natural units), where $M$ is the \bh{} mass, with the transition occurring at $\mu M \approx 1$, as shown schematically in Fig. \ref{fig:accretion_cartoon}.

We solve the Proca equation using a separation-of-variables scheme. This relies on new tools developed by the \bh{} perturbation theory community. Namely, we use both the \vsh scheme of \cite{Rosa:2011my}, and the separation scheme of \fkks, introduced in \cite{Krtous:2018bvk}. The former rests on a body of literature concerned with separating the Proca equation in Schwarzschild spacetime using various spherical harmonic decompositions \cite{Galtsov:1984ixy, Konoplya:2005hr}, and its specifics are based on the spin-2 formalism of Ref.~\cite{Barack:2005nr}.
The \fkks scheme,
instead,
is based on the study of spacetime hidden symmetries~\cite{Krtous:2008tb, Frolov:2017kze}, and the separation follows from the existence of a Killing object known as the principal tensor.
It follows from an earlier result by Lunin~\cite{Lunin:2017drx}.
A patchwork of these two methods permits a full separation of the Proca equation, allowing us to study all of the field's degrees of freedom.

The structure of this paper is as follows.
In Section \ref{sec:setup}, we review the Proca field  equation in curved spacetime.
In Section \ref{sec:massless_case}, after reviewing the Jacobson scalar hair result, we discuss its extension to the vector case.
We then describe the \vsh and \fkks methods, used to separate the Proca equation in Schwarzschild spacetime, in Section \ref{sec:separation_methods}.
We then solve the resulting equations in Section \ref{sec:solutions}, using a mixture of analytic and numerical schemes. Finally, in Section \ref{sec:astro_implications}, we explore what our solutions might imply for astrophysical observation.

\paragraph*{A note on notation:}
In this work, we
adopt a $(-,+,+,+)$ metric signature and use natural units $c=G=\hbar=1$ throughout, except when explicitly stated otherwise.
For (anti-) symmetrization
of indices, we use the standard parenthesis/bracket shorthand notation, e.g.,
\begin{align}
    T_{[\mu\nu]} = {}&\frac{1}{2!}\left(T_{\mu\nu} - T_{\nu\mu}\right)
\,,\quad
    T_{(\mu\nu)} = \frac{1}{2!}\left(T_{\mu\nu} + T_{\nu\mu}\right)\, .
\end{align}
Tensors
may be written in ``index-free" notation.
They can be written in components in a coordinate basis of  appropriate rank as,
e.g., $v = v^\mu\partial_\mu$.

\section{Proca fields coupled to gravity}\label{sec:setup}
\subsection{Proca field action and equations of motion}
We consider a $4$-dimensional Lorentzian manifold $\mathcal{M}$, equipped with metric $g_{\mu\nu}$ which is  minimally coupled to a complex Proca field $A_\mu$ of mass $\mu$.
We may associate the latter with a field-strength tensor
\begin{align}
\label{eq:FmunuForm}
F_{\mu\nu} & \coloneq (\dif A)_{\mu\nu}
    = \nabla_{\mu} A_{\nu} - \nabla_{\nu} A_{\mu}
\,.
\end{align}

Assuming
Einstein-Hilbert dynamics for the metric,
the Einstein--Proca action takes the form
\begin{align}
\label{eq:ActionEinsteinProca}
S & = \int \dif^{4}x \sqrt{-g} \left(
    \frac{R}{2 \kappa} + \Lie_{\rm{Proca}}
\right)
\,,
\end{align}
where
$R$ is the Ricci scalar of the metric $g_{\mu\nu}$
and the gravitational coupling is $\kappa=8\pi = M_{\text{pl}}^{-2}$, where $M_{\text{pl}}$ is the Planck mass.
The Proca Lagrangian is given by
\begin{align}
\label{eq:ProcaLagrangian}
\Lie_{\rm{Proca}} & =
    -\frac{1}{4} F^{\ast}_{\rho\sigma}F^{\rho\sigma} - \frac{\mu^2}{2}A^{\ast}_\rho A^\rho
\,.
\end{align}
From the matter sector of the Einstein--Proca action in Eq.~\eqref{eq:ActionEinsteinProca},
we find the energy-momentum tensor
\begin{align}
\label{eq:ProcaTmunu}
T_{\mu\nu} & =
    F_{(\mu}{}^{\rho} F^{\ast}_{\nu)\rho}
+ \mu^2 A^{\ast}_{(\mu} A^{}_{\nu)} + g_{\mu\nu} \Lie_{\rm{Proca}}
\,.
\end{align}
Extremizing the action in Eq.~\eqref{eq:ActionEinsteinProca}
with respect to the vector field
yields the Proca equation
\begin{align}
\label{eq:ProcaEoM}
\nabla_{\nu} F^{\mu\nu} + \mu^2 A^{\mu} & = 0
\,.
\end{align}
If the mass vanishes, $\mu=0$, we are free
to choose a gauge condition for the vector field.
This gauge freedom is lost for a nonvanishing mass, $\mu\neq0$,
and the Lorenz condition
\begin{align}
    \label{eq:Lorenzgauge}
    \nabla_{\mu}A^{\mu} = {}&0
    \, ,
\end{align}
must be satisfied.
To see this, one may take a covariant derivative of the Proca equation~\eqref{eq:ProcaEoM} and use the identity $\nabla_{\mu} \nabla_{\nu} F^{\mu\nu} = 0$.
This point is worth emphasizing; with its additional mass term, the Proca action loses the invariance under $\mathit{U}(1)$ gauge transformations $A_{\nu} \to A_{\nu} + \partial_{\nu} \phi$ enjoyed by Maxwell's theory
(i.e. the $\mu=0$ Proca case).
Consequently, the redundant gauge degree of freedom of the latter becomes, here, a physical degree of freedom.
Thus, the Proca field has three polarizations.
Treating each of these properly is a central task of this work.

\subsection{Classification of Proca field polarizations}
Several schemes for classifying the Proca field's polarizations are used in the literature. The most common are ``spin projection" $S$, parity, and ``tensorial type". We summarize them in Table~\ref{tab:polarizations_table}.

The spin projection $S$ is related to each mode's total angular momentum $j$ by $j=\ell + S$.
To make this definition consistent with the requirement for a spin-1 field that $|\ell - 1| \leq j \leq \ell + 1$,
the spin
$S$ must take values $\{0, \pm 1\}$.
Polarizations with $S=\pm1$ have even parity and are, therefore, referred to as ``electric'' modes,
while the $S=0$ polarization is a parity-odd or ``magnetic'' mode.
While the spin projection $S$
is a more concise classification, for clarity we more often refer to each polarization by its parity and tensorial type, with the latter given by the lowest-$\ell$ mode in its angular spectrum.\footnote{I.e., a spectrum starting at $\ell = 0$ is ``scalar-type", one starting at $\ell = 1$ is ``vector-type", and so on.}

\begin{table}[!ht]
    \centering
    \caption{Summary of the Proca field's polarization modes}
    \renewcommand{\arraystretch}{1.4}
    \begin{tabular}{@{}p{0.23\linewidth}|p{0.23\linewidth}p{0.23\linewidth}p{0.23\linewidth}@{}}
        \toprule
        Spin Projection &$S=0$ & $S=-1$ & $S=1$\\
        \midrule
        Parity & Odd & Even & Even\\
        Name & Magnetic (B) & Electric (E) & Electric (E)\\
        Lowest multipole & $\ell=1$  & $\ell = 1$  & $\ell = 0$ \\
        Tensorial type & Vector (V)& Vector (V)& Scalar (S)\\
        $\mu \to 0$ limit & Physical & Physical & Pure gauge\\
        \bottomrule
    \end{tabular}
    \label{tab:polarizations_table}
\end{table}

\subsection{Background spacetime}
\label{ssec:backgroundspacetime}
In this paper, we consider a test Proca field propagating on a background \bh{} spacetime.
That is, we work in the decoupling limit, which assumes that the backreaction of the vector field onto the geometry is negligible
and the metric $g_{\mu\nu}$ is a solution to vacuum Einstein equations, $R_{\mu\nu} = 0$.
The self-consistency of this approximation is verified post hoc in App.~\ref{appsec:decoupling_approx}.

We may use the Lorenz condition in Eq.~\eqref{eq:Lorenzgauge},
together with the vacuum Einstein equations,
to rewrite the Proca equation~\eqref{eq:ProcaEoM}
as a wave equation
\begin{align}
\label{eq:procaEqWave}
\left(\nabla_\nu \nabla^\nu - \mu^2\right) A^{\mu} & = 0
\,.
\end{align}
We now clearly see the resemblance between our Proca field model and the better-studied Klein-Gordon scalar field.
The procedure to solve the equation in a \bh{} spacetime follows along much the same lines,
but the additional degrees of freedom introduce complexity into the separation-of-variables procedure.
This is discussed in sections~\ref{sec:vsh} and~\ref{sec:fkks}.

For the remainder of this paper we fix the background spacetime to be the Schwarzschild solution.
The metric in Schwarzschild coordinates $\{t,r,\theta,\phi\}$ is
\begin{align}
\label{eq:sch_metric}
\dif s^2 & =
    - f(r) \dif t^2
    + \frac{1}{f(r)} \dif r^2 + r^2 \dif\Omega^2
\,,
\end{align}
where $f(r) = 1 - \frac{\rS}{r}$,
$\rS = 2M$ is
the Schwarzschild radius,
and $\dif\Omega^{2}=\dif\theta^{2}+\sin^{2}\theta\dif\phi^{2}$.

\section{Prelude: black-hole hair from a time-varying Maxwell field}
\label{sec:massless_case}
\subsection{The Jacobson scalar hair result}
In what is now a classic paper~\cite{Jacobson:1999vr},
Jacobson showed that a scalar test field $\phi$ obeying the massless Klein-Gordon equation in Schwarzschild (as well as Kerr) spacetime admits a nontrivial solution which is regular on the horizon
if it is time-dependent at (spatial) infinity.
Imposing the boundary condition
$\lim_{r\to\infty} \phi = c_\infty t$, with arbitrary constant $c_\infty$,
the scalar field
solution is
\begin{align}
\phi & =  c_\infty \left(t + \rS\log\left(1- \frac{\rS}{r} \right)\right)
\,.
\end{align}
In the $r \to \infty$ limit the field behaves as
\begin{align}
\phi & \approx c_\infty \left(t - \frac{\rS^2}{r}\right)
\, ,
\end{align}
suggesting the interpretation that the \bh{} is endowed with a scalar charge of magnitude $c_\infty \rS^2$.

This solution was suggested in the context of scalar-tensor gravity, where, for certain models, such a scalar-field configuration can exist while maintaining the spacetime's stationarity and asymptotic flatness.
This is not true for regular Einstein-Hilbert gravity, as the field sources a nontrivial energy-momentum tensor. The true final state is thus not accessible by a perturbative analysis, precluding us from fully establishing whether this mechanism leads to a ``hairy" \bh.
However, the time before the spacetime backreaction effects become important can be arbitrarily long, depending on the field's amplitude. This means the Jacobson solution may constitute ``hair" in a weaker sense---this may be effectively the final state of a \bh{} on galactic or even cosmological timescales.

\subsection{Maxwell field}
\label{sec:massless_vector_hair}
In the massless limit $\mu \to 0$, the Proca equation \eqref{eq:ProcaEoM} reduces to the standard Maxwell equation
\begin{equation}
    \label{eq:maxwell_eq}
    \nabla_\nu F^{\mu\nu} = 0\, .
\end{equation}
The most well-known \bh{} solution to the Einstein-Maxwell system is the Reissner-Nordstr\"{o}m metric, describing a charged spherically symmetric \bh{}. Following \cite{Herdeiro:2015waa}, we may recover a perturbative version of the Reissner-Nordstr\"{o}m charge by taking the ansatz $A_\mu dx^\mu = A_t(r)dt$, yielding
\begin{equation}
    \left(\partial_r^2+\frac{2}{r}\partial_r\right)A_t = 0\, .
\end{equation}
This has one constant solution which is pure-gauge, and one physical solution
\begin{equation}
    \label{eq:point_charge_em_sol}
    A_t(r) = -\frac{Q}{r}\, ,
\end{equation}
which is simply the electromagnetic field sourced by a point-charge $Q$. We have thus found the electromagnetic potential of a weakly charged \bh{}, the perturbative limit of the Reissner-Nordstr\"{o}m solution with charge $Q$. To strengthen this claim, we recall that our field satisfies a Gauss law with conserved charge $Q_\Sigma$ on hypersurface $\Sigma$, which takes the form
\begin{align}
    \label{eq:gauss_law_em}
    Q_\Sigma = {}&\frac{1}{8\pi}\oint_{\partial \Sigma} \dif S_{\mu\nu} F^{\mu\nu}\\\nonumber
     = {}&\lim_{r\to\infty}\frac{r^2}{4\pi}\oint_{S^2}\dif \Omega \left(\partial_r A_t - \partial_t A_r\right)\\
      ={}& Q\, .
\end{align}
We thus see that the notions of charge as the coefficient of the $1/r$ term in the field's expansion and as the conserved quantity in a Gauss' law coincide, and our solution indeed describes a charged \bh{}.

We may think this exhausts the set of non-transient, nontrivial solutions for an electromagnetic field in Schwarzschild spacetime, a statement supported by the no-hair conjecture.
In theories that include higher-order interactions, such as generalized Proca theories \cite{Heisenberg:2014rta}, other static solutions can be found with a procedure like this~\cite{Minamitsuji:2016ydr}. However, for the regular Einstein-Maxwell system, Eq. \eqref{eq:point_charge_em_sol} is indeed the only static solution.

More solutions are available, however, if we relax the condition of time independence. In the spirit of the Jacobson scalar hair result, we now take an ansatz $A_\mu dx^\mu = A_\phi(t,r)\dif \phi$, which reduces the Maxwell equation \eqref{eq:maxwell_eq} to
\begin{equation}
    \label{eq:Aphi_EoM_massless}
    \left(-\partial^2_t + \partial^2_{r_*}\right) A_\phi = 0\, ,
\end{equation}
where we have introduced the tortoise coordinate, defined by $\dif r= f \dif r_*$, with Schwarzschild metric function $f(r)$. This is just the standard wave equation, and it admits a general solution in terms of characteristic curves
\begin{equation}
    \label{eq:Aphi_gen_sol}
    A_\phi = h_\text{in}(v) + h_\text{out}(u)\, ,
\end{equation}
where $h_\text{in/out}$ are arbitrary functions and we have introduced the null coordinates
\begin{subequations}
\begin{align}
    v = {}&t + r_*\\
    u = {}&t - r_*\, .
\end{align}
\end{subequations}
The former is constant along ingoing null curves, meaning it is automatically regular on the horizon. The solution $h_\text{in}(v)$ is thus also automatically regular on the horizon. Importantly, we can check that it sources a regular energy-momentum tensor. For a timelike observer freefalling from rest at spatial infinity, the density and radial pressure are
\begin{align}
    \rho_\text{in} = P_{r,\text{in}} = {}&\frac{\csc^2\theta}{r\left(\sqrt{\rS} + \sqrt{r}\right)^2}|h_\text{in}(v)|^2
\end{align}
These are clearly nonzero and regular on the horizon, meaning the solution is physically meaningful in some sense, though the divergence at the spherical poles limits their applicability for anything beyond toy models. For additional insight, we may calculate the charge associated with these solutions, finding for all $h_\text{in}(v)$ that
\begin{equation}
    Q_\Sigma = \frac{1}{8\pi} \oint_{\partial\Sigma} \dif S_{\mu\nu}F^{\mu\nu} = 0\, .
\end{equation}
The solutions in this class thus corresponds to long-lasting nontrivial buildups of massless vector field in Schwarzschild spacetime which, nonetheless, contribute no additional electromagnetic charge.

The point of this exercise is that there exist long-lived nontrivial solutions for the massless vector beyond the usual Reissner-Nordstr\"{o}m charge which are regular on the horizon. The trick here was to allow the field a nontrivial dependence on $t$, manifesting in a time-dependent boundary condition at spatial infinity. In this sense, we may consider these solutions a vector analog of Jacobson's time-dependent massless scalar Schwarzschild hair~\cite{Jacobson:1999vr}. While they do not carry additional electromagnetic charge, they source a nontrivial energy-momentum tensor that is described by an arbitrarily large number of parameters, depending on the choice for the function $h_\text{in}(v)$. They thus constitute ``hairy" solutions in a perturbative sense. One may then ask whether a system with this setup will settle into a hairy final configuration once backreaction effects are included.
The answer to this question is not yet clear. As we continue on to the more complicated, more interesting, and more realistic massive case, the same questions persists.

\section{Separation methods}
\label{sec:separation_methods}

While simple in form, the Proca equation~\eqref{eq:procaEqWave} in Schwarzschild spacetime is really a complicated system of partial differential equations and yields few immediate solutions outside of direct numerical integration.
To solve the system
semi-analytically, we turn to separation-of-variables schemes, of which several are well-developed for \bh{} spacetimes.

\begin{figure}
    \centering
    \includegraphics[width=\linewidth]{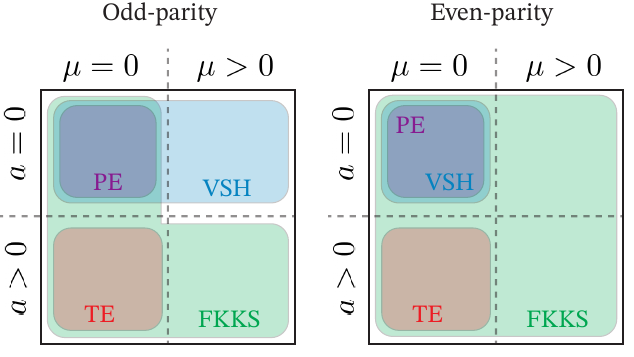}
    \caption{Cases in which each of the four separation methods discussed---the Price equation (PE), the Teukolsky equation (TE), vector spherical harmonics (VSH), and the Frolov-Krtou\v{s}-Kubiz\v{n}\'{a}k-Santos (FKKS) method---separate the Proca equation into a set of fully decoupled equations. Here, $a$ is the \bh{} spin parameter from the Kerr metric, which is $0$ for Schwarzschild, and $\mu$ is the Proca field mass, meaning $\mu=0$ is the electromagnetic case. In the massive case, the two parity sectors have differing behavior, changing the applicability of the \vsh and \fkks approaches.}
    \label{fig:separation_diagram}
\end{figure}

The study of separation schemes for vector perturbations of \bh{s} goes back to at least the early 1970s. Using the null-tetrad scheme of Newman and Penrose~\cite{Newman:1961qr}, Price and Teukolsky, in close succession to each other, presented separation schemes for electromagnetic fields in Schwarzschild \cite{Price:1972pw} and Kerr \cite{Teukolsky:1972my, Teukolsky:1973ha} spacetimes, respectively. These methods, referred to as the Price equation and the Teukolsky equation respectively, work well for massless fields, capturing all electromagnetic degrees of freedom. However, they break in the massive case. This has precipitated the development of more modern schemes; chief among them are the \vsh approach of Refs. \cite{Konoplya:2005hr,Rosa:2011my} and the \fkks approach of Refs. \cite{Krtous:2018bvk, Frolov:2018ezx}. A key complicating factor in the massive case is that the even- and odd-parity sectors have differing dynamics. This is reflected clearly in the \vsh approach, which separates the odd-parity sector into a single equation but fails to separate the electric modes, which we show. Conversely, the \fkks approach, while shown in Ref. \cite{Dolan:2018dqv} to capture all polarizations in the spinning \bh{} case, fails to recover the odd-parity polarization in exact Schwarzschild. The ``successes" and ``failures" of each of these approaches are shown schematically in Fig. \ref{fig:separation_diagram}.

Since we are interested in a massive field in Schwarzschild spacetime, we use a patchwork of the \vsh and \fkks approaches, using the former for the magnetic and the latter for the electric polarizations. This allows us to recover all of the field's degrees of freedom. We describe 
the two approaches in the following sections.

\subsection{Vector spherical harmonics approach}
\label{sec:vsh}

The first we consider is the \vsh approach developed in Refs. \cite{Konoplya:2005hr,Rosa:2011my}. This method is applicable for a vector field perturbation in any spherically symmetric spacetime. It has been used extensively for \qnm and \qbs calculations in Schwarzschild spacetime and related geometries like Schwarzschild-(anti-)de Sitter spacetimes~\cite{Rosa:2011my,Fernandes:2021qvr}.

We introduce four vector spherical harmonics
\begin{subequations}
\label{eq:vector_spherical_harm}
\begin{align}
    \label{eq:z1_harmonic}
    Z^{1, \ell m}_\mu = {}&\left(1,0,0,0\right) Y^{\ell m}(\theta, \phi)\, ,\\
    \label{eq:z2_harmonic}
    Z^{2, \ell m}_\mu = {}&\left(0, \frac{1}{f},0,0\right) Y^{\ell m}(\theta, \phi)\, ,\\
    \label{eq:z3_harmonic}
    Z^{3, \ell m}_\mu = {}&\frac{r}{\sqrt{\ell(\ell+1)}}\left(0,0,\partial_\theta, \partial_\phi\right) Y^{\ell m}(\theta, \phi)\, ,\\
    \label{eq:z4_harmonic}
    Z^{4, \ell m}_\mu = {}&\frac{r}{\sqrt{\ell(\ell+1)}}\left(0,0, \frac{1}{\sin{\theta}}\partial_\phi, -\sin{\theta}\partial_\theta\right) Y^{\ell m}(\theta, \phi)
    \,,
\end{align}
\end{subequations}
where $Y^{\ell m}(\theta, \phi)$ are spin-0 spherical harmonics.
The vector spherical harmonics satisfy the orthonormality condition
\begin{equation}
    \int \dif \Omega\, \mathcal{T}^{\mu\nu} \left(Z^{i, \ell m}_\mu\right)^* Z^{i',\ell'm'}_\nu = \delta^{i i'} \delta^{\ell \ell'} \delta^{m m'}\, ,
\end{equation}
where
$\mathcal{T}^{\mu \nu} = \text{diag}\left(1, f^2, r^{-2}, r^{-2} \sin^{-2}\theta \right)$.
They form a basis for all $4$-vectors, and thus constitute a complete and orthonormal set.

It is helpful to consider the behavior of these harmonics under parity inversion. The action of the parity operator $\mathcal{P}$ on an arbitrary tensor $T$ is $\mathcal{P}:T(x) \to \mathcal{P}T(x)=T(x')$. In spherical coordinates, $x^\mu=(t,r,\theta,\phi)$ and $x'^\mu = (t,r,\pi-\theta,\phi+\pi)$. Under parity inversion, the harmonics in Eq. (\ref{eq:vector_spherical_harm}) transform as
\begin{subequations}
\begin{align}
    \left(Z_\mu^{1,\ell m}, Z_\mu^{2,\ell m}, Z_\mu^{3,\ell m}\right) \to {}&(-1)^\ell \left(Z_\mu^{1,\ell m}, Z_\mu^{2,\ell m}, Z_\mu^{3,\ell m}\right)\\
    Z_\mu^{4,\ell m} \to {}& (-1)^{\ell + 1} Z_\mu^{4, \ell m}\, .
\end{align}
\end{subequations}
The vector basis thus naturally splits into an even-parity $(i=1,2,3)$ and an odd-parity $(i=4)$ sector. 
Any vector field may be split into its even- and odd-parity components, and each sector may be fully captured by its decomposition in the \vsh of the same parity.
We categorize degrees of freedom by their parity, referring to even-parity, ``electric", or ``E" modes and odd-parity ``magnetic", or ``B" modes.

In the massless ($\mu \to 0$) limit, the \vsh approach reproduces the expected dynamics of a Maxwell field~\cite{Rosa:2011my}, complete with the scalar-type electric polarization becoming removable via a gauge transformation.

Having demonstrated the completeness and orthonormality of our basis, we may now expand an arbitrary vector field $A_\mu$ in \vsh as
\begin{equation}
    \label{eq:vsh_ansatz}
    A_\mu(t,r,\theta,\phi) = \frac{1}{r}\sum^4_{i=1} \sum_{\ell,m} \gamma_i u^{\ell m}_{i}(t,r) Z^{i,\ell m}_{\mu}(\theta, \phi)
    \, ,
\end{equation}
where the constants $\gamma_i$ are given by
\begin{align}
    \gamma_1 = {}&\gamma_2 = 1\,,\quad
    \gamma_3 = \gamma_4 = \frac{1}{\sqrt{\ell(\ell+1)}}\, .
\end{align}
Taking this expansion as an ansatz separates the Proca equation (\ref{eq:procaEqWave}) into an even- and an odd-parity sector, each of which is decoupled from the other.

\subsubsection{Separated equations}

Taking the \vsh expansion in Eq.~\eqref{eq:vsh_ansatz} as an ansatz for the vector field $A_\mu$ in the Proca equation (\ref{eq:procaEqWave}) yields four equations for the mode functions $u_i^{\ell m}(t,r)$:
\begin{subequations}
\begin{align}
    \label{eq:u1_mode_eq}
    0 = {}&\hat{\mathcal{D}}_2 u_1 + \frac{\rS}{r^2} \left(\partial_t u_2 - \partial_{r_*} u_1 \right)\\
    \label{eq:u2_mode_eq}
    0 = {}&\hat{\mathcal{D}}_2 u_2 + \frac{2 f^2}{r^2}\left(u_3 - u_2\right) + \frac{\rS}{r^2}\left(\partial_t u_1 - \partial_{r_*}u_2\right)\\
    \label{eq:u3_mode_eq}
    0 = {}&\hat{\mathcal{D}}_2 u_3 + 2f\frac{\ell(\ell+1)}{r^2} u_2\\
    \label{eq:u4_mode_eq}
    0 = {}&\hat{\mathcal{D}}_2 u_4\, .
\end{align}
\end{subequations}
Here, we have introduced the operator
\begin{equation}
    \label{eq:D2_operator}
    \hat{\mathcal{D}}_2 = -\partial_t^2 + \partial_{r_*}^2 - f\left(\frac{\ell(\ell+1)}{r^2} + \mu^2\right)\, ,
\end{equation}
and the tortoise coordinate $r_*$, defined by $\dif r=f \dif r_*$. We may supplement these with the Lorenz condition (\ref{eq:Lorenzgauge}), which takes the form
\begin{align}
    \label{eq:uLorenz}
    0 = {}& -\partial_t u_1 + \partial_{r_*}u_2 + \frac{f}{r}\left(u_2 - u_3\right)\, .
\end{align}
Multiplying this equation by $\rS/r^2$ and adding it to the $u_2$ mode equation (\ref{eq:u2_mode_eq}) yields
\begin{align}
    \label{eq:u2_lorenz_eq}
    0 = {}& \hat{\mathcal{D}}_2 u_2 + \frac{f }{r^3}(3\rS - 2r)\left(u_2 - u_3\right)\, ,
\end{align}
which forms a closed system with (\ref{eq:u3_mode_eq}). We have thus reduced the Proca equation to two mutually decoupled systems: an even-parity sector, Eqs. (\ref{eq:u2_lorenz_eq}, \ref{eq:u3_mode_eq}), and an odd-parity sector, Eq. \eqref{eq:u4_mode_eq}. Note that $u_1$ is now manifestly non-independent, and it can be determined from $u_2$ using Eq. \eqref{eq:u1_mode_eq}. In the limit $\mu \to 0$, we notice that the odd-parity equation with $\ell=0$ reduces to the massless monopole $A_\phi$ equation \eqref{eq:Aphi_EoM_massless}.\footnote{Formally, the $\ell=0$ mode for $Z_\mu^{4, \ell m}$ is not defined, as we can see in \eqref{eq:z4_harmonic}. However, it is still useful to see that the massless monopole dynamics can be reproduced from the $u_4$ mode equation.} In the general case, the odd-parity equation may be trivially reduced to an \ode by taking the ansatz $u_4 = e^{-i \omega t} R(r)$, giving
\begin{equation}
    \label{eq:odd-par_radial_eq}
    \left(\partial_{r_*}^2 + \omega^2 - f\left(\frac{\ell(\ell+1)}{r^2} + \mu^2\right)\right)R(r) = 0\, .
\end{equation}
Using the \vsh approach, we can thus describe the entire odd-parity sector with a single \ode, which is precisely what we desire. For the even-parity sector, however, \vsh yields two coupled equations, meaning we need an alternative treatment to achieve full separation.

\subsection{Frolov-Krtou\v{s}-Kubiz\v{n}\'{a}k-Santos approach}
\label{sec:fkks}

The \vsh approach, while very useful for the field's odd-parity sector in spherically symmetric settings, is very limited for more general spacetimes or for close study of the even-parity sector. Since we are interested in the latter, we turn now to the separation-of-variables approach of Frolov, Krtou\v{s}, Kubiz\v{n}\'{a}k, and Santos (FKKS)~\cite{Krtous:2018bvk, Frolov:2018ezx}. This approach reduces the Proca equation to a set of decoupled \ode{s} for any background spacetime in the highly general Kerr-NUT-(A)dS class. It has been used to study quasi-normal modes and superradiant instabilities in Kerr background spacetimes (e.g.,  \cite{Dolan:2018dqv,Dolan:2019hcw,Frolov:2018ezx, Percival:2020skc}). Here, we further specialize it to a Schwarzschild background.

\subsubsection{The principal tensor}

The \fkks approach relies on the existence of a Killing object known as the principal tensor, $h$, a non-degenerate closed conformal Killing-Yano 2-form. For a thorough discussion, see Ref.~\cite{Frolov:2017kze}.

A closed conformal Killing-Yano $k$-form $\omega$ is a differential form whose covariant derivative $\nabla \omega$ is given exclusively by its divergence part, or
\begin{equation}
    \label{eq:close_conf_kill_yano_form_def}
    \nabla_\nu\omega_{\mu_1 \ldots \mu_k} = \frac{k}{D-k+1} g_{\nu[\mu_1}\nabla^\rho \omega_{|\rho|\mu_2 \ldots \mu_k]}\, ,
\end{equation}
where $D$ is the spacetime dimension. The form is called ``conformal" because its derivative $\nabla \omega$ lies in the kernel of the twistor operator. Such forms also have a vanishing antisymmetric part, $\nabla_{[\nu} \omega_{\mu_1,\ldots,\mu_k]}=0$, from which it follows that $\dif \omega=0$, hence the label ``closed".

The principal tensor $h$ is a non-degenerate closed conformal Killing-Yano 2-form. As such, it obeys Eq. (\ref{eq:close_conf_kill_yano_form_def}), which now takes the form
\begin{equation}
\nabla_\sigma h_{\mu\nu} = \frac{1}{D-1} \left(g_{\sigma \mu} \nabla^\rho h_{\rho \nu} - g_{\sigma \nu} \nabla^\rho h_{\rho \mu}\right)\, .
\end{equation}
Additionally, ``non-degenerate" means its matrix form $h^{\mu}{}_\nu$ should have no repeated eigenvalues.

The key property of the principal tensor is that it generates a complete set of explicit and hidden symmetries in the spacetimes for which it exists. The former are symmetries associated with Killing vectors, while the latter are associated with higher-rank Killing tensors. A complete ``tower" of both can be built using only the principal tensor and the metric.

The principal tensor of Kerr spacetime is well-defined, and in Boyer-Lindquist coordinates it takes the form
\begin{align}
    \nonumber
    h = {}&r \dif t \wedge \dif r+ a r \sin^2\theta\, \dif r\wedge \dif \phi\\
    {}&- a\cos\theta \sin\theta\, \dif \theta \wedge \left(a \dif t - \left(a^2 + r^2\right) \dif \phi\right)\, ,
    \label{eq:princ_tens_bl_coords}
\end{align}
where $a$ is the Kerr spin parameter. This has eigenvalues $\{\pm r, \pm i a \cos \theta \}$. In the $a \to 0$ limit, it takes the form
\begin{equation}
    \label{eq:schwarzschild_princ_tens}
    h = r \dif t \wedge\dif r\, ,
\end{equation}
with corresponding eigenvalues $\{\pm r, 0 \}$. The repeated $0$-eigenvalue means this tensor is now degenerate. Equivalently, one can check that $h\wedge h = 0$. Having lost the non-degeneracy condition, $h$ is no longer a principal tensor in the proper sense. This restricts which Proca degrees of freedom can be captured with the \fkks approach in Schwarzschild spacetime. In particular, this treatment recovers only the field's electric polarizations, not its magnetic polarization. However, as shown in section \ref{sec:vsh}, the latter can be recovered with the \vsh approach, allowing us to find solutions for all polarizations.

\subsubsection{Separable ansatz}

We now demonstrate that the \fkks approach can indeed reduce the Proca equation to a set of separated \ode{s} in Schwarzschild spacetime. We adopt the ansatz
\begin{equation}
    \label{eq:fkks_ansatz}
    A^\mu = B^{\mu\nu}\nabla_\nu Z\, .
\end{equation}
We refer to $B^{\mu\nu}$ as the polarization tensor, defined by
\begin{equation}
    \label{eq:polariz_tensor_eq}
    B^{\mu\rho}\left(g_{\rho\nu} + i\nu h_{\rho\nu}\right) = \delta^\mu_\nu\, ,
\end{equation}
where $\nu$ is an undetermined separation constant. This reduces the problem to a single PDE for the scalar $Z$. To see this, we borrow a result from \cite{Krtous:2018bvk} that
\begin{align}
    \nabla_\nu F^{\mu\nu} = -B^{\mu\nu}\nabla_\nu \left(\nabla_\rho \nabla^\rho Z - 2 i \nu \xi^\rho A_\rho\right)\, ,
    \label{eq:proca_fkks}
\end{align}
where we've introduced the timelike Killing vector $\xi^\mu\partial_\mu = \partial_t$. The Proca equation (\ref{eq:ProcaEoM}) is thus satisfied if
\begin{equation}
    \label{eq:Z_eq}
    \left(\nabla_\mu \nabla^\mu - 2 i \nu \xi_\mu B^{\mu\nu} \nabla_\nu - \mu^2\right) Z = 0\, .
\end{equation}
We have thus reduced the Proca equation to a wave-type equation for the scalar $Z$.

\subsubsection{Separated equations}

We may use Eq. \eqref{eq:polariz_tensor_eq} with the Schwarzschild principal tensor from Eq. \eqref{eq:schwarzschild_princ_tens} to solve for the polarization tensor in Schwarzschild spacetime. It takes the form
\begin{equation}
    B^{\mu\nu} =
    \begin{pmatrix}
        \frac{-1}{f q_r} {}& \frac{i r \nu}{q_r} {}& 0 {}& 0\\
        \frac{-i r \nu}{q_r} {}& \frac{f}{q_r} {}& 0 {}& 0\\
        0 {}& 0 {}& \frac{1}{r^2} {}& 0\\
        0 {}& 0 {}& 0 {}& \frac{\csc^2\theta}{r^2}
    \end{pmatrix}\, ,
\end{equation}
where $q_r = 1+\nu^2r^2$. Since Schwarzschild spacetime does not admit a non-degenerate principal tensor, it is not a priori clear that the wave equation (\ref{eq:Z_eq}) will separate like in Kerr. We now show that it does, in fact, yield separated equations even if the entire calculation is done in Schwarzschild, with the restriction that the magnetic polarization is not recovered.

We take an ansatz for $Z$,
\begin{align}
    \label{eq:Z_ansatz}
    Z = {}& R(r) S(\theta) e^{-i\omega t} e^{i m \phi}\, ,
\end{align}
with parameters $\omega$ and $m$, which we interpret as the mode's (complex) frequency and magnetic quantum number respectively. We seek solutions which are regular everywhere, so we restrict to $m \in \mathbb{Z}$. This allows us to separate Eq. (\ref{eq:Z_eq}) into a radial and an angular equation
\begin{subequations}
\begin{align}
    \nonumber
    0 = {}&q_r \frac{\dif}{\dif r}\left(\frac{r^2 f}{q_r} R'(r)\right)\\
    {}&+ \left(\frac{r^2}{f}\left(\omega^2 - \frac{2 \nu \omega f}{q_r}\right) - \mu^2 r^2 -\kappa_1\right)R (r)
    \label{eq:R_proca_eq_unfixed}\\
    0 = {}&\csc\theta \frac{\dif}{\dif \theta}\left(\sin \theta S'(\theta)\right) + (\kappa_1 - m^2 \csc^2\theta) S(\theta)\, ,
    \label{eq:S_proca_eq_unfixed}
\end{align}
\end{subequations}
where $\kappa_1$ is a separation constant. To fix the separation constant, we consider the (automatically enforced) Lorenz condition, Eq. \eqref{eq:Lorenzgauge}, which separates into
\begin{subequations}
\begin{align}
    \nonumber
    0 = {}&\frac{\dif}{\dif r}\left(\frac{r^2 f}{q_r} R'(r)\right)\\
    {}& + \left(\frac{r^2}{f q_r} \left(\omega^2 - \frac{\nu \omega f}{q_r}\left(q_r+2\right)\right) - \kappa_2 \right) R(r)
    \label{eq:R_lorenz_eq}\\
    0 = {}&\csc\theta \frac{\dif}{\dif \theta}\left(\sin \theta S'(\theta)\right) + (\kappa_2 - m^2 \csc^2\theta) S(\theta)\, .
    \label{eq:s_lorenz_eq}
\end{align}
\end{subequations}
Subtracting the radial Lorenz equation (\ref{eq:R_lorenz_eq}) (multiplied by $q_r$) from the radial part of the Proca equation (\ref{eq:R_proca_eq_unfixed}) yields the consistency relation
\begin{equation}
    \kappa_2 - \kappa_1 + r^2\nu^2\left(\kappa_2 - \frac{\mu^2}{\nu^2} + \frac{\omega}{\nu}\right) = 0\, ,
\end{equation}
from which we may read off the separation constants
\begin{align}
    \label{eq:sep_constants}
    \kappa_1 {}&= \kappa_2 = \frac{\mu^2}{\nu^2} - \frac{\omega}{\nu}\, .
\end{align}
Imposing Eq. \eqref{eq:sep_constants}, we reduce both the Proca system (\ref{eq:R_proca_eq_unfixed}, \ref{eq:S_proca_eq_unfixed}) and the Lorenz system (\ref{eq:R_lorenz_eq}, \ref{eq:s_lorenz_eq}) to the same set of equations, which take the form
\begin{subequations}
\begin{align}
    \label{eq:fkks_theta_eq}
    0 = {}&\frac{q_r f}{r^2} \frac{\dif}{\dif r}\left(\frac{r^2 f}{q_r} R'(r)\right)\\\nonumber
    \label{eq:fkks_r_eq}
    {}&+ \left[\omega^2 - f\left(\frac{\nu \omega}{q_r}\left(q_r+2\right) - \frac{q_r}{r^2} \left(\frac{\mu^2}{\nu^2} - \frac{\omega}{\nu}\right)\right) \right]R(r)\\
    0 = {}&\csc\theta \frac{\dif}{\dif \theta}\left(\sin \theta S'(\theta)\right) + \left( \frac{\mu^2}{\nu^2}-\frac{\omega}{\nu}-m^2\csc^2\theta \right) S(\theta)\, .
\end{align}
\end{subequations}
We identify the latter as the standard spherical harmonic equation with
\begin{equation}
    \label{eq:nu_charac_eq}
    \ell(\ell+1) = \frac{\mu^2}{\nu^2}-\frac{\omega}{\nu}\, ,
\end{equation}
which has solutions $S(\theta) = P^{\ell m}(\cos\theta)$, where $P^{\ell m}$ are the Legendre polynomials. The characteristic polynomial in Eq. (\ref{eq:nu_charac_eq}) has $\ell \neq 0$ solutions
\begin{equation}
    \label{eq:nu_sol_S}
    \nu_\pm = -\frac{\omega \pm \sqrt{4\mu^2\ell(\ell+1) + \omega^2}}{2\ell(\ell+1)}\, ,
\end{equation}
and a monopole ($\ell = 0$) solution
\begin{equation}
    \label{eq:nu_sol_monopole}
    \nu_0 = \frac{\mu^2}{\omega}\, .
\end{equation}
We interpret the two branches of $\ell \neq 0$ solutions in Eq. (\ref{eq:nu_sol_S}) as each corresponding to a polarization of the field. As shown in Refs. \cite{Frolov:2018ezx, Dolan:2018dqv, Fernandes:2021qvr}, they are both even-parity (``electric") polarizations. To understand them, we consider the massless ($\mu \to 0$) limit. The system should reduce to a Maxwell field with two vector-type (i.e. $\ell \geq 1$) polarizations. In this limit, $\nu_+$ remains well-defined, so we identify it with the Maxwell field's electric polarization. Conversely, $\nu_-$ vanishes, so we identify it with the scalar-type polarization which is physical in the Proca case but becomes pure-gauge when $\mu=0$.

Combining Eqs.~\eqref{eq:nu_charac_eq} and~\eqref{eq:fkks_r_eq}, we find a radial equation for all even-parity modes
\begin{align}
    \label{eq:fkks_radial_eq_even}
    0 = {}& \frac{q_r f}{r^2} \frac{\dif}{\dif r}\left(\frac{r^2 f}{q_r} R'(r)\right)\\\nonumber
    {}&+ \left(\omega^2 - f\left(q_r \frac{\ell(\ell+1)}{r^2} + \frac{\nu \omega}{q_r}\left(q_r + 2\right) \right)\right)R (r)\, .
\end{align}
We now proceed with solving this equation for all values of $\nu$.

\section{Solutions}
\label{sec:solutions}

We now proceed to solve Eqs.~\eqref{eq:odd-par_radial_eq} and~\eqref{eq:fkks_radial_eq_even} using a mixture of analytical and numerical schemes.

\subsection{Boundary conditions}

We seek to model a \bh{} surrounded by a cosmological condensate of cold vector dark matter. Due to the separation of scales between the \bh{'s} region of influence and the timescale of cosmological expansion, we may simply treat the condensate as an infinite bath of particles with steady time dependence $A_\mu \sim e^{-i \mu t}$. We thus take this as our boundary condition far from the horizon. Near the horizon, we use a ``pure-ingoing" boundary condition, as is standard in \bh{} perturbation theory.

\subsubsection{Near horizon}

We have now reduced the Proca equation in Schwarzschild spacetime to that of solving a second-order linear \ode for each angular mode of each field polarization. For the odd-parity sector this equation is (\ref{eq:odd-par_radial_eq}), while for the even-parity sector this equation is (\ref{eq:fkks_radial_eq_even}). In the near-horizon ($r_* \to -\infty$) limit, both equations have the general solution
\begin{equation}
    \label{eq:R_near_horizon_general}
    R(r) = c_1 e^{i \omega r_*} + c_2 e^{-i \omega r_*}\, .
\end{equation}
We may interpret the two terms as representing outgoing and ingoing waves respectively. As a basic physical requirement, we require solutions to be regular across the \bh{} horizon. As such, we must disregard the first term (i.e. set $c_1 = 0$) which blows up at the horizon for any timelike or null observer. More heuristically, nothing comes out of the horizon of a classical \bh, so we disregard the outgoing wave solution.

\subsubsection{Far from horizon}

We have now fixed the field's horizon boundary condition. In a standard \qnm or \qbs calculation, 
one would then fix a pure-outgoing or pure-ingoing boundary condition at spatial infinity ($r_* \to \infty$). This step overconstrains the system, and solutions cease to exist for all except a discrete set of (complex) frequencies $\{\omega_n^{\ell m}\}$ \cite{Berti_2009}. The problem is thus reduced to an eigenvalue problem for this frequency spectrum, and properties like the stability and oscillatory period of each mode can be read off from the imaginary and real parts of its frequency.

In this work, we do something different. If the \bh{} is submerged in a universe filled with vector dark matter, then it can pull from an effectively infinite bath of particles. Rather than decaying like the usual Schwarzschild \qbs, we thus expect our solutions to reach a steady-state which oscillates in tandem with the surrounding dark matter at a real frequency. Following this, we set the imaginary part of $\omega$ to zero, enforcing that the field neither grows nor decays, and we set the real part to match the oscillation frequency of the surrounding dark matter condensate.

We have a freedom here to set the real oscillation frequency of the dark matter bath. In a background that is approximately flat, the field in general propagates as a superposition of waves with frequencies $\omega = \sqrt{\vec{k}^2+\mu^2}$, where $\vec{k}$ is the wavenumber of each mode. As a realistic model, however, it is standard in cosmology to consider a ``homogeneous" configuration $A_\mu = A_\mu(t)$, or a field with only its $\vec{k} = 0,\, \omega=\mu$ mode populated \cite{Turner:1983he, Arias:2012az, Alonso-Alvarez:2019ixv, Nakayama:2019rhg}. This model causes the field to behave as cold dark matter, and it can be seen as a fully non-relativistic approximation. For our model, we take the field to asymptotically approach this vector cold dark matter model, meaning we set the oscillation frequency to the mass of the field, $\omega = \mu$.
This in general produces a superposition of ingoing and outgoing waves far from the \bh. We explore this in more detail later in this section.

\subsection{Odd-parity sector}

We first turn to the field's magnetic polarization. Recall that, using the \vsh approach, we reduced the Proca equation to a single \ode for the radial variable $R(r)$ in Eq. (\ref{eq:odd-par_radial_eq}). 
It also applies for the entire odd-parity sector.

\subsubsection{Exact solution}

We now solve the magnetic radial equation \eqref{eq:odd-par_radial_eq} exactly. This equation is linear and second order, and it is characterized by three singular points\footnote{Linear \ode{s} can be characterized by the singular points in their coefficient functions. For instance, a hypergeometric equation has three regular singular points, and a Heun equation has four regular singular points.}: two regular singularities at $r=0$ and $r=\rS$, and an irregular singularity at $r = \infty$. As such, it is of the confluent Heun class of equations, and we expect it to have solutions in terms of confluent Heun functions. To see this more clearly, we introduce the coordinate transformation $r \to z$ and the function redefinition $R \to W$ defined by
\begin{align}
    R(r) \eqcolon r^2 (r-\rS)^{i\omega \rS} e^{i k r} W(r), \quad z\coloneq 1-r/\rS\, ,
\end{align}
where $k = \sqrt{\omega^2-\mu^2}$. This reduces Eq. \eqref{eq:odd-par_radial_eq} to
\begin{equation}
    \label{eq:odd-par_radial_eq_trans2}
    W'' + \left( A + \frac{C + 1}{z-1} + \frac{B + 1}{z} \right) W' + \left(\frac{N}{z-1} + \frac{M}{z} \right) W = 0\, ,
\end{equation}
which is the confluent Heun equation in standard form~\cite{2010JPhA...43c5203F, Hortacsu:2011rr}. Here,
\begin{subequations}
\begin{align}
    M = {}&\frac{1}{2} (A -B -C + AB - BC)-E\, ,\\
    N = {}&\frac{1}{2} (A + B + C + AC + BC) + D + E \, ,
\end{align}
\end{subequations}
and
\begin{subequations}
\begin{align}
    A = {}&-2 i k \rS\, ,\\
    B = {}&2 i \omega \rS\, ,\\
    C = {}&2\, ,\\
    D = {}&- \left(\omega^2 + k^2\right) \rS^2\, ,\\
    E = {}& \left(\omega^2 + k^2\right) \rS^2 - \ell(\ell+1) + 1\, .
\end{align}
\end{subequations}
From here immediately follows the exact general solution
\begin{align}
    W(z) = {}&c_1 h_\text{out}(z) + c_2 z^{-B}e^{-A z} h_\text{in}(z)\, ,
\end{align}
where
\begin{subequations}
\begin{align}
    h_\text{out}(z) \coloneq &\text{HeunC}(A, B, C, D, E, z)\\
    h_\text{in}(z)\coloneq &\text{HeunC}(-A, -B, C, D, E, z)\, .
\end{align}
\end{subequations}
Converting back to $R(r)$, for the  original odd-parity radial equation (\ref{eq:odd-par_radial_eq}) we find the general solution
\begin{align}
    \label{eq:R_odd-par_gen_sol}
    R(r) = {}&c_1 r^2 (r-\rS)^{i \omega \rS} e^{i k r} h_\text{out}(1-r/\rS)\\\nonumber
    {}& + c_2 r^2 (r-\rS)^{-i \omega \rS} e^{-i k r} h_\text{in}(1-r/\rS)\, .
\end{align}
By definition, in the vicinity of $r=\rS$ the HeunC function approaches $1$. As such, in the near-horizon region $\left(r-\rS\right) \ll \mu \rS^2$, we find
\begin{equation}
    \label{eq:odd-par_near_hor_approx}
    R = \tilde{c}_1(r-\rS)^{i \omega \rS} + \tilde{c}_2 (r-\rS)^{-i \omega \rS}\, ,
\end{equation}
where $\tilde{c}_1$ and $\tilde{c}_2$ are each nontrivial, but depend only on $c_1$ and $c_2$ respectively. These are precisely the ingoing and outgoing waves from Eq. (\ref{eq:R_near_horizon_general}), and we discard the outgoing solution like before, setting $c_1 = 0$. We thus arrive at a final exact solution
\begin{equation}
    \label{eq:R_heun_sol_ingoing}
    R(r) = c_2 r^2 (r-\rS)^{-i \omega \rS} e^{-i k r} h_\text{in}(1-r/\rS)\, .
\end{equation}
We stress that up to to this point, we have made no assumptions about $\omega$, so the set of solutions (\ref{eq:R_heun_sol_ingoing}) for all $\omega$ and $\ell$ thus form a complete basis for all odd-parity solutions to the Proca equation which are regular on the horizon. Included in these are the \qnm and \qbs solutions, corresponding to values of $\omega$ for which this HeunC solution is purely outgoing or purely ingoing respectively in the large-$r$ region.

We may reconstruct the original vector field $A_\mu$ using the \vsh ansatz (\ref{eq:vsh_ansatz}). Keeping only the odd-parity ($i=4$) mode function, we find for each mode $(\ell,m,\omega)$ a solution
\begin{align}
    A_\mu = c_2 r &(r-\rS)^{-i \omega \rS} \\\nonumber
    &e^{-i(\omega t + k r)} h_\text{in}(1-r/\rS) Z_{\mu}^{4, \ell m}\, ,
\end{align}
where we have absorbed a factor of $(\ell(\ell+1))^{-1/2}$ into $c_2$. We have thus formally solved the system exactly for the odd-parity sector. However, this formula is rather opaque. To make sense of it, we both continue our study of its asymptotics and analyze it numerically.

\begin{figure*}
    \centering
    \includegraphics[width=\textwidth]{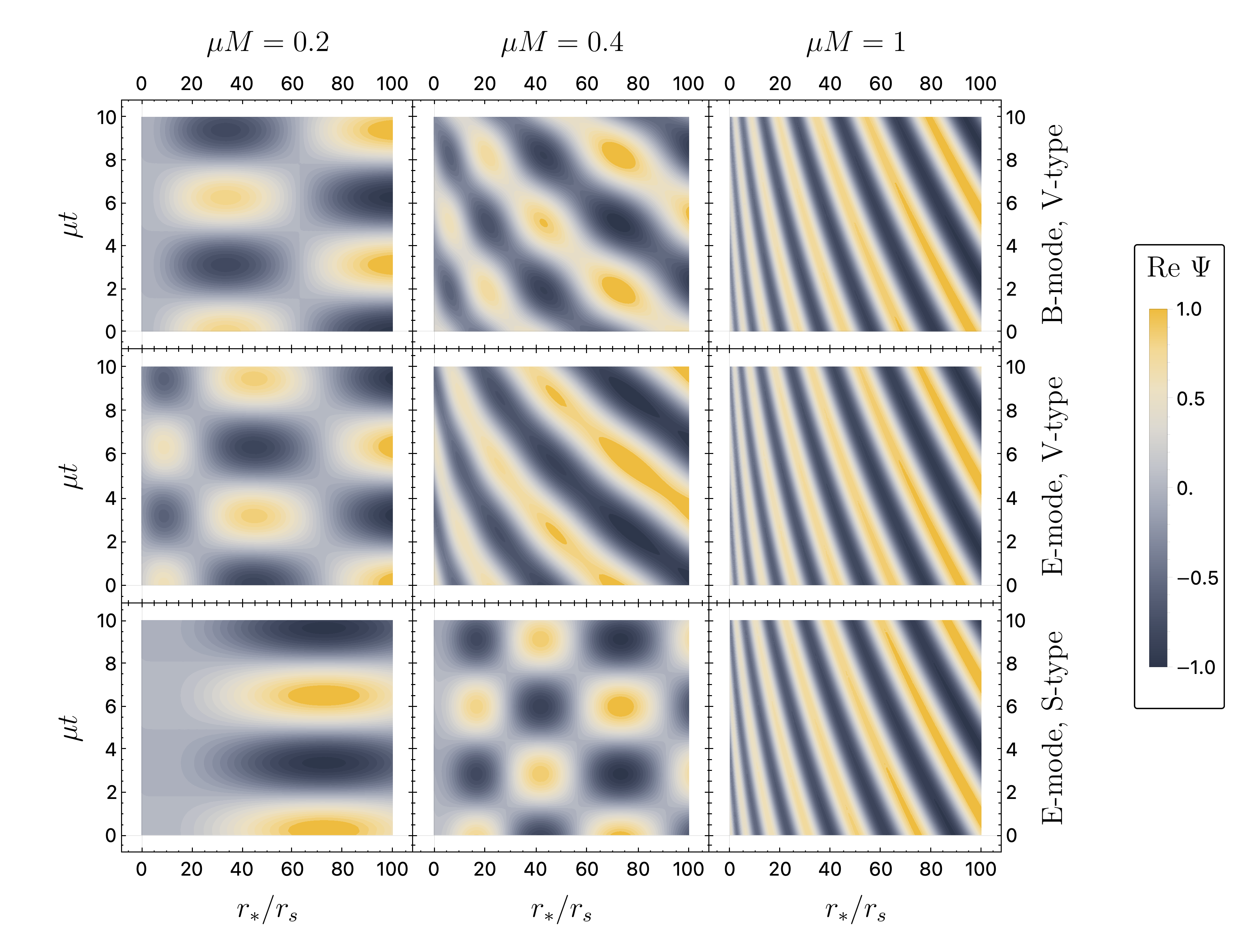}
    \caption{Time evolution of the real part of the radial function $\Psi(t,r) \coloneq R(r)e^{-i \mu t}$ for all three polarizations at $\ell=1$. {Note that $\Psi(t,r)$ is normalized by its maximum value over the plotting region.} The rows show the evolution for the vector-type magnetic mode, the vector-type electric mode, and the scalar-type electric mode respectively. Three values of the mass parameter $\mu M$ are selected, one in the standing-wave regime ($\mu M = 0.2$) regime, one in the intermediate regime ($\mu M = 0.4$) regime, and one in the infalling-wave regime ($\mu M = 1$).}
    \label{fig:psi_evol_plots}
\end{figure*}

\subsubsection{Asymptotic behavior}
To understand the behavior of the field outside the near-horizon region, it is helpful to develop an asymptotic approximation. The precise details of this are given in App. \ref{sec:odd-par_asymp}, but the salient point is that for our cold dark matter bath, with $\omega=\mu$, the radial function has behavior
\begin{equation}
    \label{eq:odd-par_large_r_approx}
    R(r) \approx c_{\text{in}}(\mu)r^{1/4}e^{-2 i \mu \sqrt{r \rS}} + c_{\text{out}}(\mu) r^{1/4} e^{2i\mu \sqrt{r \rS}}\,
\end{equation}
in the $r \gg 1/\left(\rS \mu^2\right)$ region. The coefficients $c_{\text{in}}(\mu)$ and $c_{\text{out}}(\mu)$ depend only on the mass $\mu$ and the exact solution's coefficient $c_2$, but the precise functional form of this dependence is not known. We may, however, approximate it in two principal regimes.

We have derived a small-$r$ approximation in Eq.~\eqref{eq:odd-par_near_hor_approx}, valid in the region $(r-\rS) \ll \mu \rS^2 \approx (\mu M)\rS$,
and a large-$r$ approximation in Eq.~\eqref{eq:odd-par_large_r_approx}, valid in the region $r \gg 1/\rS\mu^2 \approx \rS/(\mu M)^2$.
For $\mu M \gtrsim 1$, these regions overlap.
We may thus directly match the two approximations (\ref{eq:odd-par_near_hor_approx}) and (\ref{eq:odd-par_large_r_approx}), forcing us to keep only the ingoing solution far from the horizon.
For $\mu M < 1$ the two regions cease to overlap, and the large-$r$ behavior becomes more complicated.
However, we find numerically that for smaller mass parameters $\mu M \approx 0.2$, the coefficients $c_{\text{in}}(\mu)$ and $c_{\text{out}}(\mu)$ equilibrate and produce standing waves.
We may thus summarize these behaviors as
\begin{equation}
    \label{eq:odd-par_R_approx_behavior}
    R(r) \sim \begin{cases}
        r^{1/4}e^{-2 i \mu \sqrt{r \rS}} & \mu M \gtrsim 1\\
        r^{1/4}\cos \left(2\mu\sqrt{r \rS}\right) & \mu M < 1
    \end{cases}\, .
\end{equation}
We now understand the qualitative behavior of the odd-parity radial equation. Notably, at this level of approximation the behavior is the same as for the analogous scalar-field model~\cite{Hui:2019aqm}.

\subsection{Even-parity sector}

In Section \ref{sec:fkks}, we reduced the even-parity part of the Proca equation
to a single \ode in the radial function $R(r)$ in Eq.~\eqref{eq:fkks_radial_eq_even}.
Fixing/choosing the multipole $\ell$ fixes the separation constant $\nu$ in Eqs.~\eqref{eq:nu_sol_S} and~\eqref{eq:nu_sol_monopole}.

As in the odd-parity case, the resulting \ode in Eq.~\eqref{eq:odd-par_radial_eq} is linear and second order in $r$.
It also shares the two regular singular points at $r=0$ and $r=\rS$, as well as the irregular singular point at $r=\infty$.
However, it is complicated by the presence of two additional singular points at $r=\pm i/\nu$.
It is thus not of confluent Heun type, and we do not find an exact solution for it.
Instead we construct numerical solutions.
While this may appear to make the even-parity analysis considerably harder than its odd-parity counterpart, we find that the principal behaviors of solutions are similarly straightforward to extract from their asymptotic expansions.
Additionally, despite the complicated singularity structure, the even-parity equation is easy to solve numerically for a wide parameter range, allowing for a precise analysis of solutions and their implications.


\subsubsection{Asymptotic behavior}

At next-to-leading order in $1/r$, the expansion of the even-parity equation (\ref{eq:fkks_radial_eq_even}) matches the expansion of the odd-parity equation (\ref{eq:odd-par_radial_eq}) up to some constant coefficients. It therefore shares the same large-$r$ behavior, given by Eq.~\eqref{eq:odd-par_large_r_approx}.
The details are worked out in App. \ref{sec:even-parity_asymptotics}.

For the $\omega = \mu$ case, the additional singular points $r=\pm i/\nu$ in the even-parity equation do not lie between $r=\rS$ and $r=\infty$. Thus, we may use the same asymptotic matching procedure as in the odd-parity sector to argue that solutions are purely ingoing in the $\mu M \gtrsim 1$ regime. For $\mu M < 1$, we also find that the coefficients of the ingoing and outgoing solutions equilibrate, producing standing waves.
As such, the approximate general behavior in Eq.~\eqref{eq:odd-par_R_approx_behavior} for
$r \gg 1/\left(\rS\mu^2\right)$ applies for the even-parity sector as well as the odd-parity one.

\subsubsection{Numerical solution}

To understand the precise behavior of even-parity solutions, we solve Eq. (\ref{eq:fkks_radial_eq_even}) numerically. For the $\omega = \mu$ case, the radial function has the behavior $R \sim e^{-i \mu r_*}$ near the horizon. As our initial conditions for the numerical problem, we may thus impose that $R(r)$ and its first derivative match this approximation somewhere in the regime $(r-\rS) \ll (\mu M)^2 \rS$. This calculation is straightforward using any standard \ode solver, and we implement it using \textit{Mathematica}'s ``NDSolve". Solutions for a range of masses are shown in the lower two rows of Figure \ref{fig:psi_evol_plots}.

\subsection{Time evolution of solutions}

To further illustrate the behavior of our solutions, it is helpful to plot the behavior of the function $\Psi(t,r) \coloneq R(r) e^{-i \mu t}$, which qualitatively captures the time dependence of $A_\mu$.
We do this in Fig. \ref{fig:psi_evol_plots}, which shows the real part of $\Psi$ over a few radial and temporal periods in several cases {for $\ell = 1$.}
Each row plots one polarization of the field, and each column uses a different mass parameter $\mu M$.

{We select mass parameters
$\mu M = 0.2$, $0.4$, and $1.0$, representing the wave, intermediate, and particle regimes respectively.
In the $\mu M=1.0$ (particle regime) case, we find nearly identical behavior across the three different polarizations.
This shows that effects from the polarizations' differing spin angular momenta are suppressed, or ``washed out". 
Conversely, in the wave regime, the field exhibits a standing-wave pattern with a wavelength depending on the mode's angular momentum, though all modes converge to the behavior from Eq.} \eqref{eq:odd-par_R_approx_behavior} {for $r \gg 1/\left(\rS \mu^2\right)$.}

\section{Astrophysical implications}
\label{sec:astro_implications}

We now consider the astrophysical implications of the solutions found in section \ref{sec:solutions}. We take the field to model a cosmological condensate of vector dark matter. The phenomenology is characterized by two qualitatively different regimes, $\mu M \gtrsim 1$ and $\mu M < 1$. For $\mu M \gtrsim 1$, the field's Compton wavelength is smaller than the \bh{} horizon radius $\rS$, so it acts like a cloud of particles. Conversely, for $\mu M < 1$, the Compton wavelength is similar or greater than $\rS$, so it behaves like a wave. The transition point $\mu M \sim 1$ occurs at $\mu \sim 10^{-10}$ eV for a \bh{} with mass $M \sim 1 M_\odot$ at $\mu \sim 10^{-20}$ eV for a \bh{} with mass $M \sim 10^{10} M_\odot$. These particle masses are all within experimental constraints and viable given multiple proposed vector dark matter generation channels such as the misalignment mechanism and inflationary fluctuations \cite{Arias:2012az, Nelson:2011sf, Graham:2015rva}. Thus, for known astrophysical \bh{s}, which span the mass range $M \in [1 M_\odot, 10^{10} M_\odot]$, both regimes are viable.

Our primary concerns are the density profile of the cloud, the enclosed mass of the cloud, and the \bh{} mass accretion rate. We define these precisely in App. \ref{sec:accretion_rate_def}.

\subsection{Density of cloud}
\label{sec:density_profile}

The most prominent common feature of these Proca-field solutions is the presence of an over-dense region around the \bh{}. We show that the dark matter density is amplified over its ambient value by a factor of $10^6-10^7$ within a few Schwarzschild radii of the horizon. At the edge of the cloud, the field density should match the ambient dark matter density. To understand this transition, we must first estimate the radius of the \bh{'s} sphere of influence.

\subsubsection{Black hole sphere of influence}

Our approximation treats the \bh{} as the only gravitating body in the universe. This is a valid assumption near the horizon, but it breaks down once the gravitational potential of the ambient dark matter rivals that of the \bh{}. To understand roughly where this point occurs, we turn to a virial theorem argument \cite{Hui:2019aqm, Ghez:1998ph}.

The virial theorem of Newtonian gravity states that the time-averaged kinetic energy $\langle T \rangle$ and the time-averaged potential energy $\langle V \rangle$ of a stable self-gravitating system are related by
\begin{equation}
    \label{eq:virial_theorem}
    2\langle T \rangle + \langle V \rangle = 0\, .
\end{equation}
We consider a cloud of $N$ particles of mass $\mu$, each at distance $r_i$ from a central \bh{} of mass $M$. If we assume the \bh{} dominates the total potential energy, then (restoring Newton's constant) we have
\begin{align}
    \label{eq:virial_avg_V}
    \langle V \rangle \approx G\mu M \sum_{i = 1}^N \frac{1}{r_i} \eqcolon \frac{G M M_c}{r_c}\,
\end{align}
where $M_c = N \mu$ is the total mass of the cloud and we define the cloud's characteristic radius $r_c = \langle 1/r_i \rangle^{-1}$. The cloud has velocity dispersion $\langle v^2 \rangle$, and its average kinetic energy is
\begin{equation}
    \label{eq:virial_avg_T}
    \langle T \rangle = \frac{1}{2} M_c \langle v^2 \rangle\, .
\end{equation}
It follows from Eqs. \eqref{eq:virial_theorem}-(\ref{eq:virial_avg_T}) that
\begin{equation}
    \langle v^2 \rangle \approx \frac{2 G M}{r_c} = \frac{\rS}{r_c}\, .
\end{equation}
The quantity of interest here is $r_c$, which is roughly the radius of the cloud if effects of surrounding matter are neglected. We interpret it as the distance from the \bh{} at which the \bh{'s} gravitational potential rivals that of the surrounding matter, i.e., the radius of the \bh{'s} sphere of influence. Thus, we match the cloud density to that of the surrounding dark matter at $r=r_c$.

Realistically, the dark matter velocity dispersion can take on a range of values, but in a typical galactic halo it is of order 100 km/s $\sim 10^{-3}c$
\cite{Hoeft:2003ea}. This corresponds to a \bh{} sphere of influence of $r_c \sim 10^6 \rS$, so this is the $r_c$ value we use in this work.\footnote{In Ref. \cite{Hui:2022sri}, it is argued in the analogous scalar-field model that the \bh{} sphere of influence for the $\ell=1$ mode should be modeled by considering a surrounding dark matter vortex rather than using the virial theorem argument. This yields an $\ell=1$ estimate of $r_c \sim 10^3 \rS$, which significantly reduces the predicted density amplification and accretion rate. This complexity is neglected here, but, if desired, one can scale by appropriate factors the values reported here for modes with nonzero angular momentum. It should also be emphasized that this would not affect the monopole mode, so the broad astrophysical implications would be unaffected. For a study of these vortices in dark photon systems, see Ref. \cite{East:2022rsi}.}

\subsubsection{Density profile}

We now have the tools necessary to study the density profile that the field forms around the \bh{}.

As seen by an observer with 4-velocity $u^\mu$, the Proca field has energy density
\begin{equation}
    \label{eq:rho_def}
    \rho = u^\mu u^\nu T_{\mu\nu}\, ,
\end{equation}
where $T_{\mu\nu}$ is given in Eq. (\ref{eq:ProcaTmunu}). We consider an observer stationary with respect to the \bh{}, i.e. $u^\mu \partial_\mu = f^{-1/2}\partial_t$. Since $T_{\mu\nu}$ is constant for the steady-state solutions considered here, it follows that $\rho$ is independent of $t$.

The dark matter density should be smooth across the boundary of the \bh{'s} sphere of influence. As such, we must normalize $\rho$ such that $\rho|_{r_c} = \rho_c$. This process is subtly complicated by the presence of oscillations in the radial profile of $\rho$, mirroring those seen for the radial function $\Psi(t,r)$, shown in Fig. \ref{fig:psi_evol_plots}. It also varies with $\theta$ and $\phi$ for modes other than $\ell=0$. To account for these effects, we define the $\langle\rho\rangle$, which is averaged over angle and smeared over a few radial periods, i.e.,
\begin{equation}
    \label{eq:rho_avg_def}
    \langle\rho\rangle|_r = \frac{\oint_{S^2} \dif \Omega \int \dif r' r'^2\rho(r',\theta,\phi) q(r,r')}{\oint_{S^2} \dif \Omega \int \dif r' r'^2 q(r,r')}\, ,
\end{equation}
with smearing function $q(r,r')$. We then impose $\langle\rho\rangle|_{r_c} = \rho_c$, which fixes the amplitude of $\rho$.

A question of obvious astrophysical relevance is that of the value of the ambient dark matter density, $\rho_c$. We find that all effects analyzed in this work are most prominent for large values of $\rho_c$. As such, we seek astrophysical scenarios in which it is maximized. One such scenario is that of a \bh{} submerged in a dark matter soliton. It has been shown that for sufficiently low dark matter masses, wave dark matter generically forms solitons (i.e. boson stars) at the centers of most galaxies with constant-density cores of radius $\sim 100 \pc$ \cite{Schive:2014hza}. In optimal circumstances, the central densities of these solitons can reach $10 M_{\odot}/\pc^3$. We thus calculate values using $\rho_c \sim 10 M_{\odot}/\pc^3$, as this is a natural environment where the effects we consider would be most relevant.

\begin{figure}
    \centering
    \includegraphics[width=1.08\linewidth]{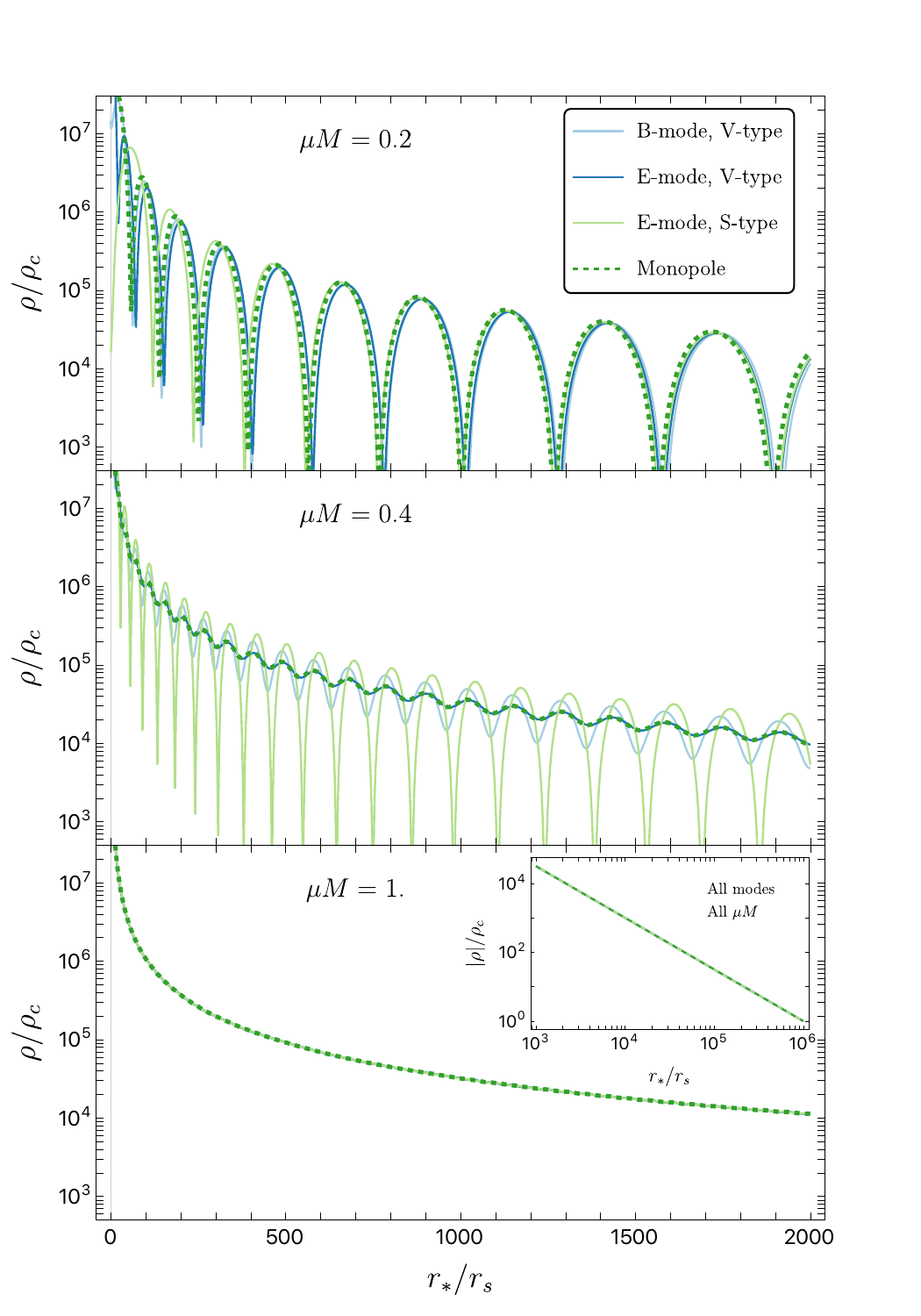}
    \caption{Radial profiles of the field's (angle-averaged) density for $\ell=1$ and $\ell = 0$. The three values of the mass parameter $\mu M$ chosen show, from top to bottom, (1) the standing-wave regime ($\mu M = 0.2$), with nodes at $r_n \approx \frac{n^2 \pi^2 \rS}{16(\mu M)^2}$ for all modes, (2) the intermediate regime ($\mu M = 0.4$), with each mode displaying behavior between standing and infalling waves, and (3) the infalling-wave regime ($\mu M = 1$), with a smooth density profile identical across all three modes. {The lower-panel inset shows the asymptotic behavior of the averaged density, given by Eq.} \eqref{eq:rho_avg_def}, {which goes as $|\rho| \sim r^{-3/2}$ for all modes in all $\mu M$ regimes.}}
    \label{fig:rho_profiles}
\end{figure}

We now investigate the behavior of $\rho$ for our solutions, understanding it to represent the density of vector dark matter in an overdense cloud as it accretes onto an astrophysical \bh{}. Radial profiles of $\rho$ for the $\ell=1$ and $\ell=0$ modes are shown in Fig. \ref{fig:rho_profiles}. The obvious qualitative difference between wave ($\mu M < 1$) and particle ($\mu M \gtrsim 1$) cases is the presence of periodic zero-density nodes in the former. These correspond precisely to the standing-wave nodes seen in the radial function (e.g., see the left-most panels in Fig. \ref{fig:psi_evol_plots}). Using the field's asymptotic behavior, Eq. (\ref{eq:odd-par_R_approx_behavior}), we may estimate the positions of these nodes as $r_n \approx \frac{n^2 \pi^2 \rS}{16 (\mu M)^2}$, with integer label $n$, which is accurate for $r \gg 1/\rS\mu^2$.

The high-mass ($\mu M \gtrsim 1$) regime, shown in the bottom panel of Fig. \ref{fig:rho_profiles}, is characterized by a smooth node-less density profile. This results from the field's purely ingoing-wave behavior (see the right-most panels in Fig. \ref{fig:psi_evol_plots}). Since $\rho \sim |A|^2+|\partial A|^2$ (see Eqs. \eqref{eq:ProcaTmunu} and \eqref{eq:rho_def}), the field's high-frequency complex oscillations disappear, leaving only a slow variation closely related to the field's amplitude.

We derive a large-$r$ approximation of $\rho$ using the radial functions' large-$r$ asymptotic behavior, given by Eq. (\ref{eq:odd-par_R_approx_behavior}). Keeping only the leading contribution, we find, for all modes, the dependence $\rho \sim r^{-3/2}$. This is the same behavior as has been found for analogous scalar-field solutions \cite{Hui:2019aqm, Clough:2019jpm}, and it matches the well-understood behavior of cold dust accreting onto a central \bh{}~\cite{Bertschinger:1985pd, Gondolo:1999ef}.  We may use this to approximate the averaged density as
\begin{equation}
    \label{eq:rho_avg_approx}
    \langle\rho\rangle \approx \rho_c \left(\frac{r}{r_c}\right)^{-3/2}\, .
\end{equation}
This approximation is valid for $r \gg 1/\left(\rS\mu^2\right)$, and closely matches the numerical density profile for the node-less $\mu M \gtrsim 1$ case, shown in the bottom panel of Fig. \ref{fig:rho_profiles}.

\subsection{Mass of cloud}

\begin{figure}
    \centering
    \includegraphics[width=1.08\linewidth]{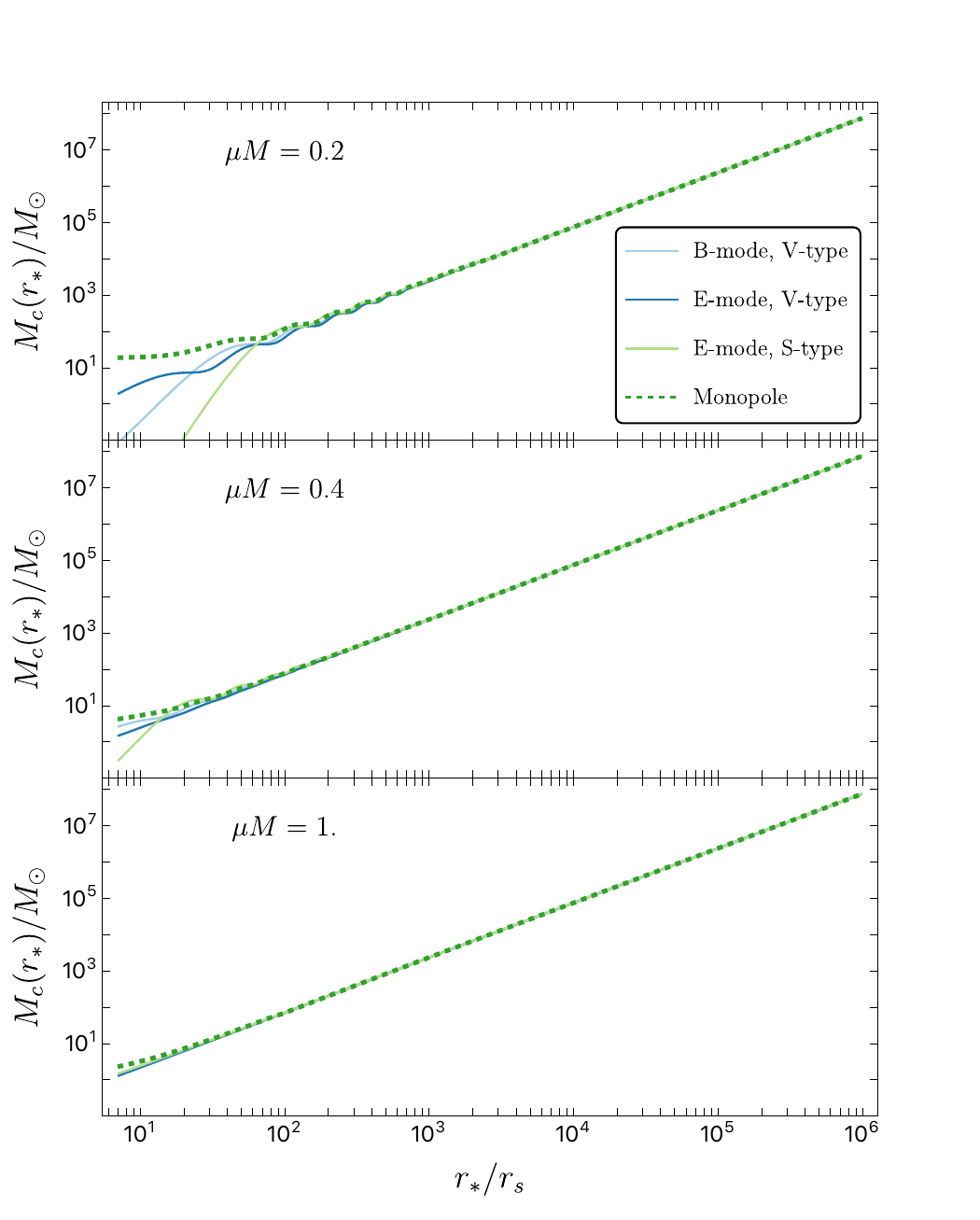}
    \caption{Dark matter mass enclosed between the horizon and a spherical surface situated at $r_*$. The $\ell=1$ mode for each polarization is shown, along with the monopole mode. Numbers shown on the $y$-axis are for a \bh{} of mass $10^9 M_\odot$ and an ambient dark matter density of $\rho_c = 10 M_\odot / \text{pc}^3$. For all modes, the enclosed mass converges to a $M_{c}(r) \sim r^{3/2}$ powerlaw behavior for $r \gg 1/{\left(\rS \mu^2\right)}$.}
    \label{fig:mass_profiles}
\end{figure}

In order to understand the gravitational effect of the dark matter cloud, it is useful to study the amount of mass it concentrates in the vicinity of the \bh. We may identify the cloud mass with the ``total matter energy" defined in (\ref{eq:tot_energy_def}), which is a Noether charge associated with the timelike killing vector $\xi = \partial_t$ (see App. \ref{sec:accretion_rate_def}). In precise terms, this is the total matter energy enclosed on a constant-$t$ hypersurface between the \bh{} horizon and a spherical surface at $r$. Specializing to Schwarzschild coordinates, we may write this definition as
\begin{equation}
    M_{c}(r) = \oint_{S^2} \dif \Omega \int_{\rS}^{r} \dif r' r'^2 \rho(r',\theta,\phi)\, ,
\end{equation}
which we interpret as the mass of dark matter enclosed between the horizon and the surface at $r$. {This quantity scales with the \bh{} mass $M$ by $M_c \sim M^3$, so it should be most relevant for the largest \bh{s}. We thus plot it for a $10^9 M_\odot$ supermassive \bh{} over a range of field masses in Fig.} \ref{fig:mass_profiles}. To convert the numbers in Fig. \ref{fig:mass_profiles} to those for a \bh{} with a mass $M$, one should multiply them by $\left(M/10^9 M_\odot\right)^3$. As discussed previously, we choose the ambient dark matter density to be $\rho_c \approx 10 M_\odot / \text{pc}^3$, modeling a dark matter soliton in a galactic center.

A clear and nontrivial takeaway from this analysis is that the field's mass has very little effect on the mass of the cloud it forms, save for small variations which are washed out at large $r$. Changing the field's mass affects its oscillatory behavior and ultimately how much energy it transfers, but it has little effect on how it distributes itself across the \bh's gravitational potential well.

Using the $\rho \sim r^{-3/2}$ approximation found using the field's asymptotics, we may approximate the cloud mass by $M_{c}(r) \sim r^{3/2}$. In physical coordinates, this yields the useful relation
\begin{equation}
    \label{eq:cloud_mass_approx}
    M_c(r) \approx 10^{-2} M_{\odot} \left(\frac{\rho_c}{1 M_\odot/ \text{pc}^3}\right)\left(\frac{M}{10^9 M_\odot}\right)^3\left(\frac{r}{\rS}\right)^{3/2}\, .
\end{equation}
We see that the cloud mass is in general small relative to the mass of the \bh{}. Taking the ``boundary" of the cloud to be the surface at which $\langle \rho\rangle \approx 10 \rho_c$, or $r \approx 0.2 r_c$, we find that the cloud's mass is a mere $M_{c} \sim 10^{-20} M_\odot = 10^{-20} M$ for a $M = 1 M_\odot$ \bh, but this rises to $M_{c} \sim 10^7 M_\odot = 10^{-2}M$ for a $M=10^9 M_\odot$ \bh.

\subsubsection{Mass excess}

\begin{figure}
    \centering
    \includegraphics[width=1.08\linewidth]{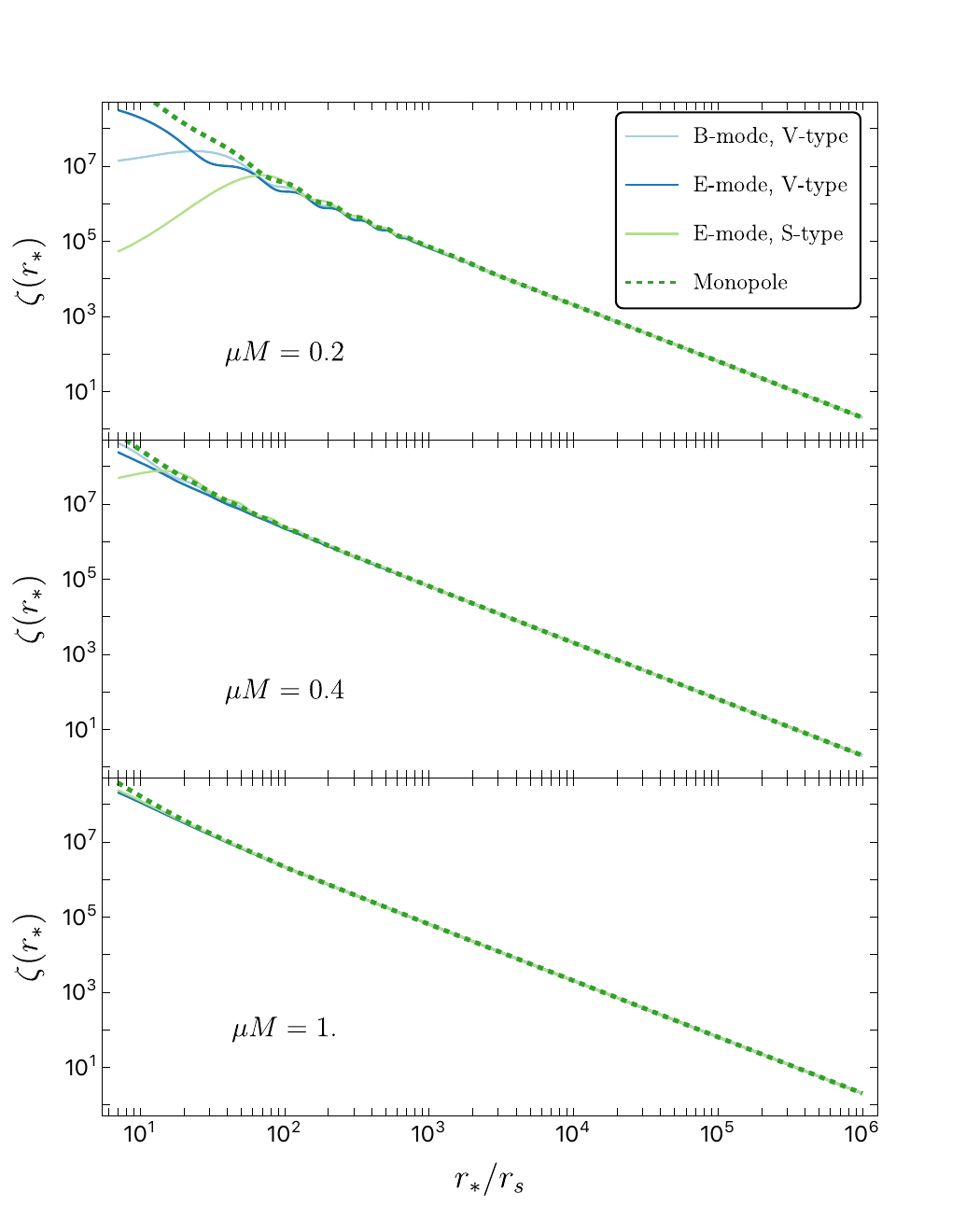}
    \caption{Radial dependence of the mass excess $\zeta(r)$ for all $\ell=0$ and $\ell=1$ modes. For $r \gg 1/\rS \mu^2 \approx \rS/(\mu M)^2$, all modes approach a $\zeta \sim r^{-3/2}$ behavior.}
    \label{fig:mass_excess_profiles}
\end{figure}

Another way to understand the cloud's mass is to consider the degree to which the \bh{} increases the mass of dark matter within a given radius over the mass of dark matter that would otherwise have been present in the same region. We codify this notion in the quantity
\begin{equation}
    \zeta(r) = \frac{M_{c}(r)}{M_{c,0}(r)}\, ,
\end{equation}
which we refer to as the ``mass excess", and where
\begin{equation}
    M_{c,0}(r) = \frac{4\pi}{3} (r^3 -\rS^3)\rho_c\, .
\end{equation}
A useful aspect of this quantity is that the ambient dark matter density and the \bh{} mass cancel and scale out respectively, making it agnostic to astrophysical peculiarities. Plots of the mass excess for a range of $\mu M$ are given in Fig. \ref{fig:mass_excess_profiles}. Like the cloud mass, the behavior is consistent for all $\mu M$ values, save for some near-horizon fluctuations in the $\mu M < 1$ cases.

Using the $M(r) \sim r^{3/2}$ approximation, we may capture the principal behavior of $\zeta$ in the approximation
\begin{equation}
    \zeta(r) \approx 2 \left(\frac{r}{r_c}\right)^{-3/2}\, .
\end{equation}
We thus see that within $r=10^3 \rS$ the effect of the \bh{} is tremendous, with a mass amplification upwards of $10^5$. Farther away from the horizon, however, the \bh-induced mass amplification is much less prominent, with $\zeta \sim \mathcal{O}(1)$ at $r\sim r_c$. In other words, a test particle at $r=r_c$ feels virtually no enhancement in the amount of dark matter present due to the \bh{}.

\subsection{Mass accretion rate}
\label{sec:bh_mass_accretion}

A natural further question is how much mass the \bh{} accretes from the surrounding dark matter. One should expect that the presence of a dark matter condensate filling the universe would cause a long-term secular growth in all \bh{} masses not attributable to other phenomena. Understanding this behavior precisely allows us both to predict its detectability and to place bounds on mass and density of the field using populations of astrophysical \bh.

\begin{figure}
    \centering
    \includegraphics[width=0.99\linewidth]{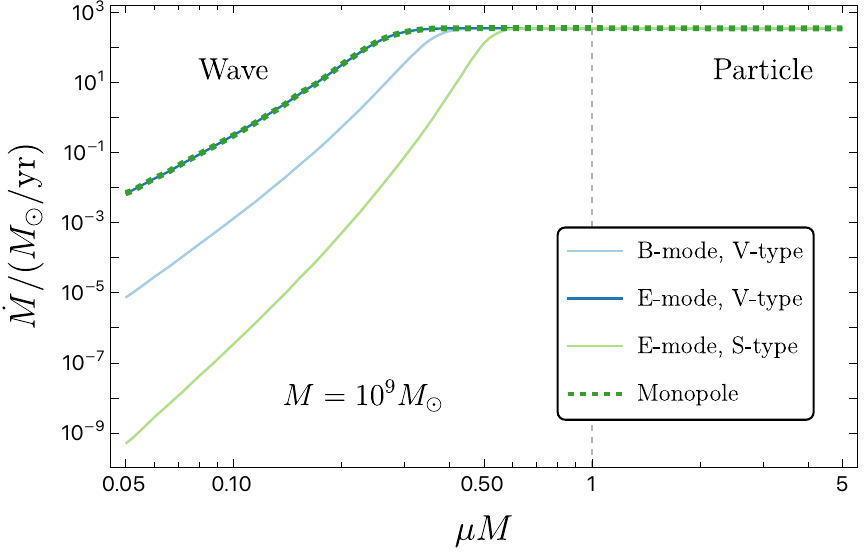}
    \caption{Mass accretion rate for a $M=10^9 M_{\odot}$ \bh{} submerged in an ambient dark matter density of $\rho_c=10 M_\odot/ \pc^3$. For $\mu M \gtrsim 1$, the rate plateaus, corresponding to the ``particle" regime. For $\mu M < 1$, the field enters the wave regime, and accretion is powerlaw suppressed by diffraction effects.}
    \label{fig:accretion_plot}
\end{figure}

To define the accretion rate properly, we consider a spherical surface at a radius $r$. It follows from the presence of the timelike Killing vector $\xi = \partial_t$ and the time independence of $T_{\mu\nu}$ for our solutions that there exists an energy flux which is independent of $r$. This is proven in App. \ref{sec:accretion_rate_def}. Given that this must also match the energy flux across the horizon, we may identify it with the mass accretion rate $\dot{M}$, from which follows the definition
\begin{equation}
    \dot{M} = f r^2 \oint_{S^2} \dif \Omega \langle k^\mu u^\nu T_{\mu\nu}\rangle\, ,
\end{equation}
where $k^\mu\partial_\mu = f^{1/2} \partial_r$ and $u^\mu\partial_\mu = f^{-1/2}\partial_t$. This quantity is plotted for a range of $\mu M$ values for a \bh{} of mass $M=10^9 M_\odot$ in Fig. \ref{fig:accretion_plot}. {Like with the cloud mass, this \bh{} mass is chosen because the accretion rate is maximized for the largest \bh{s}.}

The difference between the ingoing-wave and standing-wave regimes appears clearly in the behavior of $\dot{M}$. The former corresponds to the ``particle" limit, where the field appears to the \bh{} as a cloud of non-interacting particles of mass $\mu$. The rate at which mass is transferred across a surface of constant $r$ is unaffected by whether it's carried by many small particles or a few large particles. Conversely, in the $\mu M < 1$ regime, the field acts as a wave, and diffraction effects off the horizon become large. The superposition of infalling waves and outgoing diffracted waves creates a standing-wave pattern. As $\mu M$ is decreased, horizon absorption is reduced and diffraction amplified, causing an increasing suppression of the mass accretion rate.

{We additionally see, as in Fig.} \ref{fig:psi_evol_plots}, {that angular momentum effects are irrelevant in the particle regime but become dominant in the wave regime. The curves plotted in Fig.} \ref{fig:accretion_plot} {are the monopole ($\ell = 0$) mode and the $\ell=1$ modes of the three polarizations. 
The vector-type E-mode has total angular momentum $j = \ell + S = 0$, which is the same as the monopole mode. This gives the two modes nearly identical behavior in the wave regime. The B-mode and the scalar-type E-mode have total angular momenta of $j=1$ and $j=2$, respectively, and their accretion rates are exponentially suppressed in the wave regime. We thus see that in the wave regime, accretion is most relevant for the modes of lowest angular momentum, and it is
suppressed for higher-$j$ modes. }

In the particle regime ($\mu M \gtrsim 1$), we find an approximate accretion rate of
\begin{equation}
    \label{eq:bh_accretion_rate}
    \dot{M} \approx \left(40 \frac{M_\odot}{\text{yr}}\right)\left(\frac{\rho_c}{1 M_\odot/\pc^3}\right) \left(\frac{M}{10^9 M_\odot}\right)^2\, ,
\end{equation}
which is compatible with accretion rates found for supermassive \bh{s} with nonrelativistic particle dark matter spikes in the literature \cite{Shapiro:2023gpe}.
The $\dot{M} \sim M^2$ behavior naturally makes accretion most relevant for large \bh{s}.
A $10^9 M_\odot$ \bh{} submerged in a dark matter soliton with $\rho_c \sim 10 M_\odot/\pc^3$ has an accretion rate $\dot{M} \sim 100 M_\odot/\text{yr}$, corresponding to a growth timescale of $\tau \coloneq M/\dot{M} \sim 10^7$ years. Conversely, for a $1 M_\odot$ \bh{} in the same environment, the growth timescale is $10^{16}$ years, making accretion far less interesting for stellar-mass \bh{s}.

\section{Summary and discussion}

In this work, we constructed steady-state solutions for a massive vector field in Schwarzschild spacetime allowed to accrete from an infinite bath of particles far from the \bh{}. We performed this calculation as a semi-analytical problem under the assumption of negligible backreaction of the field onto the spacetime. We showed that the Proca field's even-parity and odd-parity sectors could each be fully separated, allowing us to reduce the problem to a set of decoupled \ode{s}. We found an exact solution to the odd-parity equation for all frequencies in terms of confluent Heun functions. For the choice $\omega = \mu$, we solved the even-parity equation numerically. We then calculated the density profiles, cloud masses, and accretion rates corresponding to these solutions and discussed their astrophysical relevance. We now discuss the implications of our results in more detail.

\subsection{Effect on binary black hole mergers}

Much attention has been paid to the impact of dark matter environments on binary \bh{} mergers. Dynamical friction from the dark matter should naturally slow \bh{s} during inspiral, introducing a phase shift into the waveform \cite{Baryakhtar:2022hbu}. The accretion-induced dark matter ``spike" discussed in this paper is a natural mechanism to enhance this effect. One might expect that the dephasing would be most prevalent in the particle regime, where the steady-state density enhancement is strongest, with the $\rho \sim r^{-3/2}$ profile extending all the way to the \bh{} horizon. However, N-body simulations suggest that equal-mass \bh{} binaries tend to ``shake off" their dark matter spikes during the early inspiral process in the case of particle dark matter \cite{Kavanagh:2018ggo}. Equal-mass binary \bh{} mergers are thus an unappealing setting to study particle dark matter, but there is more hope in the extreme mass ratio inspiral case, where the larger body may maintain its spike, opening up the possibility of detection with LISA~\cite{Barausse:2020rsu, LISA:2022kgy, Duque:2023seg, Gliorio:2025cbh}.

In the wave dark matter case, the story is less clear. Recent numerical-relativity studies have explored the effects of ultralight scalar dark matter on binary \bh{} mergers, focusing on the wave regime \cite{Bamber:2022pbs}. There, as in our work, the steady-state density profile around each \bh{} is smeared, reducing the central density. Na\"{i}vely, one might expect this would reduce the dark matter's impact on the merger dynamics. Results for equal-mass \bh{} mergers in wave dark matter environments suggest, however, that effects other than the usual dynamical friction can come to dominate, producing significant dephasing, especially when the field's compton wavelength matches the binary separation \cite{Aurrekoetxea:2023jwk}. Additional work has attempted to further constrain this effect, but understanding is still limited due to the cloud's complicated dependence on inspiral history \cite{Aurrekoetxea:2024cqd, Bamber:2022pbs,Cheng:2025wac}.
These questions are likely generic to wave dark matter, and the additional polarizations in the vector case may introduce further complexity.

With these motivations, we are safe to consider binary \bh{} mergers a primary setting for observational study of the vector dark matter clouds considered in this work, both in the wave and the particle regimes. Quantitatively constraining its impact on merger waveforms, namely by dephasing, is the subject of future work.

\subsection{Contribution to black hole growth}

Another astrophysical effect we highlight is the contribution of vector dark matter accretion to \bh{} growth. As shown in Section \ref{sec:bh_mass_accretion}, for the particle regime ($\mu M \gtrsim 1$), the central \bh{} accretes mass at a rate dependent only on its mass $M$ and the ambient dark matter density $\rho_c$, given by Eq. (\ref{eq:bh_accretion_rate}). For a $M \sim 10^9 M_\odot$ supermassive \bh, this regime is reached for any field mass $\mu \gtrsim 10^{-19}$ eV---practically the entire mass range of interest. Moreover, some cosmological models, such as the inflationary production mechanism of Ref. \cite{Graham:2015rva}, prefer vector dark matter masses roughly in the range $10^{-10} \ev < \mu \lesssim 10^{-4}\ev$. In this scenario, even stellar-mass \bh{s} would sit in the particle regime, meaning every known \bh{} in the universe would accrete vector dark matter at the rate given by Eq. \eqref{eq:bh_accretion_rate}.

The $M^2$ dependence of $\dot{M}$ in Eq. (\ref{eq:bh_accretion_rate}) implies that large \bh{s} experience the fastest proportionate growth via dark matter accretion. To illustrate the relevance of this phenomenon, we thus turn to a $\sim 10^9 M_\odot$ \bh{} at the center of a $\sim 10^{12} M_\odot$ galaxy. Such galaxies at large redshifts $z \sim 8$ should generically host wave dark-matter solitons of mass $\sim 10^9 M_\odot$ \cite{Schive:2014hza}. The central \bh{} thus devours the entire soliton, doubling its size, within its accretion timescale, which is $10^7$ years. This is similar, if not slightly shorter than, the well-studied growth timescale of a supermassive \bh{} accreting baryonic matter at its Eddington limit \cite{1964ApJ...140..796S}. The precise soliton condensation timescale is not fully understood, but studies indicate it may be as short as $10^7$ years for some field masses \cite{Hui:2021tkt}. If this timescale is shorter than the \bh{} accretion timescale, then the soliton can continuously re-establish itself as it loses mass into the \bh{}. This may allow the \bh{} to double its mass multiple times over, leaving a much larger \bh{} than we would expect from baryonic accretion alone.

We thus have good reason to suspect that vector dark matter accretion, if present, may be a leading contribution to the mass of supermassive \bh{s} in some galaxies, especially the largest ones. Validating this statement more precisely is the task of future work. Namely, we must better understand under what conditions a galaxy's central wave dark matter soliton is fully devoured by its central \bh{} and under what conditions it continuously regrows itself, serving as a steady funnel of mass into the central \bh{}.

\subsection{Proca cloud as black hole hair}

The no-hair conjecture is a cornerstone of \bh{} physics. In the 1960s and 1970s, the general relativity community produced a patchwork of proofs showing, on the one hand, the uniqueness of certain \bh{} spacetimes \cite{Israel:1967wq, Israel:1967za, Carter:1971zc, Wald:1971iw} and, on the other, that these spacetimes cannot indefinitely support a nontrivial buildup of various fields \cite{Bekenstein:1971hc, Bekenstein:1972ky, Bekenstein:1972ny, Teitelboim:1972qx, Adler:1978dp}. From these results emerged a more radical proposal purporting to tie them together: that of the no-hair conjecture \cite{Ruffini:1971bza}. It proposes that the endpoint of any gravitational collapse is a Kerr-Newman \bh{}, characterized only by its mass, spin, and electromagnetic charge.

In typical physicist fashion, the statement straddles the domains of formal mathematics and observational astronomy without fully committing to either. On the one hand, since it goes beyond simply a uniqueness statement about stationary spacetimes, referring instead to the endpoint of a dynamical collapse involving \textit{any} types of matter one could imagine, it is difficult to even formulate in a mathematically rigorous way, much less to prove. What is more, astrophysical \bh{s} frequently host long-lived features such as accretion disks and jets, and even the quiescent ones live in an expanding universe and accrete matter from their surroundings. The stationary ``final state" \bh{} of the no-hair conjecture thus only exists as an approximation to the astrophysical endstates of gravitational collapse.

While this discussion may seem pedantic, these details become important when understanding what constitutes a ``violation" of the no-hair conjecture. The literature is filled with claims of such violations, ranging from perturbative results like ours \cite{Jacobson:1999vr, Hui:2019aqm, Richards:2025ows}, to novel stationary \bh{} solutions of Einstein's equations \cite{Herdeiro:2014goa, Chodosh:2015oma, Herdeiro:2016tmi, Santos:2020pmh}, to full dynamical evolutions of the spacetime with endpoints outside the Kerr-Newman family \cite{Clough:2019jpm}. Each of these cases uses a subtly different working definition of hair, whether it be long-lived configurations of some field, stationary \bh{} spacetimes outside the Kerr-Newman family, or, indeed, extra conserved global charges. While these notions often coincide, they do not always, as, e.g., in the case of stationary perturbations which become non-stationary with the inclusion of backreaction, or of stable nontrivial field configurations which do not correspond to a new global charge, as in our example in Section \ref{sec:massless_vector_hair}. To circumvent this ambiguity, rather than claiming to fully invalidate the no-hair conjecture, we instead choose to make precise our working definition of hair.

We thus narrow our focus just to the classic no-boson-hair result of Bekenstein, introduced in Ref. \cite{Bekenstein:1971hc} and extended in Refs. \cite{Bekenstein:1972ky, Bekenstein:1972ny}. Bekenstein summarizes it in the following statement:
\begin{quote}
``A black hole in its final (static or stationary) state cannot be endowed with any exterior massive scalar, vector, or spin-2 meson fields." \cite{Bekenstein:1972ny}
\end{quote}
In other words, any \bh{} system with a coupled massive bosonic field will necessarily decay to either a Kerr or a Schwarzschild \bh{} with a trivial field profile in the exterior region. We may thus define a ``hairy" solution for such a system as a steady-state configuration hosting a nontrivial buildup of the bosonic field outside the \bh{}. For model simplicity and physicality respectively, we maintain Bekenstein's assumptions that all fields are minimally coupled to the geometry and regular on the horizon.

At first glance, it appears that our massive vector cloud satisfies all of these conditions unambiguously; it is regular on the horizon, minimally coupled, and persists for all time. However, we must remember that this calculation was done in linear perturbation theory, and the inclusion of backreaction onto the spacetime may complicate the story significantly. In Ref. \cite{Clough:2019jpm}, the authors considered the analogous scalar-field model with full backreaction and found several novel effects, including persistent \bh{} growth due to mass accretion, as well as cosmological expansion of the spacetime due to the field's nonzero asymptotic density. These would likely persist in the vector case. Bekenstein's statement concerns the ``final" state of the spacetime, and these phenomena suggest a \textit{final} configuration in which the field is either fully diluted by cosmological expansion or completely devoured by the \bh. Understanding the conditions under which each of these scenarios occurs, or indeed whether the system can settle into a truly ``hairy" final configuration with a nontrivial field and deformed spacetime, is the subject of future work.

For now, it suffices to say our solution, as well as the analogous scalar-field results of Refs. \cite{Clough:2019jpm, Hui:2019aqm}, constitutes ``hair" in a weaker sense; our field forms an eternal nontrivial profile external to the \bh{} in the perturbative regime. \bh{} perturbations of any spin generically admit mode decompositions, but upon imposing physical boundary conditions one finds that only modes with a strictly complex frequency survive, with the imaginary part demonstrating either precipitous decay or unstable growth of the mode. This can be seen as the perturbative reflection of the no-hair conjecture. Here, we have shown that this can be circumvented for a Proca field by allowing the field a time-dependent boundary condition at spatial infinity, yielding an arbitrarily long-lived steady-state configuration. We thus conclude that accretion from a particle bath is a viable path towards hairy Proca field solutions, using the definition supplied above.

\subsection{Outlook and open questions}

In this work, we limited ourselves to a single vector field with only a mass term in its potential, and we treated it using a perturbative analysis, neglecting backreaction onto the spacetime. A clear path for future work is to study the effects of relaxing any of these conditions. For example, studying the dynamical growth of the cloud (particularly with full backreaction), as in Ref. \cite{Clough:2019jpm}, would shed light on its stability and allow us to quantify its growth timescale. This would also open the door towards numerical simulation of an inspiraling \bh{} binary in the vector dark matter environment. Conversely, extending this study to the rotating \bh{} case, as in Refs. \cite{Hui:2022sri, Bamber:2020bpu}, would allow us to study how this accretion effect mixes with superradiance. The story in each of these cases may be additionally complicated by the inclusion of higher terms in the field's potential. Lastly, by including mixing with standard model matter, as has begun to be explored for superradiance \cite{Siemonsen:2022ivj, Xin:2024trp}, we could understand the observable signatures these vector dark matter spikes might produce outside the gravitational sector. We plan to study these questions in future work.

\section*{Acknowledgements}
We thank
J.~Shelton
and H.~O.~Silva
for insightful discussions and comments.
H.~W. acknowledges support provided by the National Science Foundation under NSF Award No.~OAC-2004879,
No.~OAC-2411068 and No.~PHY-2409726.
F.H. acknowledges support from the National Science Foundation Graduate Research Fellowship Program under Grant No. DGE 21-46756.

\appendix
\section{Asymptotics of odd-parity radial equation}\label{sec:odd-par_asymp}

The odd-parity sector of the Proca equation in the background of a Schwarzschild \bh{} can be reduced to a single \ode~\eqref{eq:odd-par_radial_eq}
for the radial function $R(r)$.
We now study the asymptotics of solutions to this equation. For simplicity we set $\rS = 1$ in this section.

\subsection{General \texorpdfstring{$r\to\infty$}{} case}

To study the large-$r$ asymptotics of
Eq.~\eqref{eq:odd-par_radial_eq},
we introduce the transformation $R \to y$ defined by
\begin{equation}
\label{appeq:OddParityRadialTrafo}
    R(r) = \frac{y(r)}{\sqrt{1-1/r}}\, ,
\end{equation}
which puts Eq.~\eqref{eq:odd-par_radial_eq}
in the form
\begin{equation*}
\left(\frac{\dif^2}{\dif r^2}
+ \frac{r^4 k^2 + r^3 \mu^2
- r(r-1)\ell(\ell+1)
+r -3/4}{r^2(r-1)^2}\right)y = 0
\, .
\end{equation*}
Expanding the latter to second order in $1/r$, we find the asymptotic approximation
\begin{equation}
    \label{eq:oddpar_asymp_eq}
    \left(\frac{\dif^2}{\dif r^2} + k^2 +\frac{2 k^2+\mu^2}{r} + \frac{3 k^2-\ell(\ell+1) + 2 \mu^2}{r^2}\right)y = 0\, .
\end{equation}
This has a general solution
\begin{align}
    y(r) = {}&c_3 e^{-i k r} r^{(1-\beta_{\ell,k})/2} W^{-}(r)
\nonumber \\ {}&
    + c_4 e^{-i k r} r^{(1+\beta_{\ell,k})/2} W^{+}(r)
\, ,
\end{align}
where
\begin{equation}
    \beta_{\ell,k} \coloneq \sqrt{4\ell(\ell+1) + 1 -12k^2 - 8\mu^2}\, ,
\end{equation}
and we have defined
\begin{equation}
W^{\pm}(r) = {_1F_1}\left(\frac{1 \pm \beta_{\ell,k}}{2} + \frac{i(2k^2+\mu^2)}{2k}, 1 \pm \beta_{\ell,k}, 2ikr \right)\, ,
\end{equation}
where $_1F_1$ is a confluent hypergeometric function of the first kind.
The transformation in Eq.~\eqref{appeq:OddParityRadialTrafo}
yields
\begin{align}
    R(r) = {}& \frac{c_3}{\sqrt{r-1}} e^{-i k r} r^{1-\beta_{\ell,k}/2} W^-(r)
\nonumber \\ {}&
    + \frac{c_3}{\sqrt{r-1}} e^{-i k r} r^{1+\beta_{\ell,k}/2} W^+(r)\, .
\end{align}
Asymptotically, for large r, this solution behaves as
ingoing and outgoing waves
\begin{align}
    R(r) \approx {}&c_{\text{in}} r^{-\frac{i}{2k}\left(2k^2+\mu^2\right)} e^{-i k r}
\nonumber \\ {}&
    + c_{\text{out}} r^{\frac{i}{2k}\left(2k^2+\mu^2\right)} e^{i k r}\,
\end{align}
where the relation between the coefficients $\{c_{\text{in}}, c_{\text{out}}\}$ and $\{c_3, c_4\}$ is lengthy but can be found exactly using the asymptotic expansion of the $_1F_1$ function.

\subsubsection{\texorpdfstring{$r \to \infty, k=0$}{} case}

In the nonrelativistic limit, $k \to 0$,
the asymptotic equation \eqref{eq:oddpar_asymp_eq} reduces to
\begin{equation}
    \left(\frac{\dif^2}{\dif r^2} + \frac{\mu^2}{r} + \frac{2\mu^2 - \ell(\ell+1)}{r^2} \right)y = 0\, .
\end{equation}
This has a general solution in terms of Bessel $J$ functions
\begin{align}
    \nonumber
    y(r) = {}&c_3\sqrt{r} J_{\beta_{\ell,0}}\left(2 \mu \sqrt{r}\right) \\
    {}&+ c_4\sqrt{r} J_{-\beta_{\ell,0}}\left(2 \mu \sqrt{r}\right)\, .
\end{align}
This corresponds to the general solution,
\begin{align}
    \label{eq:oddpar_bessel_sol}
    R(r) = {}&\frac{c_3 r}{\sqrt{r-1}} J_{\beta_{\ell,0}}\left(2 \mu \sqrt{r}\right) + \frac{c_4 r}{\sqrt{r-1}} J_{-\beta_{\ell,0}}\left(2 \mu \sqrt{r}\right)\, .
\end{align}
We can consider this approximation valid roughly for $r\gg 1/\mu^2$. The $r\to \infty$ asymptotic expansion of this solution yields an approximation in terms of ingoing and outgoing waves of the form
\begin{equation}
    \label{eq:odd_large_r_behav}
    R(r) \approx c_{\text{in}} r^{1/4} e^{-2i\mu \sqrt{r}} + c_{\text{out}}r^{1/4}e^{2i\mu \sqrt{r}}\, ,
\end{equation}
where the coefficients $\{c_{\text{in}}, c_{\text{out}}\}$ are related to $\{c_3, c_4\}$ by
\begin{align}
    c_{\text{in}} = {}&\frac{(-1)^{1/4}}{2\sqrt{\mu \pi}} e^{-i \pi \beta_{\ell,0}/2}\left(c_4 + c_3 e^{i\pi \beta_{\ell,0}}\right)\\
    c_{\text{out}} = {}&\frac{-1}{2\sqrt{\mu \pi}} e^{-i \pi (\beta_{\ell,0}/2-3/4)}\left(c_3 + c_4 e^{i\pi \beta_{\ell,0}}\right)
\end{align}
Thus, at large $r$, $R(r)$ is an oscillating function with an envelope which goes as $r^{1/4}$.

\section{Asymptotics of even-parity radial equation}
\label{sec:even-parity_asymptotics}

To derive asymptotic approximations for the even-parity equations, we proceed along the same lines as the odd-parity sector. We again set $\rS = 1$ for simplicity.

\subsection{\texorpdfstring{$r \to \rS$}{} case}

In the limit $r \to \rS$, solutions to the radial equation~\eqref{eq:fkks_radial_eq_even} reduce to ingoing and outgoing waves
\begin{equation}
    R = c_1 e^{i \omega r_*} + c_2 e^{-i \omega r_*}\, .
\end{equation}
To impose regularity, we discard the outgoing solution ($c_1 \to 0$).
In this paper we focus on the case in which the frequency is fixed to $\omega=\mu$,
leaving
\begin{equation}
    \label{eq:even_near_horizon_bc}
    R = c_2 e^{-i \mu r_*}\, .
\end{equation}
This approximation is valid in the regime {$(r-\rS) \ll 1/\mu$}.

\subsection{\texorpdfstring{$r \to \infty, k = 0$}{} case}

To study the large-$r$ asymptotics, we first introduce the variable redefinition $R \to y$ defined by
\begin{equation}
    \label{eq:even_large_r_redef}
    R(r) = \sqrt{\frac{q_r}{r^2f}} y(r)\, ,
\end{equation}
where we set $\rS = 1$, $f(r)$ is the Schwarzschild metric function, $q_r = 1+\nu^2r^2$, and $\nu$ is the separation constant.
Applying the transformation to Eq.~\eqref{eq:fkks_radial_eq_even} yields
\begin{align}
    \label{eq:y_eq_even}
    \biggl(\frac{\dif^2}{\dif r^2} + \frac{\omega^2}{f^2}+\frac{1}{4r^4 f^2} + \frac{3\nu^2}{q_r^2} + \frac{\nu^2}{r f q_r}{}&\\\nonumber
     - \frac{\nu \omega \left(q_r + 2\right)}{f q_r} - \frac{\ell(\ell+1) q_r}{r^2 f} {}&\biggr)y = 0\, .
\end{align}
We may expand this equation for large r. This yields, to first sub-leading order,
\begin{equation}
    y''(r) + \frac{\mu^2(r+2) - \ell(\ell-1)}{r^2} y(r) = 0
\end{equation}
for the vector-type polarization, and
\begin{equation}
    y''(r) + \frac{\mu^2 (r+2)- \ell (\ell+3) - 2}{r^2} y(r) = 0
\end{equation}
for the scalar-type polarization.
The latter also applies to the monopole mode with $\ell=0$.
Both equations have
a general solution of the form
\begin{equation}
    y = c_3 \sqrt{r} J_{\gamma_{\ell}}(2\mu \sqrt{r}) + c_4 \sqrt{r} J_{-\gamma_\ell}(2\mu \sqrt{r})\, ,
\end{equation}
where
\begin{equation}
    \gamma_{\ell} = \begin{cases}
        4 \ell(\ell-1) + 1 - 8 \mu^2 {}& \text{(vector)} \\
        4 \ell(\ell+3) + 9 - 8 \mu ^2 {}& \text{(scalar)}\, .
    \end{cases}
\end{equation}
This closely resembles the behavior in the odd-parity case. For large $r$, both solutions behave as
\begin{equation}
    \label{eq:even_large_r_behav}
    R(r) \approx c_{\text{in}} r^{1/4} e^{-2i\mu \sqrt{r}} + c_{\text{out}}r^{1/4}e^{2i\mu \sqrt{r}}\, ,
\end{equation}
just like in the odd-parity case. The precise relationship between $\{c_3, c_4\}$ and $\{c_{\text{in}}, c_{\text{out}}\}$ is nontrivial but straightforward to determine. This solution is a valid approximation in the regime $r \gg 1/\rS\mu^2$.

\section{Cloud mass and accretion rate}
\label{sec:accretion_rate_def}

Here, we define notions of the vector-field cloud's mass and its rate of accretion across the horizon. While we specialize these to the Schwarzschild case under consideration, the method is valid for any stationary spacetime and any matter with a symmetric and divergence-free energy-momentum tensor.

Since we work in the decoupling limit, we do not compute
\bh{} growth directly.
At the same time, the symmetric nature of the background allows us to construct a notion of matter-energy which is rigorous, in the sense that it is associated with a precise conservation law.
We begin by recalling that the Schwarzschild background admits a timelike Killing vector $\xi^\mu\partial_\mu = \partial_t$.
We may use this and the energy-momentum tensor to construct the current
\begin{equation}
    J_\mu = T_{\mu\nu}\xi^\nu\, .
\end{equation}
The current is conserved, $\nabla_\mu J^\mu = 0$, as long as $T_{\mu\nu}$ is symmetric and divergence-free.
$J^\mu$ is a Noether current associated with timelike translations, so we may think of it as an ``energy" current.

We now
translate
from a differential conservation law to an integral one. To do so, we introduce four 3-surfaces: the horizon surface $\mathcal{H} = \{\mathcal{M}: r = \rS \}$, a timelike surface $\chi = \{\mathcal{M}: r = r_0 \}$, where $r_0$ is an arbitrary radius outside the horizon, and finally two spacelike surfaces at $t=t_1, t_2$, {$\Sigma_{i} = \{\mathcal{M}: t = t_{i} \}$}. We call the spacetime region enclosed between these surfaces $\mathcal{V}$.
Integrating the conservation law over the bulk and applying Stokes' theorem, we find
\begin{align}
    0 = {}&\int_{\mathcal{V}} \dif^4 x \sqrt{-g} \nabla_\mu J^\mu
    = \oint_{\partial \mathcal{V}} \dif\Sigma_\mu J^\mu\, ,
    \label{eq:current_stokes}
\end{align}
where $\dif\Sigma_\mu$ is the directed area element of the boundary hypersurface.
It is given by $\dif\Sigma_\mu = \dif^3x \sqrt{|\gamma|} n_\mu$, where $\gamma$ is the determinant of the induced metric, $n_\mu$ is the normal, and the overall sign is chosen based on the geometry of each segment.
We may split $\partial \mathcal{V}$ into the surfaces $\partial\mathcal{V} = \tilde{\Sigma}_{1} \cup \tilde{\mathcal{H}}\cup \tilde{\Sigma}_{2} \cup \tilde{\chi}$, where the tilde denotes the segment of each surface intersecting $\partial \mathcal{V}$, or $\tilde{\Sigma} = \Sigma \cap \partial \mathcal{V}$ for generic surface $\Sigma$.
We may thus rewrite the boundary integral equation \eqref{eq:current_stokes} as
\begin{align}
    \nonumber
    0 = {}&\int_{\tilde{\Sigma}_2} \dif\Sigma_\mu J^\mu - \int_{\tilde{\Sigma}_1} \dif\Sigma_\mu J^\mu\\
    {}&+ \int_{\tilde{\mathcal{H}}} \dif\Sigma_\mu J^\mu - \int_{\tilde{\chi}} \dif\Sigma_\mu J^\mu\, ,
\end{align}
where we have chosen timelike and spacelike area elements to point ``forward" and ``outward" respectively. These terms each have a natural interpretation; the $\tilde{\Sigma}_{(i)}$ integrals represent the total matter energy outside the horizon at time $t_i$, while the $\tilde{\mathcal{H}}$ and $\tilde{\chi}$ integrals give the energy lost between $t_1$ and $t_2$ across $\mathcal{H}$ and $\chi$ respectively. We may thus rewrite our conservation law as
\begin{equation}
    \label{eq:energy_cons_law}
    \mathcal{E}_{2} - \mathcal{E}_1 + (\Delta \mathcal{E})_{\mathcal{H}} - (\Delta \mathcal{E})_{\chi} = 0\, ,
\end{equation}
where
\begin{subequations}
\begin{align}
    \label{eq:tot_energy_def}
    \mathcal{E}_{i} = {}&\int_{\tilde{\Sigma}_i} \dif\Sigma_\mu J^\mu
\,,\\
    (\Delta \mathcal{E})_{\mathcal{H}} = {}&\int_{\tilde{\mathcal{H}}} \dif\Sigma_\mu J^\mu
\,,\\
    (\Delta \mathcal{E})_{\chi} = {}&\int_{\tilde{\chi}} \dif\Sigma_\mu J^\mu
\, .
\end{align}
\end{subequations}
We may associate the horizon energy loss with an averaged flux $\mathcal{F}_\mathcal{H}$ by the simple definition
\begin{equation}
    \mathcal{F}_{\mathcal{H}} = \frac{(\Delta \mathcal{E})_{\mathcal{H}}}{\Delta t}\, ,
\end{equation}
with $\Delta t = t_2 - t_1$. The horizon's time-averaged energy accretion rate can thus always be calculated in this way, regardless of the specifics of the field's evolution.
In this work, we consider a steady-state solution,
for which $\mathcal{E}_1 =\mathcal{E}_2$,
and the conservation law~\eqref{eq:energy_cons_law} reduces to
\begin{equation}
    \label{eq:energy_cons_law_steady}
    (\Delta \mathcal{E})_{\mathcal{H}} = (\Delta \mathcal{E})_{\chi} = \int_{t_1}^{t_2} \dif t \oint_{S^2} \dif \Omega\, r^2 \sqrt{f}\, k_\mu J^\mu|_{r=r_0}\, ,
\end{equation}
with spacelike normal vector $k^\mu\partial_\mu = \sqrt{f} \partial_r$. It follows that
\begin{align}
    \nonumber
    \mathcal{F}_{\mathcal{H}} = \frac{(\Delta \mathcal{E})_{\chi}}{\Delta t} = {}&r^2 \sqrt{f}  \oint_{S^2} \dif \Omega \langle k^\mu \xi^\nu T_{\mu\nu} \rangle |_{r=r_0}\\
    = {}&r^2 f \oint_{S^2} \dif \Omega \langle \Phi_r \rangle |_{r=r_0}\, ,
    \label{eq:accretion_energy_flux}
\end{align}
where $\Phi_r = k^\mu u^\nu T_{\mu\nu}$, with timelike normal $u^\mu \partial_\mu = f^{-1/2}\partial_t$, and where $\langle \cdot \rangle$ denotes time-averaging over $\Delta t$.
We notice that, by virtue of the conservation law in Eq.~\eqref{eq:energy_cons_law_steady},
this quantity is the same regardless of the extraction radius $r_0$ selected.
In practice, this is the most practical way to calculate $\mathcal{F}_{\mathcal{H}}$, while its theoretical lack of $r_0$ dependence provides a useful consistency check for numerical solutions.

We have thus found well-motivated expressions for quantities of interest in understanding the vector-field cloud:
the total mass-energy outside the horizon~\eqref{eq:tot_energy_def} and the energy accretion rate across the horizon~\eqref{eq:accretion_energy_flux}.

\section{Self-consistency of decoupling approximation}
\label{appsec:decoupling_approx}

Throughout this work, we have assumed the backreaction of the Proca field onto the background spacetime is negligible.
In other words, we considered perturbations of a vacuum spacetime in a theory of Einstein-Hilbert gravity with a massive vector field, truncated the expansion at leading order, and neglected spin-2 perturbations.
The validity of our results thus rely on the applicability of this approximation.

The perturbative expansion applies when the Proca test field is ``small" compared to the gravitational background.
To define a notion of ``smallness'' we turn to Einstein's equations
\begin{equation}
    R_{\mu\nu} - \frac{1}{2}R g_{\mu\nu} = \kappa T_{\mu\nu}\, ,
\end{equation}
where $T_{\mu\nu}$ is the Proca field energy-momentum tensor given in Eq.~\eqref{eq:ProcaTmunu}. Taking the trace of Einstein's equations, we find
\begin{align}
-R = {}&\kappa\left(-\rho + \sum_i P_i\right)
\,.
\end{align}
Assuming a nonrelativistic fluid $\rho \gg P_i$, as is appropriate for a cold dark matter model,
we obtain
\begin{align}
R \approx \kappa \rho
\,.
\end{align}
We may interpret $R$ here as the scalar curvature sourced by the Proca field. Using the approximation developed in Section~\ref{sec:density_profile}, we find
\begin{equation}
    R_{\text{Matt}} \approx \kappa \rho_c
    \left(\frac{r}{r_c} \right)^{-3/2}
\,,
\end{equation}
where $\rho_c$ is the ambient dark matter density and $r_c$ is the radius of the \bh{} sphere of influence.

As a vacuum spacetime, the background has a vanishing Ricci scalar. However, we may measure its curvature using the Kretschmann scalar
\begin{equation}
    K = R^{\mu \nu \rho\sigma} R_{\mu\nu\rho\sigma} = \frac{12 \rS^2}{r^6}\, ,
\end{equation}
where the second equality is true only for Schwarzschild spacetime. Since $R \sim \text{Riemann}$ and $K \sim \text{Riemann}^2$, we may define the dimensionless parameter
\begin{equation}
    \epsilon \coloneq \frac{R_{\text{Matt}}}{\sqrt{K}}
\end{equation}
which measures the strength of the curvature sourced by the matter field relative to the curvature of the background. For our case, this gives
\begin{align}
    \nonumber
    \epsilon \approx {}&\frac{r^3}{M} \rho\\
    \approx {}&10^{-30} \left(\frac{\rho_c}{1 M_\odot/\pc^3}\right)\left(\frac{M}{M_\odot}\right)^2\left(\frac{r}{\rS}\right)^{3/2}\, .
\end{align}
We see that the matter-sourced curvature remains small near the horizon, with $\epsilon \lesssim 10^{-9}$ for any astrophysical \bh{} mass and realistic dark matter density. Given $\epsilon \sim M^2$, we expect gravitational backreaction to be most relevant for supermassive \bh{s}, with $\epsilon$ growing to $\mathcal{O}(1)$ within $r=10^6 \rS$ for a
supermassive \bh{}
with mass
$M=10^{10} M_\odot$
and
$\rho_c \gtrsim1 M_{\odot}/\pc^{3}$.
For smaller BHs, $\epsilon$ remains small in the entirety of the \bh{} sphere of influence.

\bibliographystyle{apsrev4-2}
\bibliography{Refs_ProcaHair.bib}

\begin{thebibliography}{93}%
\makeatletter
\providecommand \@ifxundefined [1]{%
 \@ifx{#1\undefined}
}%
\providecommand \@ifnum [1]{%
 \ifnum #1\expandafter \@firstoftwo
 \else \expandafter \@secondoftwo
 \fi
}%
\providecommand \@ifx [1]{%
 \ifx #1\expandafter \@firstoftwo
 \else \expandafter \@secondoftwo
 \fi
}%
\providecommand \natexlab [1]{#1}%
\providecommand \enquote  [1]{``#1''}%
\providecommand \bibnamefont  [1]{#1}%
\providecommand \bibfnamefont [1]{#1}%
\providecommand \citenamefont [1]{#1}%
\providecommand \href@noop [0]{\@secondoftwo}%
\providecommand \href [0]{\begingroup \@sanitize@url \@href}%
\providecommand \@href[1]{\@@startlink{#1}\@@href}%
\providecommand \@@href[1]{\endgroup#1\@@endlink}%
\providecommand \@sanitize@url [0]{\catcode `\\12\catcode `\$12\catcode `\&12\catcode `\#12\catcode `\^12\catcode `\_12\catcode `\%12\relax}%
\providecommand \@@startlink[1]{}%
\providecommand \@@endlink[0]{}%
\providecommand \url  [0]{\begingroup\@sanitize@url \@url }%
\providecommand \@url [1]{\endgroup\@href {#1}{\urlprefix }}%
\providecommand \urlprefix  [0]{URL }%
\providecommand \Eprint [0]{\href }%
\providecommand \doibase [0]{https://doi.org/}%
\providecommand \selectlanguage [0]{\@gobble}%
\providecommand \bibinfo  [0]{\@secondoftwo}%
\providecommand \bibfield  [0]{\@secondoftwo}%
\providecommand \translation [1]{[#1]}%
\providecommand \BibitemOpen [0]{}%
\providecommand \bibitemStop [0]{}%
\providecommand \bibitemNoStop [0]{.\EOS\space}%
\providecommand \EOS [0]{\spacefactor3000\relax}%
\providecommand \BibitemShut  [1]{\csname bibitem#1\endcsname}%
\let\auto@bib@innerbib\@empty
\bibitem [{\citenamefont {{Proca}}(1936)}]{1936JPhyR...7..347P}%
  \BibitemOpen
  \bibfield  {author} {\bibinfo {author} {\bibfnamefont {A.}~\bibnamefont {{Proca}}},\ }\href@noop {} {\bibfield  {journal} {\bibinfo  {journal} {J. Phys. Radium 7}\ }\textbf {\bibinfo {volume} {7}},\ \bibinfo {pages} {347} (\bibinfo {year} {1936})}\BibitemShut {NoStop}%
\bibitem [{\citenamefont {Antypas}\ \emph {et~al.}(2022)\citenamefont {Antypas} \emph {et~al.}}]{Antypas:2022asj}%
  \BibitemOpen
  \bibfield  {author} {\bibinfo {author} {\bibfnamefont {D.}~\bibnamefont {Antypas}} \emph {et~al.},\ }\href@noop {} {\bibinfo {title} {{New Horizons: Scalar and Vector Ultralight Dark Matter}}} (\bibinfo {year} {2022}),\ \Eprint {https://arxiv.org/abs/2203.14915} {arXiv:2203.14915 [hep-ex]} \BibitemShut {NoStop}%
\bibitem [{\citenamefont {Arias}\ \emph {et~al.}(2012)\citenamefont {Arias}, \citenamefont {Cadamuro}, \citenamefont {Goodsell}, \citenamefont {Jaeckel}, \citenamefont {Redondo},\ and\ \citenamefont {Ringwald}}]{Arias:2012az}%
  \BibitemOpen
  \bibfield  {author} {\bibinfo {author} {\bibfnamefont {P.}~\bibnamefont {Arias}}, \bibinfo {author} {\bibfnamefont {D.}~\bibnamefont {Cadamuro}}, \bibinfo {author} {\bibfnamefont {M.}~\bibnamefont {Goodsell}}, \bibinfo {author} {\bibfnamefont {J.}~\bibnamefont {Jaeckel}}, \bibinfo {author} {\bibfnamefont {J.}~\bibnamefont {Redondo}},\ and\ \bibinfo {author} {\bibfnamefont {A.}~\bibnamefont {Ringwald}},\ }\href {https://doi.org/10.1088/1475-7516/2012/06/013} {\bibfield  {journal} {\bibinfo  {journal} {JCAP}\ }\textbf {\bibinfo {volume} {2012}}\bibfield  {number} {\bibinfo  {number} { (06)},\ \bibinfo {pages} {013}},\ }\Eprint {https://arxiv.org/abs/1201.5902} {arXiv:1201.5902 [hep-ph]} \BibitemShut {NoStop}%
\bibitem [{\citenamefont {Nelson}\ and\ \citenamefont {Scholtz}(2011)}]{Nelson:2011sf}%
  \BibitemOpen
  \bibfield  {author} {\bibinfo {author} {\bibfnamefont {A.~E.}\ \bibnamefont {Nelson}}\ and\ \bibinfo {author} {\bibfnamefont {J.}~\bibnamefont {Scholtz}},\ }\href {https://doi.org/10.1103/PhysRevD.84.103501} {\bibfield  {journal} {\bibinfo  {journal} {Phys. Rev. D}\ }\textbf {\bibinfo {volume} {84}},\ \bibinfo {pages} {103501} (\bibinfo {year} {2011})},\ \Eprint {https://arxiv.org/abs/1105.2812} {arXiv:1105.2812 [hep-ph]} \BibitemShut {NoStop}%
\bibitem [{\citenamefont {Graham}\ \emph {et~al.}(2016)\citenamefont {Graham}, \citenamefont {Mardon},\ and\ \citenamefont {Rajendran}}]{Graham:2015rva}%
  \BibitemOpen
  \bibfield  {author} {\bibinfo {author} {\bibfnamefont {P.~W.}\ \bibnamefont {Graham}}, \bibinfo {author} {\bibfnamefont {J.}~\bibnamefont {Mardon}},\ and\ \bibinfo {author} {\bibfnamefont {S.}~\bibnamefont {Rajendran}},\ }\href {https://doi.org/10.1103/PhysRevD.93.103520} {\bibfield  {journal} {\bibinfo  {journal} {Phys. Rev. D}\ }\textbf {\bibinfo {volume} {93}},\ \bibinfo {pages} {103520} (\bibinfo {year} {2016})},\ \Eprint {https://arxiv.org/abs/1504.02102} {arXiv:1504.02102 [hep-ph]} \BibitemShut {NoStop}%
\bibitem [{\citenamefont {Fabbrichesi}\ \emph {et~al.}(2020)\citenamefont {Fabbrichesi}, \citenamefont {Gabrielli},\ and\ \citenamefont {Lanfranchi}}]{Fabbrichesi:2020wbt}%
  \BibitemOpen
  \bibfield  {author} {\bibinfo {author} {\bibfnamefont {M.}~\bibnamefont {Fabbrichesi}}, \bibinfo {author} {\bibfnamefont {E.}~\bibnamefont {Gabrielli}},\ and\ \bibinfo {author} {\bibfnamefont {G.}~\bibnamefont {Lanfranchi}},\ }\bibfield  {journal} {\bibinfo  {journal} {SpringerBriefs in Physics}\ }\href {https://doi.org/10.1007/978-3-030-62519-1} {10.1007/978-3-030-62519-1} (\bibinfo {year} {2020}),\ \Eprint {https://arxiv.org/abs/2005.01515} {arXiv:2005.01515 [hep-ph]} \BibitemShut {NoStop}%
\bibitem [{\citenamefont {Caputo}\ \emph {et~al.}(2021)\citenamefont {Caputo}, \citenamefont {Millar}, \citenamefont {O'Hare},\ and\ \citenamefont {Vitagliano}}]{Caputo:2021eaa}%
  \BibitemOpen
  \bibfield  {author} {\bibinfo {author} {\bibfnamefont {A.}~\bibnamefont {Caputo}}, \bibinfo {author} {\bibfnamefont {A.~J.}\ \bibnamefont {Millar}}, \bibinfo {author} {\bibfnamefont {C.~A.~J.}\ \bibnamefont {O'Hare}},\ and\ \bibinfo {author} {\bibfnamefont {E.}~\bibnamefont {Vitagliano}},\ }\href {https://doi.org/10.1103/PhysRevD.104.095029} {\bibfield  {journal} {\bibinfo  {journal} {Phys. Rev. D}\ }\textbf {\bibinfo {volume} {104}},\ \bibinfo {pages} {095029} (\bibinfo {year} {2021})},\ \Eprint {https://arxiv.org/abs/2105.04565} {arXiv:2105.04565 [hep-ph]} \BibitemShut {NoStop}%
\bibitem [{\citenamefont {Jaeckel}\ and\ \citenamefont {Ringwald}(2010)}]{Jaeckel:2010ni}%
  \BibitemOpen
  \bibfield  {author} {\bibinfo {author} {\bibfnamefont {J.}~\bibnamefont {Jaeckel}}\ and\ \bibinfo {author} {\bibfnamefont {A.}~\bibnamefont {Ringwald}},\ }\href {https://doi.org/10.1146/annurev.nucl.012809.104433} {\bibfield  {journal} {\bibinfo  {journal} {Ann. Rev. Nucl. Part. Sci.}\ }\textbf {\bibinfo {volume} {60}},\ \bibinfo {pages} {405} (\bibinfo {year} {2010})},\ \Eprint {https://arxiv.org/abs/1002.0329} {arXiv:1002.0329 [hep-ph]} \BibitemShut {NoStop}%
\bibitem [{\citenamefont {Goodsell}\ \emph {et~al.}(2009)\citenamefont {Goodsell}, \citenamefont {Jaeckel}, \citenamefont {Redondo},\ and\ \citenamefont {Ringwald}}]{Goodsell:2009xc}%
  \BibitemOpen
  \bibfield  {author} {\bibinfo {author} {\bibfnamefont {M.}~\bibnamefont {Goodsell}}, \bibinfo {author} {\bibfnamefont {J.}~\bibnamefont {Jaeckel}}, \bibinfo {author} {\bibfnamefont {J.}~\bibnamefont {Redondo}},\ and\ \bibinfo {author} {\bibfnamefont {A.}~\bibnamefont {Ringwald}},\ }\href {https://doi.org/10.1088/1126-6708/2009/11/027} {\bibfield  {journal} {\bibinfo  {journal} {JHEP}\ }\textbf {\bibinfo {volume} {2009}}\bibfield  {number} {\bibinfo  {number} { (11)},\ \bibinfo {pages} {027}},\ }\Eprint {https://arxiv.org/abs/0909.0515} {arXiv:0909.0515 [hep-ph]} \BibitemShut {NoStop}%
\bibitem [{\citenamefont {Holdom}(1986)}]{Holdom:1985ag}%
  \BibitemOpen
  \bibfield  {author} {\bibinfo {author} {\bibfnamefont {B.}~\bibnamefont {Holdom}},\ }\href {https://doi.org/10.1016/0370-2693(86)91377-8} {\bibfield  {journal} {\bibinfo  {journal} {Phys. Lett. B}\ }\textbf {\bibinfo {volume} {166}},\ \bibinfo {pages} {196} (\bibinfo {year} {1986})}\BibitemShut {NoStop}%
\bibitem [{\citenamefont {Curtin}\ \emph {et~al.}(2015)\citenamefont {Curtin}, \citenamefont {Essig}, \citenamefont {Gori},\ and\ \citenamefont {Shelton}}]{Curtin:2014cca}%
  \BibitemOpen
  \bibfield  {author} {\bibinfo {author} {\bibfnamefont {D.}~\bibnamefont {Curtin}}, \bibinfo {author} {\bibfnamefont {R.}~\bibnamefont {Essig}}, \bibinfo {author} {\bibfnamefont {S.}~\bibnamefont {Gori}},\ and\ \bibinfo {author} {\bibfnamefont {J.}~\bibnamefont {Shelton}},\ }\href {https://doi.org/10.1007/JHEP02(2015)157} {\bibfield  {journal} {\bibinfo  {journal} {JHEP}\ }\textbf {\bibinfo {volume} {2015}}\bibfield  {number} {\bibinfo  {number} { (157)}},\ }\Eprint {https://arxiv.org/abs/1412.0018} {arXiv:1412.0018 [hep-ph]} \BibitemShut {NoStop}%
\bibitem [{\citenamefont {Adshead}\ \emph {et~al.}(2022)\citenamefont {Adshead}, \citenamefont {Ralegankar},\ and\ \citenamefont {Shelton}}]{Adshead:2022ovo}%
  \BibitemOpen
  \bibfield  {author} {\bibinfo {author} {\bibfnamefont {P.}~\bibnamefont {Adshead}}, \bibinfo {author} {\bibfnamefont {P.}~\bibnamefont {Ralegankar}},\ and\ \bibinfo {author} {\bibfnamefont {J.}~\bibnamefont {Shelton}},\ }\href {https://doi.org/10.1088/1475-7516/2022/09/056} {\bibfield  {journal} {\bibinfo  {journal} {JCAP}\ }\textbf {\bibinfo {volume} {09}},\ \bibinfo {pages} {056}},\ \Eprint {https://arxiv.org/abs/2206.13530} {arXiv:2206.13530 [hep-ph]} \BibitemShut {NoStop}%
\bibitem [{\citenamefont {Detweiler}(1980)}]{Detweiler:1980uk}%
  \BibitemOpen
  \bibfield  {author} {\bibinfo {author} {\bibfnamefont {S.~L.}\ \bibnamefont {Detweiler}},\ }\href {https://doi.org/10.1103/PhysRevD.22.2323} {\bibfield  {journal} {\bibinfo  {journal} {Phys. Rev. D}\ }\textbf {\bibinfo {volume} {22}},\ \bibinfo {pages} {2323} (\bibinfo {year} {1980})}\BibitemShut {NoStop}%
\bibitem [{\citenamefont {Dolan}(2007)}]{Dolan:2007mj}%
  \BibitemOpen
  \bibfield  {author} {\bibinfo {author} {\bibfnamefont {S.~R.}\ \bibnamefont {Dolan}},\ }\href {https://doi.org/10.1103/PhysRevD.76.084001} {\bibfield  {journal} {\bibinfo  {journal} {Phys. Rev. D}\ }\textbf {\bibinfo {volume} {76}},\ \bibinfo {pages} {084001} (\bibinfo {year} {2007})},\ \Eprint {https://arxiv.org/abs/0705.2880} {arXiv:0705.2880 [gr-qc]} \BibitemShut {NoStop}%
\bibitem [{\citenamefont {Shlapentokh-Rothman}(2014)}]{Shlapentokh-Rothman:2013ysa}%
  \BibitemOpen
  \bibfield  {author} {\bibinfo {author} {\bibfnamefont {Y.}~\bibnamefont {Shlapentokh-Rothman}},\ }\href {https://doi.org/10.1007/s00220-014-2033-x} {\bibfield  {journal} {\bibinfo  {journal} {Commun. Math. Phys.}\ }\textbf {\bibinfo {volume} {329}},\ \bibinfo {pages} {859} (\bibinfo {year} {2014})},\ \Eprint {https://arxiv.org/abs/1302.3448} {arXiv:1302.3448 [gr-qc]} \BibitemShut {NoStop}%
\bibitem [{\citenamefont {Brito}\ \emph {et~al.}(2015)\citenamefont {Brito}, \citenamefont {Cardoso},\ and\ \citenamefont {Pani}}]{Brito:2015oca}%
  \BibitemOpen
  \bibfield  {author} {\bibinfo {author} {\bibfnamefont {R.}~\bibnamefont {Brito}}, \bibinfo {author} {\bibfnamefont {V.}~\bibnamefont {Cardoso}},\ and\ \bibinfo {author} {\bibfnamefont {P.}~\bibnamefont {Pani}},\ }\href {https://doi.org/10.1007/978-3-319-19000-6} {\bibfield  {journal} {\bibinfo  {journal} {Lect. Notes Phys.}\ }\textbf {\bibinfo {volume} {906}},\ \bibinfo {pages} {pp.1} (\bibinfo {year} {2015})},\ \Eprint {https://arxiv.org/abs/1501.06570} {arXiv:1501.06570 [gr-qc]} \BibitemShut {NoStop}%
\bibitem [{\citenamefont {Pani}\ \emph {et~al.}(2012{\natexlab{a}})\citenamefont {Pani}, \citenamefont {Cardoso}, \citenamefont {Gualtieri}, \citenamefont {Berti},\ and\ \citenamefont {Ishibashi}}]{Pani:2012bp}%
  \BibitemOpen
  \bibfield  {author} {\bibinfo {author} {\bibfnamefont {P.}~\bibnamefont {Pani}}, \bibinfo {author} {\bibfnamefont {V.}~\bibnamefont {Cardoso}}, \bibinfo {author} {\bibfnamefont {L.}~\bibnamefont {Gualtieri}}, \bibinfo {author} {\bibfnamefont {E.}~\bibnamefont {Berti}},\ and\ \bibinfo {author} {\bibfnamefont {A.}~\bibnamefont {Ishibashi}},\ }\href {https://doi.org/10.1103/PhysRevD.86.104017} {\bibfield  {journal} {\bibinfo  {journal} {Phys. Rev. D}\ }\textbf {\bibinfo {volume} {86}},\ \bibinfo {pages} {104017} (\bibinfo {year} {2012}{\natexlab{a}})},\ \Eprint {https://arxiv.org/abs/1209.0773} {arXiv:1209.0773 [gr-qc]} \BibitemShut {NoStop}%
\bibitem [{\citenamefont {Pani}\ \emph {et~al.}(2012{\natexlab{b}})\citenamefont {Pani}, \citenamefont {Cardoso}, \citenamefont {Gualtieri}, \citenamefont {Berti},\ and\ \citenamefont {Ishibashi}}]{Pani:2012vp}%
  \BibitemOpen
  \bibfield  {author} {\bibinfo {author} {\bibfnamefont {P.}~\bibnamefont {Pani}}, \bibinfo {author} {\bibfnamefont {V.}~\bibnamefont {Cardoso}}, \bibinfo {author} {\bibfnamefont {L.}~\bibnamefont {Gualtieri}}, \bibinfo {author} {\bibfnamefont {E.}~\bibnamefont {Berti}},\ and\ \bibinfo {author} {\bibfnamefont {A.}~\bibnamefont {Ishibashi}},\ }\href {https://doi.org/10.1103/PhysRevLett.109.131102} {\bibfield  {journal} {\bibinfo  {journal} {Phys. Rev. Lett.}\ }\textbf {\bibinfo {volume} {109}},\ \bibinfo {pages} {131102} (\bibinfo {year} {2012}{\natexlab{b}})},\ \Eprint {https://arxiv.org/abs/1209.0465} {arXiv:1209.0465 [gr-qc]} \BibitemShut {NoStop}%
\bibitem [{\citenamefont {Dolan}(2018)}]{Dolan:2018dqv}%
  \BibitemOpen
  \bibfield  {author} {\bibinfo {author} {\bibfnamefont {S.~R.}\ \bibnamefont {Dolan}},\ }\href {https://doi.org/10.1103/PhysRevD.98.104006} {\bibfield  {journal} {\bibinfo  {journal} {Phys. Rev. D}\ }\textbf {\bibinfo {volume} {98}},\ \bibinfo {pages} {104006} (\bibinfo {year} {2018})},\ \Eprint {https://arxiv.org/abs/1806.01604} {arXiv:1806.01604 [gr-qc]} \BibitemShut {NoStop}%
\bibitem [{\citenamefont {Siemonsen}\ and\ \citenamefont {East}(2020)}]{Siemonsen:2019ebd}%
  \BibitemOpen
  \bibfield  {author} {\bibinfo {author} {\bibfnamefont {N.}~\bibnamefont {Siemonsen}}\ and\ \bibinfo {author} {\bibfnamefont {W.~E.}\ \bibnamefont {East}},\ }\href {https://doi.org/10.1103/PhysRevD.101.024019} {\bibfield  {journal} {\bibinfo  {journal} {Phys. Rev. D}\ }\textbf {\bibinfo {volume} {101}},\ \bibinfo {pages} {024019} (\bibinfo {year} {2020})},\ \Eprint {https://arxiv.org/abs/1910.09476} {arXiv:1910.09476 [gr-qc]} \BibitemShut {NoStop}%
\bibitem [{\citenamefont {Cardoso}\ \emph {et~al.}(2018)\citenamefont {Cardoso}, \citenamefont {Dias}, \citenamefont {Hartnett}, \citenamefont {Middleton}, \citenamefont {Pani},\ and\ \citenamefont {Santos}}]{Cardoso:2018tly}%
  \BibitemOpen
  \bibfield  {author} {\bibinfo {author} {\bibfnamefont {V.}~\bibnamefont {Cardoso}}, \bibinfo {author} {\bibfnamefont {O.~J.~C.}\ \bibnamefont {Dias}}, \bibinfo {author} {\bibfnamefont {G.~S.}\ \bibnamefont {Hartnett}}, \bibinfo {author} {\bibfnamefont {M.}~\bibnamefont {Middleton}}, \bibinfo {author} {\bibfnamefont {P.}~\bibnamefont {Pani}},\ and\ \bibinfo {author} {\bibfnamefont {J.~E.}\ \bibnamefont {Santos}},\ }\href {https://doi.org/10.1088/1475-7516/2018/03/043} {\bibfield  {journal} {\bibinfo  {journal} {JCAP}\ }\textbf {\bibinfo {volume} {2018}}\bibfield  {number} {\bibinfo  {number} { (03)},\ \bibinfo {pages} {043}},\ }\Eprint {https://arxiv.org/abs/1801.01420} {arXiv:1801.01420 [gr-qc]} \BibitemShut {NoStop}%
\bibitem [{\citenamefont {Baumann}\ \emph {et~al.}(2019)\citenamefont {Baumann}, \citenamefont {Chia},\ and\ \citenamefont {Porto}}]{Baumann:2018vus}%
  \BibitemOpen
  \bibfield  {author} {\bibinfo {author} {\bibfnamefont {D.}~\bibnamefont {Baumann}}, \bibinfo {author} {\bibfnamefont {H.~S.}\ \bibnamefont {Chia}},\ and\ \bibinfo {author} {\bibfnamefont {R.~A.}\ \bibnamefont {Porto}},\ }\href {https://doi.org/10.1103/PhysRevD.99.044001} {\bibfield  {journal} {\bibinfo  {journal} {Phys. Rev. D}\ }\textbf {\bibinfo {volume} {99}},\ \bibinfo {pages} {044001} (\bibinfo {year} {2019})},\ \Eprint {https://arxiv.org/abs/1804.03208} {arXiv:1804.03208 [gr-qc]} \BibitemShut {NoStop}%
\bibitem [{\citenamefont {Witek}\ \emph {et~al.}(2013)\citenamefont {Witek}, \citenamefont {Cardoso}, \citenamefont {Ishibashi},\ and\ \citenamefont {Sperhake}}]{Witek:2012tr}%
  \BibitemOpen
  \bibfield  {author} {\bibinfo {author} {\bibfnamefont {H.}~\bibnamefont {Witek}}, \bibinfo {author} {\bibfnamefont {V.}~\bibnamefont {Cardoso}}, \bibinfo {author} {\bibfnamefont {A.}~\bibnamefont {Ishibashi}},\ and\ \bibinfo {author} {\bibfnamefont {U.}~\bibnamefont {Sperhake}},\ }\href {https://doi.org/10.1103/PhysRevD.87.043513} {\bibfield  {journal} {\bibinfo  {journal} {Phys. Rev. D}\ }\textbf {\bibinfo {volume} {87}},\ \bibinfo {pages} {043513} (\bibinfo {year} {2013})},\ \Eprint {https://arxiv.org/abs/1212.0551} {arXiv:1212.0551 [gr-qc]} \BibitemShut {NoStop}%
\bibitem [{\citenamefont {East}(2017)}]{East:2017mrj}%
  \BibitemOpen
  \bibfield  {author} {\bibinfo {author} {\bibfnamefont {W.~E.}\ \bibnamefont {East}},\ }\href {https://doi.org/10.1103/PhysRevD.96.024004} {\bibfield  {journal} {\bibinfo  {journal} {Phys. Rev. D}\ }\textbf {\bibinfo {volume} {96}},\ \bibinfo {pages} {024004} (\bibinfo {year} {2017})},\ \Eprint {https://arxiv.org/abs/1705.01544} {arXiv:1705.01544 [gr-qc]} \BibitemShut {NoStop}%
\bibitem [{\citenamefont {East}\ and\ \citenamefont {Pretorius}(2017)}]{East:2017ovw}%
  \BibitemOpen
  \bibfield  {author} {\bibinfo {author} {\bibfnamefont {W.~E.}\ \bibnamefont {East}}\ and\ \bibinfo {author} {\bibfnamefont {F.}~\bibnamefont {Pretorius}},\ }\href {https://doi.org/10.1103/PhysRevLett.119.041101} {\bibfield  {journal} {\bibinfo  {journal} {Phys. Rev. Lett.}\ }\textbf {\bibinfo {volume} {119}},\ \bibinfo {pages} {041101} (\bibinfo {year} {2017})},\ \Eprint {https://arxiv.org/abs/1704.04791} {arXiv:1704.04791 [gr-qc]} \BibitemShut {NoStop}%
\bibitem [{\citenamefont {East}(2018)}]{East:2018glu}%
  \BibitemOpen
  \bibfield  {author} {\bibinfo {author} {\bibfnamefont {W.~E.}\ \bibnamefont {East}},\ }\href {https://doi.org/10.1103/PhysRevLett.121.131104} {\bibfield  {journal} {\bibinfo  {journal} {Phys. Rev. Lett.}\ }\textbf {\bibinfo {volume} {121}},\ \bibinfo {pages} {131104} (\bibinfo {year} {2018})},\ \Eprint {https://arxiv.org/abs/1807.00043} {arXiv:1807.00043 [gr-qc]} \BibitemShut {NoStop}%
\bibitem [{\citenamefont {Clough}\ \emph {et~al.}(2022)\citenamefont {Clough}, \citenamefont {Helfer}, \citenamefont {Witek},\ and\ \citenamefont {Berti}}]{Clough:2022ygm}%
  \BibitemOpen
  \bibfield  {author} {\bibinfo {author} {\bibfnamefont {K.}~\bibnamefont {Clough}}, \bibinfo {author} {\bibfnamefont {T.}~\bibnamefont {Helfer}}, \bibinfo {author} {\bibfnamefont {H.}~\bibnamefont {Witek}},\ and\ \bibinfo {author} {\bibfnamefont {E.}~\bibnamefont {Berti}},\ }\href {https://doi.org/10.1103/PhysRevLett.129.151102} {\bibfield  {journal} {\bibinfo  {journal} {Phys. Rev. Lett.}\ }\textbf {\bibinfo {volume} {129}},\ \bibinfo {pages} {151102} (\bibinfo {year} {2022})},\ \Eprint {https://arxiv.org/abs/2204.10868} {arXiv:2204.10868 [gr-qc]} \BibitemShut {NoStop}%
\bibitem [{\citenamefont {Herdeiro}\ \emph {et~al.}(2016)\citenamefont {Herdeiro}, \citenamefont {Radu},\ and\ \citenamefont {R\'unarsson}}]{Herdeiro:2016tmi}%
  \BibitemOpen
  \bibfield  {author} {\bibinfo {author} {\bibfnamefont {C.}~\bibnamefont {Herdeiro}}, \bibinfo {author} {\bibfnamefont {E.}~\bibnamefont {Radu}},\ and\ \bibinfo {author} {\bibfnamefont {H.}~\bibnamefont {R\'unarsson}},\ }\href {https://doi.org/10.1088/0264-9381/33/15/154001} {\bibfield  {journal} {\bibinfo  {journal} {Class. Quant. Grav.}\ }\textbf {\bibinfo {volume} {33}},\ \bibinfo {pages} {154001} (\bibinfo {year} {2016})},\ \Eprint {https://arxiv.org/abs/1603.02687} {arXiv:1603.02687 [gr-qc]} \BibitemShut {NoStop}%
\bibitem [{\citenamefont {Santos}\ \emph {et~al.}(2020)\citenamefont {Santos}, \citenamefont {Benone}, \citenamefont {Crispino}, \citenamefont {Herdeiro},\ and\ \citenamefont {Radu}}]{Santos:2020pmh}%
  \BibitemOpen
  \bibfield  {author} {\bibinfo {author} {\bibfnamefont {N.~M.}\ \bibnamefont {Santos}}, \bibinfo {author} {\bibfnamefont {C.~L.}\ \bibnamefont {Benone}}, \bibinfo {author} {\bibfnamefont {L.~C.~B.}\ \bibnamefont {Crispino}}, \bibinfo {author} {\bibfnamefont {C.~A.~R.}\ \bibnamefont {Herdeiro}},\ and\ \bibinfo {author} {\bibfnamefont {E.}~\bibnamefont {Radu}},\ }\href {https://doi.org/10.1007/JHEP07(2020)010} {\bibfield  {journal} {\bibinfo  {journal} {JHEP}\ }\textbf {\bibinfo {volume} {07}},\ \bibinfo {pages} {010}},\ \Eprint {https://arxiv.org/abs/2004.09536} {arXiv:2004.09536 [gr-qc]} \BibitemShut {NoStop}%
\bibitem [{\citenamefont {Ruffini}\ and\ \citenamefont {Wheeler}(1971)}]{Ruffini:1971bza}%
  \BibitemOpen
  \bibfield  {author} {\bibinfo {author} {\bibfnamefont {R.}~\bibnamefont {Ruffini}}\ and\ \bibinfo {author} {\bibfnamefont {J.~A.}\ \bibnamefont {Wheeler}},\ }\href {https://doi.org/10.1063/1.3022513} {\bibfield  {journal} {\bibinfo  {journal} {Phys. Today}\ }\textbf {\bibinfo {volume} {24}},\ \bibinfo {pages} {30} (\bibinfo {year} {1971})}\BibitemShut {NoStop}%
\bibitem [{\citenamefont {Jacobson}(1999)}]{Jacobson:1999vr}%
  \BibitemOpen
  \bibfield  {author} {\bibinfo {author} {\bibfnamefont {T.}~\bibnamefont {Jacobson}},\ }\href {https://doi.org/10.1103/PhysRevLett.83.2699} {\bibfield  {journal} {\bibinfo  {journal} {Phys. Rev. Lett.}\ }\textbf {\bibinfo {volume} {83}},\ \bibinfo {pages} {2699} (\bibinfo {year} {1999})},\ \Eprint {https://arxiv.org/abs/astro-ph/9905303} {arXiv:astro-ph/9905303} \BibitemShut {NoStop}%
\bibitem [{\citenamefont {Hui}\ \emph {et~al.}(2019)\citenamefont {Hui}, \citenamefont {Kabat}, \citenamefont {Li}, \citenamefont {Santoni},\ and\ \citenamefont {Wong}}]{Hui:2019aqm}%
  \BibitemOpen
  \bibfield  {author} {\bibinfo {author} {\bibfnamefont {L.}~\bibnamefont {Hui}}, \bibinfo {author} {\bibfnamefont {D.}~\bibnamefont {Kabat}}, \bibinfo {author} {\bibfnamefont {X.}~\bibnamefont {Li}}, \bibinfo {author} {\bibfnamefont {L.}~\bibnamefont {Santoni}},\ and\ \bibinfo {author} {\bibfnamefont {S.~S.~C.}\ \bibnamefont {Wong}},\ }\href {https://doi.org/10.1088/1475-7516/2019/06/038} {\bibfield  {journal} {\bibinfo  {journal} {JCAP}\ }\textbf {\bibinfo {volume} {2019}}\bibfield  {number} {\bibinfo  {number} { (06)},\ \bibinfo {pages} {038}},\ }\Eprint {https://arxiv.org/abs/1904.12803} {arXiv:1904.12803 [gr-qc]} \BibitemShut {NoStop}%
\bibitem [{\citenamefont {Clough}\ \emph {et~al.}(2019)\citenamefont {Clough}, \citenamefont {Ferreira},\ and\ \citenamefont {Lagos}}]{Clough:2019jpm}%
  \BibitemOpen
  \bibfield  {author} {\bibinfo {author} {\bibfnamefont {K.}~\bibnamefont {Clough}}, \bibinfo {author} {\bibfnamefont {P.~G.}\ \bibnamefont {Ferreira}},\ and\ \bibinfo {author} {\bibfnamefont {M.}~\bibnamefont {Lagos}},\ }\href {https://doi.org/10.1103/PhysRevD.100.063014} {\bibfield  {journal} {\bibinfo  {journal} {Phys. Rev. D}\ }\textbf {\bibinfo {volume} {100}},\ \bibinfo {pages} {063014} (\bibinfo {year} {2019})},\ \Eprint {https://arxiv.org/abs/1904.12783} {arXiv:1904.12783 [gr-qc]} \BibitemShut {NoStop}%
\bibitem [{\citenamefont {Bamber}\ \emph {et~al.}(2021)\citenamefont {Bamber}, \citenamefont {Clough}, \citenamefont {Ferreira}, \citenamefont {Hui},\ and\ \citenamefont {Lagos}}]{Bamber:2020bpu}%
  \BibitemOpen
  \bibfield  {author} {\bibinfo {author} {\bibfnamefont {J.}~\bibnamefont {Bamber}}, \bibinfo {author} {\bibfnamefont {K.}~\bibnamefont {Clough}}, \bibinfo {author} {\bibfnamefont {P.~G.}\ \bibnamefont {Ferreira}}, \bibinfo {author} {\bibfnamefont {L.}~\bibnamefont {Hui}},\ and\ \bibinfo {author} {\bibfnamefont {M.}~\bibnamefont {Lagos}},\ }\href {https://doi.org/10.1103/PhysRevD.103.044059} {\bibfield  {journal} {\bibinfo  {journal} {Phys. Rev. D}\ }\textbf {\bibinfo {volume} {103}},\ \bibinfo {pages} {044059} (\bibinfo {year} {2021})},\ \Eprint {https://arxiv.org/abs/2011.07870} {arXiv:2011.07870 [gr-qc]} \BibitemShut {NoStop}%
\bibitem [{\citenamefont {Hui}\ \emph {et~al.}(2023)\citenamefont {Hui}, \citenamefont {Law}, \citenamefont {Santoni}, \citenamefont {Sun}, \citenamefont {Tomaselli},\ and\ \citenamefont {Trincherini}}]{Hui:2022sri}%
  \BibitemOpen
  \bibfield  {author} {\bibinfo {author} {\bibfnamefont {L.}~\bibnamefont {Hui}}, \bibinfo {author} {\bibfnamefont {Y.~T.~A.}\ \bibnamefont {Law}}, \bibinfo {author} {\bibfnamefont {L.}~\bibnamefont {Santoni}}, \bibinfo {author} {\bibfnamefont {G.}~\bibnamefont {Sun}}, \bibinfo {author} {\bibfnamefont {G.~M.}\ \bibnamefont {Tomaselli}},\ and\ \bibinfo {author} {\bibfnamefont {E.}~\bibnamefont {Trincherini}},\ }\href {https://doi.org/10.1103/PhysRevD.107.104018} {\bibfield  {journal} {\bibinfo  {journal} {Phys. Rev. D}\ }\textbf {\bibinfo {volume} {107}},\ \bibinfo {pages} {104018} (\bibinfo {year} {2023})},\ \Eprint {https://arxiv.org/abs/2208.06408} {arXiv:2208.06408 [gr-qc]} \BibitemShut {NoStop}%
\bibitem [{\citenamefont {Rosa}\ and\ \citenamefont {Dolan}(2012)}]{Rosa:2011my}%
  \BibitemOpen
  \bibfield  {author} {\bibinfo {author} {\bibfnamefont {J.~G.}\ \bibnamefont {Rosa}}\ and\ \bibinfo {author} {\bibfnamefont {S.~R.}\ \bibnamefont {Dolan}},\ }\href {https://doi.org/10.1103/PhysRevD.85.044043} {\bibfield  {journal} {\bibinfo  {journal} {Phys. Rev. D}\ }\textbf {\bibinfo {volume} {85}},\ \bibinfo {pages} {044043} (\bibinfo {year} {2012})},\ \Eprint {https://arxiv.org/abs/1110.4494} {arXiv:1110.4494 [hep-th]} \BibitemShut {NoStop}%
\bibitem [{\citenamefont {Krtou\v{s}}\ \emph {et~al.}(2018)\citenamefont {Krtou\v{s}}, \citenamefont {Frolov},\ and\ \citenamefont {Kubiz\v{n}\'ak}}]{Krtous:2018bvk}%
  \BibitemOpen
  \bibfield  {author} {\bibinfo {author} {\bibfnamefont {P.}~\bibnamefont {Krtou\v{s}}}, \bibinfo {author} {\bibfnamefont {V.~P.}\ \bibnamefont {Frolov}},\ and\ \bibinfo {author} {\bibfnamefont {D.}~\bibnamefont {Kubiz\v{n}\'ak}},\ }\href {https://doi.org/10.1016/j.nuclphysb.2018.06.019} {\bibfield  {journal} {\bibinfo  {journal} {Nucl. Phys. B}\ }\textbf {\bibinfo {volume} {934}},\ \bibinfo {pages} {7} (\bibinfo {year} {2018})},\ \Eprint {https://arxiv.org/abs/1803.02485} {arXiv:1803.02485 [hep-th]} \BibitemShut {NoStop}%
\bibitem [{\citenamefont {Gal'tsov}\ \emph {et~al.}(1984)\citenamefont {Gal'tsov}, \citenamefont {Pomerantseva},\ and\ \citenamefont {Chizhov}}]{Galtsov:1984ixy}%
  \BibitemOpen
  \bibfield  {author} {\bibinfo {author} {\bibfnamefont {D.~V.}\ \bibnamefont {Gal'tsov}}, \bibinfo {author} {\bibfnamefont {G.~V.}\ \bibnamefont {Pomerantseva}},\ and\ \bibinfo {author} {\bibfnamefont {G.~A.}\ \bibnamefont {Chizhov}},\ }\href {https://doi.org/10.1007/BF00893117} {\bibfield  {journal} {\bibinfo  {journal} {Sov. Phys. J.}\ }\textbf {\bibinfo {volume} {27}},\ \bibinfo {pages} {697} (\bibinfo {year} {1984})}\BibitemShut {NoStop}%
\bibitem [{\citenamefont {Konoplya}(2006)}]{Konoplya:2005hr}%
  \BibitemOpen
  \bibfield  {author} {\bibinfo {author} {\bibfnamefont {R.~A.}\ \bibnamefont {Konoplya}},\ }\href {https://doi.org/10.1103/PhysRevD.73.024009} {\bibfield  {journal} {\bibinfo  {journal} {Phys. Rev. D}\ }\textbf {\bibinfo {volume} {73}},\ \bibinfo {pages} {024009} (\bibinfo {year} {2006})},\ \Eprint {https://arxiv.org/abs/gr-qc/0509026} {arXiv:gr-qc/0509026} \BibitemShut {NoStop}%
\bibitem [{\citenamefont {Barack}\ and\ \citenamefont {Lousto}(2005)}]{Barack:2005nr}%
  \BibitemOpen
  \bibfield  {author} {\bibinfo {author} {\bibfnamefont {L.}~\bibnamefont {Barack}}\ and\ \bibinfo {author} {\bibfnamefont {C.~O.}\ \bibnamefont {Lousto}},\ }\href {https://doi.org/10.1103/PhysRevD.72.104026} {\bibfield  {journal} {\bibinfo  {journal} {Phys. Rev. D}\ }\textbf {\bibinfo {volume} {72}},\ \bibinfo {pages} {104026} (\bibinfo {year} {2005})},\ \Eprint {https://arxiv.org/abs/gr-qc/0510019} {arXiv:gr-qc/0510019} \BibitemShut {NoStop}%
\bibitem [{\citenamefont {Krtous}\ \emph {et~al.}(2008)\citenamefont {Krtous}, \citenamefont {Frolov},\ and\ \citenamefont {Kubiznak}}]{Krtous:2008tb}%
  \BibitemOpen
  \bibfield  {author} {\bibinfo {author} {\bibfnamefont {P.}~\bibnamefont {Krtous}}, \bibinfo {author} {\bibfnamefont {V.~P.}\ \bibnamefont {Frolov}},\ and\ \bibinfo {author} {\bibfnamefont {D.}~\bibnamefont {Kubiznak}},\ }\href {https://doi.org/10.1103/PhysRevD.78.064022} {\bibfield  {journal} {\bibinfo  {journal} {Phys. Rev. D}\ }\textbf {\bibinfo {volume} {78}},\ \bibinfo {pages} {064022} (\bibinfo {year} {2008})},\ \Eprint {https://arxiv.org/abs/0804.4705} {arXiv:0804.4705 [hep-th]} \BibitemShut {NoStop}%
\bibitem [{\citenamefont {Frolov}\ \emph {et~al.}(2017)\citenamefont {Frolov}, \citenamefont {Krtous},\ and\ \citenamefont {Kubiznak}}]{Frolov:2017kze}%
  \BibitemOpen
  \bibfield  {author} {\bibinfo {author} {\bibfnamefont {V.~P.}\ \bibnamefont {Frolov}}, \bibinfo {author} {\bibfnamefont {P.}~\bibnamefont {Krtous}},\ and\ \bibinfo {author} {\bibfnamefont {D.}~\bibnamefont {Kubiznak}},\ }\href {https://doi.org/10.1007/s41114-017-0009-9} {\bibfield  {journal} {\bibinfo  {journal} {Living Rev. Rel.}\ }\textbf {\bibinfo {volume} {20}},\ \bibinfo {pages} {6} (\bibinfo {year} {2017})},\ \Eprint {https://arxiv.org/abs/1705.05482} {arXiv:1705.05482 [gr-qc]} \BibitemShut {NoStop}%
\bibitem [{\citenamefont {Lunin}(2017)}]{Lunin:2017drx}%
  \BibitemOpen
  \bibfield  {author} {\bibinfo {author} {\bibfnamefont {O.}~\bibnamefont {Lunin}},\ }\href {https://doi.org/10.1007/JHEP12(2017)138} {\bibfield  {journal} {\bibinfo  {journal} {JHEP}\ }\textbf {\bibinfo {volume} {2017}}\bibfield  {number} {\bibinfo  {number} { (12)},\ \bibinfo {eid} {138}},\ }\Eprint {https://arxiv.org/abs/1708.06766} {arXiv:1708.06766 [hep-th]} \BibitemShut {NoStop}%
\bibitem [{\citenamefont {Herdeiro}\ and\ \citenamefont {Radu}(2015)}]{Herdeiro:2015waa}%
  \BibitemOpen
  \bibfield  {author} {\bibinfo {author} {\bibfnamefont {C.~A.~R.}\ \bibnamefont {Herdeiro}}\ and\ \bibinfo {author} {\bibfnamefont {E.}~\bibnamefont {Radu}},\ }\href {https://doi.org/10.1142/S0218271815420146} {\bibfield  {journal} {\bibinfo  {journal} {Int. J. Mod. Phys. D}\ }\textbf {\bibinfo {volume} {24}},\ \bibinfo {pages} {1542014} (\bibinfo {year} {2015})},\ \Eprint {https://arxiv.org/abs/1504.08209} {arXiv:1504.08209 [gr-qc]} \BibitemShut {NoStop}%
\bibitem [{\citenamefont {Heisenberg}(2014)}]{Heisenberg:2014rta}%
  \BibitemOpen
  \bibfield  {author} {\bibinfo {author} {\bibfnamefont {L.}~\bibnamefont {Heisenberg}},\ }\href {https://doi.org/10.1088/1475-7516/2014/05/015} {\bibfield  {journal} {\bibinfo  {journal} {JCAP}\ }\textbf {\bibinfo {volume} {05}},\ \bibinfo {pages} {015}},\ \Eprint {https://arxiv.org/abs/1402.7026} {arXiv:1402.7026 [hep-th]} \BibitemShut {NoStop}%
\bibitem [{\citenamefont {Minamitsuji}(2016)}]{Minamitsuji:2016ydr}%
  \BibitemOpen
  \bibfield  {author} {\bibinfo {author} {\bibfnamefont {M.}~\bibnamefont {Minamitsuji}},\ }\href {https://doi.org/10.1103/PhysRevD.94.084039} {\bibfield  {journal} {\bibinfo  {journal} {Phys. Rev. D}\ }\textbf {\bibinfo {volume} {94}},\ \bibinfo {pages} {084039} (\bibinfo {year} {2016})},\ \Eprint {https://arxiv.org/abs/1607.06278} {arXiv:1607.06278 [gr-qc]} \BibitemShut {NoStop}%
\bibitem [{\citenamefont {Newman}\ and\ \citenamefont {Penrose}(1962)}]{Newman:1961qr}%
  \BibitemOpen
  \bibfield  {author} {\bibinfo {author} {\bibfnamefont {E.}~\bibnamefont {Newman}}\ and\ \bibinfo {author} {\bibfnamefont {R.}~\bibnamefont {Penrose}},\ }\href {https://doi.org/10.1063/1.1724257} {\bibfield  {journal} {\bibinfo  {journal} {J. Math. Phys.}\ }\textbf {\bibinfo {volume} {3}},\ \bibinfo {pages} {566} (\bibinfo {year} {1962})}\BibitemShut {NoStop}%
\bibitem [{\citenamefont {Price}(1972)}]{Price:1972pw}%
  \BibitemOpen
  \bibfield  {author} {\bibinfo {author} {\bibfnamefont {R.~H.}\ \bibnamefont {Price}},\ }\href {https://doi.org/10.1103/PhysRevD.5.2439} {\bibfield  {journal} {\bibinfo  {journal} {Phys. Rev. D}\ }\textbf {\bibinfo {volume} {5}},\ \bibinfo {pages} {2439} (\bibinfo {year} {1972})}\BibitemShut {NoStop}%
\bibitem [{\citenamefont {Teukolsky}(1972)}]{Teukolsky:1972my}%
  \BibitemOpen
  \bibfield  {author} {\bibinfo {author} {\bibfnamefont {S.~A.}\ \bibnamefont {Teukolsky}},\ }\href {https://doi.org/10.1103/PhysRevLett.29.1114} {\bibfield  {journal} {\bibinfo  {journal} {Phys. Rev. Lett.}\ }\textbf {\bibinfo {volume} {29}},\ \bibinfo {pages} {1114} (\bibinfo {year} {1972})}\BibitemShut {NoStop}%
\bibitem [{\citenamefont {Teukolsky}(1973)}]{Teukolsky:1973ha}%
  \BibitemOpen
  \bibfield  {author} {\bibinfo {author} {\bibfnamefont {S.~A.}\ \bibnamefont {Teukolsky}},\ }\href {https://doi.org/10.1086/152444} {\bibfield  {journal} {\bibinfo  {journal} {Astrophys. J.}\ }\textbf {\bibinfo {volume} {185}},\ \bibinfo {pages} {635} (\bibinfo {year} {1973})}\BibitemShut {NoStop}%
\bibitem [{\citenamefont {Frolov}\ \emph {et~al.}(2018)\citenamefont {Frolov}, \citenamefont {Krtou\v{s}}, \citenamefont {Kubiz\v{n}\'ak},\ and\ \citenamefont {Santos}}]{Frolov:2018ezx}%
  \BibitemOpen
  \bibfield  {author} {\bibinfo {author} {\bibfnamefont {V.~P.}\ \bibnamefont {Frolov}}, \bibinfo {author} {\bibfnamefont {P.}~\bibnamefont {Krtou\v{s}}}, \bibinfo {author} {\bibfnamefont {D.}~\bibnamefont {Kubiz\v{n}\'ak}},\ and\ \bibinfo {author} {\bibfnamefont {J.~E.}\ \bibnamefont {Santos}},\ }\href {https://doi.org/10.1103/PhysRevLett.120.231103} {\bibfield  {journal} {\bibinfo  {journal} {Phys. Rev. Lett.}\ }\textbf {\bibinfo {volume} {120}},\ \bibinfo {pages} {231103} (\bibinfo {year} {2018})},\ \Eprint {https://arxiv.org/abs/1804.00030} {arXiv:1804.00030 [hep-th]} \BibitemShut {NoStop}%
\bibitem [{\citenamefont {Fernandes}\ \emph {et~al.}(2022)\citenamefont {Fernandes}, \citenamefont {Hilditch}, \citenamefont {Lemos},\ and\ \citenamefont {Cardoso}}]{Fernandes:2021qvr}%
  \BibitemOpen
  \bibfield  {author} {\bibinfo {author} {\bibfnamefont {T.~V.}\ \bibnamefont {Fernandes}}, \bibinfo {author} {\bibfnamefont {D.}~\bibnamefont {Hilditch}}, \bibinfo {author} {\bibfnamefont {J.~P.~S.}\ \bibnamefont {Lemos}},\ and\ \bibinfo {author} {\bibfnamefont {V.}~\bibnamefont {Cardoso}},\ }\href {https://doi.org/10.1103/PhysRevD.105.044017} {\bibfield  {journal} {\bibinfo  {journal} {Phys. Rev. D}\ }\textbf {\bibinfo {volume} {105}},\ \bibinfo {pages} {044017} (\bibinfo {year} {2022})},\ \Eprint {https://arxiv.org/abs/2112.03282} {arXiv:2112.03282 [gr-qc]} \BibitemShut {NoStop}%
\bibitem [{\citenamefont {Dolan}(2019)}]{Dolan:2019hcw}%
  \BibitemOpen
  \bibfield  {author} {\bibinfo {author} {\bibfnamefont {S.~R.}\ \bibnamefont {Dolan}},\ }\href {https://doi.org/10.1103/PhysRevD.100.044044} {\bibfield  {journal} {\bibinfo  {journal} {Phys. Rev. D}\ }\textbf {\bibinfo {volume} {100}},\ \bibinfo {pages} {044044} (\bibinfo {year} {2019})},\ \Eprint {https://arxiv.org/abs/1906.04808} {arXiv:1906.04808 [gr-qc]} \BibitemShut {NoStop}%
\bibitem [{\citenamefont {Percival}\ and\ \citenamefont {Dolan}(2020)}]{Percival:2020skc}%
  \BibitemOpen
  \bibfield  {author} {\bibinfo {author} {\bibfnamefont {J.}~\bibnamefont {Percival}}\ and\ \bibinfo {author} {\bibfnamefont {S.~R.}\ \bibnamefont {Dolan}},\ }\href {https://doi.org/10.1103/PhysRevD.102.104055} {\bibfield  {journal} {\bibinfo  {journal} {Phys. Rev. D}\ }\textbf {\bibinfo {volume} {102}},\ \bibinfo {pages} {104055} (\bibinfo {year} {2020})},\ \Eprint {https://arxiv.org/abs/2008.10621} {arXiv:2008.10621 [gr-qc]} \BibitemShut {NoStop}%
\bibitem [{\citenamefont {Berti}\ \emph {et~al.}(2009)\citenamefont {Berti}, \citenamefont {Cardoso},\ and\ \citenamefont {Starinets}}]{Berti_2009}%
  \BibitemOpen
  \bibfield  {author} {\bibinfo {author} {\bibfnamefont {E.}~\bibnamefont {Berti}}, \bibinfo {author} {\bibfnamefont {V.}~\bibnamefont {Cardoso}},\ and\ \bibinfo {author} {\bibfnamefont {A.~O.}\ \bibnamefont {Starinets}},\ }\href {https://doi.org/10.1088/0264-9381/26/16/163001} {\bibfield  {journal} {\bibinfo  {journal} {Classical and Quantum Gravity}\ }\textbf {\bibinfo {volume} {26}},\ \bibinfo {pages} {163001} (\bibinfo {year} {2009})}\BibitemShut {NoStop}%
\bibitem [{\citenamefont {Turner}(1983)}]{Turner:1983he}%
  \BibitemOpen
  \bibfield  {author} {\bibinfo {author} {\bibfnamefont {M.~S.}\ \bibnamefont {Turner}},\ }\href {https://doi.org/10.1103/PhysRevD.28.1243} {\bibfield  {journal} {\bibinfo  {journal} {Phys. Rev. D}\ }\textbf {\bibinfo {volume} {28}},\ \bibinfo {pages} {1243} (\bibinfo {year} {1983})}\BibitemShut {NoStop}%
\bibitem [{\citenamefont {Alonso-\'Alvarez}\ \emph {et~al.}(2020)\citenamefont {Alonso-\'Alvarez}, \citenamefont {Hugle},\ and\ \citenamefont {Jaeckel}}]{Alonso-Alvarez:2019ixv}%
  \BibitemOpen
  \bibfield  {author} {\bibinfo {author} {\bibfnamefont {G.}~\bibnamefont {Alonso-\'Alvarez}}, \bibinfo {author} {\bibfnamefont {T.}~\bibnamefont {Hugle}},\ and\ \bibinfo {author} {\bibfnamefont {J.}~\bibnamefont {Jaeckel}},\ }\href {https://doi.org/10.1088/1475-7516/2020/02/014} {\bibfield  {journal} {\bibinfo  {journal} {JCAP}\ }\textbf {\bibinfo {volume} {2020}}\bibfield  {number} {\bibinfo  {number} { (02)},\ \bibinfo {pages} {014}},\ }\Eprint {https://arxiv.org/abs/1905.09836} {arXiv:1905.09836 [hep-ph]} \BibitemShut {NoStop}%
\bibitem [{\citenamefont {Nakayama}(2019)}]{Nakayama:2019rhg}%
  \BibitemOpen
  \bibfield  {author} {\bibinfo {author} {\bibfnamefont {K.}~\bibnamefont {Nakayama}},\ }\href {https://doi.org/10.1088/1475-7516/2019/10/019} {\bibfield  {journal} {\bibinfo  {journal} {JCAP}\ }\textbf {\bibinfo {volume} {2019}}\bibfield  {number} {\bibinfo  {number} { (10)},\ \bibinfo {pages} {019}},\ }\Eprint {https://arxiv.org/abs/1907.06243} {arXiv:1907.06243 [hep-ph]} \BibitemShut {NoStop}%
\bibitem [{\citenamefont {{Fiziev}}(2010)}]{2010JPhA...43c5203F}%
  \BibitemOpen
  \bibfield  {author} {\bibinfo {author} {\bibfnamefont {P.~P.}\ \bibnamefont {{Fiziev}}},\ }\href {https://doi.org/10.1088/1751-8113/43/3/035203} {\bibfield  {journal} {\bibinfo  {journal} {Journal of Physics A Mathematical General}\ }\textbf {\bibinfo {volume} {43}},\ \bibinfo {eid} {035203} (\bibinfo {year} {2010})},\ \Eprint {https://arxiv.org/abs/0904.0245} {arXiv:0904.0245 [math-ph]} \BibitemShut {NoStop}%
\bibitem [{\citenamefont {Horta\c{c}su}(2018)}]{Hortacsu:2011rr}%
  \BibitemOpen
  \bibfield  {author} {\bibinfo {author} {\bibfnamefont {M.}~\bibnamefont {Horta\c{c}su}},\ }\href {https://doi.org/10.1142/9789814417532_0002} {\bibfield  {journal} {\bibinfo  {journal} {Adv. High Energy Phys.}\ }\textbf {\bibinfo {volume} {2018}},\ \bibinfo {pages} {23} (\bibinfo {year} {2018})},\ \Eprint {https://arxiv.org/abs/1101.0471} {arXiv:1101.0471 [math-ph]} \BibitemShut {NoStop}%
\bibitem [{\citenamefont {Ghez}\ \emph {et~al.}(1998)\citenamefont {Ghez}, \citenamefont {Klein}, \citenamefont {Morris},\ and\ \citenamefont {Becklin}}]{Ghez:1998ph}%
  \BibitemOpen
  \bibfield  {author} {\bibinfo {author} {\bibfnamefont {A.~M.}\ \bibnamefont {Ghez}}, \bibinfo {author} {\bibfnamefont {B.~L.}\ \bibnamefont {Klein}}, \bibinfo {author} {\bibfnamefont {M.}~\bibnamefont {Morris}},\ and\ \bibinfo {author} {\bibfnamefont {E.~E.}\ \bibnamefont {Becklin}},\ }\href {https://doi.org/10.1086/306528} {\bibfield  {journal} {\bibinfo  {journal} {Astrophys. J.}\ }\textbf {\bibinfo {volume} {509}},\ \bibinfo {pages} {678} (\bibinfo {year} {1998})},\ \Eprint {https://arxiv.org/abs/astro-ph/9807210} {arXiv:astro-ph/9807210} \BibitemShut {NoStop}%
\bibitem [{\citenamefont {Hoeft}\ \emph {et~al.}(2004)\citenamefont {Hoeft}, \citenamefont {Mucket},\ and\ \citenamefont {Gottlober}}]{Hoeft:2003ea}%
  \BibitemOpen
  \bibfield  {author} {\bibinfo {author} {\bibfnamefont {M.}~\bibnamefont {Hoeft}}, \bibinfo {author} {\bibfnamefont {J.~P.}\ \bibnamefont {Mucket}},\ and\ \bibinfo {author} {\bibfnamefont {S.}~\bibnamefont {Gottlober}},\ }\href {https://doi.org/10.1086/380990} {\bibfield  {journal} {\bibinfo  {journal} {Astrophys. J.}\ }\textbf {\bibinfo {volume} {602}},\ \bibinfo {pages} {162} (\bibinfo {year} {2004})},\ \Eprint {https://arxiv.org/abs/astro-ph/0311083} {arXiv:astro-ph/0311083} \BibitemShut {NoStop}%
\bibitem [{\citenamefont {East}\ and\ \citenamefont {Huang}(2022)}]{East:2022rsi}%
  \BibitemOpen
  \bibfield  {author} {\bibinfo {author} {\bibfnamefont {W.~E.}\ \bibnamefont {East}}\ and\ \bibinfo {author} {\bibfnamefont {J.}~\bibnamefont {Huang}},\ }\href {https://doi.org/10.1007/JHEP12(2022)089} {\bibfield  {journal} {\bibinfo  {journal} {JHEP}\ }\textbf {\bibinfo {volume} {2022}}\bibfield  {number} {\bibinfo  {number} { (12)},\ \bibinfo {pages} {089}},\ }\Eprint {https://arxiv.org/abs/2206.12432} {arXiv:2206.12432 [hep-ph]} \BibitemShut {NoStop}%
\bibitem [{\citenamefont {Schive}\ \emph {et~al.}(2014)\citenamefont {Schive}, \citenamefont {Liao}, \citenamefont {Woo}, \citenamefont {Wong}, \citenamefont {Chiueh}, \citenamefont {Broadhurst},\ and\ \citenamefont {Hwang}}]{Schive:2014hza}%
  \BibitemOpen
  \bibfield  {author} {\bibinfo {author} {\bibfnamefont {H.-Y.}\ \bibnamefont {Schive}}, \bibinfo {author} {\bibfnamefont {M.-H.}\ \bibnamefont {Liao}}, \bibinfo {author} {\bibfnamefont {T.-P.}\ \bibnamefont {Woo}}, \bibinfo {author} {\bibfnamefont {S.-K.}\ \bibnamefont {Wong}}, \bibinfo {author} {\bibfnamefont {T.}~\bibnamefont {Chiueh}}, \bibinfo {author} {\bibfnamefont {T.}~\bibnamefont {Broadhurst}},\ and\ \bibinfo {author} {\bibfnamefont {W.~Y.~P.}\ \bibnamefont {Hwang}},\ }\href {https://doi.org/10.1103/PhysRevLett.113.261302} {\bibfield  {journal} {\bibinfo  {journal} {Phys. Rev. Lett.}\ }\textbf {\bibinfo {volume} {113}},\ \bibinfo {pages} {261302} (\bibinfo {year} {2014})},\ \Eprint {https://arxiv.org/abs/1407.7762} {arXiv:1407.7762 [astro-ph.GA]} \BibitemShut {NoStop}%
\bibitem [{\citenamefont {Bertschinger}(1985)}]{Bertschinger:1985pd}%
  \BibitemOpen
  \bibfield  {author} {\bibinfo {author} {\bibfnamefont {E.}~\bibnamefont {Bertschinger}},\ }\href {https://doi.org/10.1086/191028} {\bibfield  {journal} {\bibinfo  {journal} {Astrophys. J. Suppl.}\ }\textbf {\bibinfo {volume} {58}},\ \bibinfo {pages} {39} (\bibinfo {year} {1985})}\BibitemShut {NoStop}%
\bibitem [{\citenamefont {Gondolo}\ and\ \citenamefont {Silk}(1999)}]{Gondolo:1999ef}%
  \BibitemOpen
  \bibfield  {author} {\bibinfo {author} {\bibfnamefont {P.}~\bibnamefont {Gondolo}}\ and\ \bibinfo {author} {\bibfnamefont {J.}~\bibnamefont {Silk}},\ }\href {https://doi.org/10.1103/PhysRevLett.83.1719} {\bibfield  {journal} {\bibinfo  {journal} {Phys. Rev. Lett.}\ }\textbf {\bibinfo {volume} {83}},\ \bibinfo {pages} {1719} (\bibinfo {year} {1999})},\ \Eprint {https://arxiv.org/abs/astro-ph/9906391} {arXiv:astro-ph/9906391} \BibitemShut {NoStop}%
\bibitem [{\citenamefont {Shapiro}(2023)}]{Shapiro:2023gpe}%
  \BibitemOpen
  \bibfield  {author} {\bibinfo {author} {\bibfnamefont {S.~L.}\ \bibnamefont {Shapiro}},\ }\href {https://doi.org/10.1103/PhysRevD.108.083037} {\bibfield  {journal} {\bibinfo  {journal} {Phys. Rev. D}\ }\textbf {\bibinfo {volume} {108}},\ \bibinfo {pages} {083037} (\bibinfo {year} {2023})},\ \Eprint {https://arxiv.org/abs/2310.13739} {arXiv:2310.13739 [astro-ph.GA]} \BibitemShut {NoStop}%
\bibitem [{\citenamefont {Baryakhtar}\ \emph {et~al.}(2022)\citenamefont {Baryakhtar} \emph {et~al.}}]{Baryakhtar:2022hbu}%
  \BibitemOpen
  \bibfield  {author} {\bibinfo {author} {\bibfnamefont {M.}~\bibnamefont {Baryakhtar}} \emph {et~al.},\ }in\ \href@noop {} {\emph {\bibinfo {booktitle} {{Snowmass 2021}}}}\ (\bibinfo {year} {2022})\ \Eprint {https://arxiv.org/abs/2203.07984} {arXiv:2203.07984 [hep-ph]} \BibitemShut {NoStop}%
\bibitem [{\citenamefont {Kavanagh}\ \emph {et~al.}(2018)\citenamefont {Kavanagh}, \citenamefont {Gaggero},\ and\ \citenamefont {Bertone}}]{Kavanagh:2018ggo}%
  \BibitemOpen
  \bibfield  {author} {\bibinfo {author} {\bibfnamefont {B.~J.}\ \bibnamefont {Kavanagh}}, \bibinfo {author} {\bibfnamefont {D.}~\bibnamefont {Gaggero}},\ and\ \bibinfo {author} {\bibfnamefont {G.}~\bibnamefont {Bertone}},\ }\href {https://doi.org/10.1103/PhysRevD.98.023536} {\bibfield  {journal} {\bibinfo  {journal} {Phys. Rev. D}\ }\textbf {\bibinfo {volume} {98}},\ \bibinfo {pages} {023536} (\bibinfo {year} {2018})},\ \Eprint {https://arxiv.org/abs/1805.09034} {arXiv:1805.09034 [astro-ph.CO]} \BibitemShut {NoStop}%
\bibitem [{\citenamefont {Barausse}\ \emph {et~al.}(2020)\citenamefont {Barausse} \emph {et~al.}}]{Barausse:2020rsu}%
  \BibitemOpen
  \bibfield  {author} {\bibinfo {author} {\bibfnamefont {E.}~\bibnamefont {Barausse}} \emph {et~al.},\ }\href {https://doi.org/10.1007/s10714-020-02691-1} {\bibfield  {journal} {\bibinfo  {journal} {Gen. Rel. Grav.}\ }\textbf {\bibinfo {volume} {52}},\ \bibinfo {pages} {81} (\bibinfo {year} {2020})},\ \Eprint {https://arxiv.org/abs/2001.09793} {arXiv:2001.09793 [gr-qc]} \BibitemShut {NoStop}%
\bibitem [{\citenamefont {Arun}\ \emph {et~al.}(2022)\citenamefont {Arun} \emph {et~al.}}]{LISA:2022kgy}%
  \BibitemOpen
  \bibfield  {author} {\bibinfo {author} {\bibfnamefont {K.~G.}\ \bibnamefont {Arun}} \emph {et~al.} (\bibinfo {collaboration} {LISA}),\ }\href {https://doi.org/10.1007/s41114-022-00036-9} {\bibfield  {journal} {\bibinfo  {journal} {Living Rev. Rel.}\ }\textbf {\bibinfo {volume} {25}},\ \bibinfo {pages} {4} (\bibinfo {year} {2022})},\ \Eprint {https://arxiv.org/abs/2205.01597} {arXiv:2205.01597 [gr-qc]} \BibitemShut {NoStop}%
\bibitem [{\citenamefont {Duque}\ \emph {et~al.}(2024)\citenamefont {Duque}, \citenamefont {Macedo}, \citenamefont {Vicente},\ and\ \citenamefont {Cardoso}}]{Duque:2023seg}%
  \BibitemOpen
  \bibfield  {author} {\bibinfo {author} {\bibfnamefont {F.}~\bibnamefont {Duque}}, \bibinfo {author} {\bibfnamefont {C.~F.~B.}\ \bibnamefont {Macedo}}, \bibinfo {author} {\bibfnamefont {R.}~\bibnamefont {Vicente}},\ and\ \bibinfo {author} {\bibfnamefont {V.}~\bibnamefont {Cardoso}},\ }\href {https://doi.org/10.1103/PhysRevLett.133.121404} {\bibfield  {journal} {\bibinfo  {journal} {Phys. Rev. Lett.}\ }\textbf {\bibinfo {volume} {133}},\ \bibinfo {pages} {121404} (\bibinfo {year} {2024})},\ \Eprint {https://arxiv.org/abs/2312.06767} {arXiv:2312.06767 [gr-qc]} \BibitemShut {NoStop}%
\bibitem [{\citenamefont {Gliorio}\ \emph {et~al.}(2025)\citenamefont {Gliorio}, \citenamefont {Berti}, \citenamefont {Maselli},\ and\ \citenamefont {Speeney}}]{Gliorio:2025cbh}%
  \BibitemOpen
  \bibfield  {author} {\bibinfo {author} {\bibfnamefont {S.}~\bibnamefont {Gliorio}}, \bibinfo {author} {\bibfnamefont {E.}~\bibnamefont {Berti}}, \bibinfo {author} {\bibfnamefont {A.}~\bibnamefont {Maselli}},\ and\ \bibinfo {author} {\bibfnamefont {N.}~\bibnamefont {Speeney}},\ }\href@noop {} {\bibinfo {title} {{Extreme mass ratio inspirals in dark matter halos: dynamics and distinguishability of halo models}}} (\bibinfo {year} {2025}),\ \Eprint {https://arxiv.org/abs/2503.16649} {arXiv:2503.16649 [gr-qc]} \BibitemShut {NoStop}%
\bibitem [{\citenamefont {Bamber}\ \emph {et~al.}(2023)\citenamefont {Bamber}, \citenamefont {Aurrekoetxea}, \citenamefont {Clough},\ and\ \citenamefont {Ferreira}}]{Bamber:2022pbs}%
  \BibitemOpen
  \bibfield  {author} {\bibinfo {author} {\bibfnamefont {J.}~\bibnamefont {Bamber}}, \bibinfo {author} {\bibfnamefont {J.~C.}\ \bibnamefont {Aurrekoetxea}}, \bibinfo {author} {\bibfnamefont {K.}~\bibnamefont {Clough}},\ and\ \bibinfo {author} {\bibfnamefont {P.~G.}\ \bibnamefont {Ferreira}},\ }\href {https://doi.org/10.1103/PhysRevD.107.024035} {\bibfield  {journal} {\bibinfo  {journal} {Phys. Rev. D}\ }\textbf {\bibinfo {volume} {107}},\ \bibinfo {pages} {024035} (\bibinfo {year} {2023})},\ \Eprint {https://arxiv.org/abs/2210.09254} {arXiv:2210.09254 [gr-qc]} \BibitemShut {NoStop}%
\bibitem [{\citenamefont {Aurrekoetxea}\ \emph {et~al.}(2024{\natexlab{a}})\citenamefont {Aurrekoetxea}, \citenamefont {Clough}, \citenamefont {Bamber},\ and\ \citenamefont {Ferreira}}]{Aurrekoetxea:2023jwk}%
  \BibitemOpen
  \bibfield  {author} {\bibinfo {author} {\bibfnamefont {J.~C.}\ \bibnamefont {Aurrekoetxea}}, \bibinfo {author} {\bibfnamefont {K.}~\bibnamefont {Clough}}, \bibinfo {author} {\bibfnamefont {J.}~\bibnamefont {Bamber}},\ and\ \bibinfo {author} {\bibfnamefont {P.~G.}\ \bibnamefont {Ferreira}},\ }\href {https://doi.org/10.1103/PhysRevLett.132.211401} {\bibfield  {journal} {\bibinfo  {journal} {Phys. Rev. Lett.}\ }\textbf {\bibinfo {volume} {132}},\ \bibinfo {pages} {211401} (\bibinfo {year} {2024}{\natexlab{a}})},\ \Eprint {https://arxiv.org/abs/2311.18156} {arXiv:2311.18156 [gr-qc]} \BibitemShut {NoStop}%
\bibitem [{\citenamefont {Aurrekoetxea}\ \emph {et~al.}(2024{\natexlab{b}})\citenamefont {Aurrekoetxea}, \citenamefont {Marsden}, \citenamefont {Clough},\ and\ \citenamefont {Ferreira}}]{Aurrekoetxea:2024cqd}%
  \BibitemOpen
  \bibfield  {author} {\bibinfo {author} {\bibfnamefont {J.~C.}\ \bibnamefont {Aurrekoetxea}}, \bibinfo {author} {\bibfnamefont {J.}~\bibnamefont {Marsden}}, \bibinfo {author} {\bibfnamefont {K.}~\bibnamefont {Clough}},\ and\ \bibinfo {author} {\bibfnamefont {P.~G.}\ \bibnamefont {Ferreira}},\ }\href {https://doi.org/10.1103/PhysRevD.110.083011} {\bibfield  {journal} {\bibinfo  {journal} {Phys. Rev. D}\ }\textbf {\bibinfo {volume} {110}},\ \bibinfo {pages} {083011} (\bibinfo {year} {2024}{\natexlab{b}})},\ \Eprint {https://arxiv.org/abs/2409.01937} {arXiv:2409.01937 [gr-qc]} \BibitemShut {NoStop}%
\bibitem [{\citenamefont {Cheng}\ \emph {et~al.}()\citenamefont {Cheng}, \citenamefont {Ficarra},\ and\ \citenamefont {Witek}}]{ChengInPrep}%
  \BibitemOpen
  \bibfield  {author} {\bibinfo {author} {\bibfnamefont {C.-H.}\ \bibnamefont {Cheng}}, \bibinfo {author} {\bibfnamefont {G.}~\bibnamefont {Ficarra}},\ and\ \bibinfo {author} {\bibfnamefont {H.}~\bibnamefont {Witek}},\ }\href@noop {} {\bibinfo {title} {{Scalar field dynamics around binary black holes}}},\ \bibinfo {note} {in prep.}\BibitemShut {Stop}%
\bibitem [{\citenamefont {{Salpeter}}(1964)}]{1964ApJ...140..796S}%
  \BibitemOpen
  \bibfield  {author} {\bibinfo {author} {\bibfnamefont {E.~E.}\ \bibnamefont {{Salpeter}}},\ }\href {https://doi.org/10.1086/147973} {\bibfield  {journal} {\bibinfo  {journal} {\apj}\ }\textbf {\bibinfo {volume} {140}},\ \bibinfo {pages} {796} (\bibinfo {year} {1964})}\BibitemShut {NoStop}%
\bibitem [{\citenamefont {Hui}(2021)}]{Hui:2021tkt}%
  \BibitemOpen
  \bibfield  {author} {\bibinfo {author} {\bibfnamefont {L.}~\bibnamefont {Hui}},\ }\href {https://doi.org/10.1146/annurev-astro-120920-010024} {\bibfield  {journal} {\bibinfo  {journal} {Ann. Rev. Astron. Astrophys.}\ }\textbf {\bibinfo {volume} {59}},\ \bibinfo {pages} {247} (\bibinfo {year} {2021})},\ \Eprint {https://arxiv.org/abs/2101.11735} {arXiv:2101.11735 [astro-ph.CO]} \BibitemShut {NoStop}%
\bibitem [{\citenamefont {Israel}(1967)}]{Israel:1967wq}%
  \BibitemOpen
  \bibfield  {author} {\bibinfo {author} {\bibfnamefont {W.}~\bibnamefont {Israel}},\ }\href {https://doi.org/10.1103/PhysRev.164.1776} {\bibfield  {journal} {\bibinfo  {journal} {Phys. Rev.}\ }\textbf {\bibinfo {volume} {164}},\ \bibinfo {pages} {1776} (\bibinfo {year} {1967})}\BibitemShut {NoStop}%
\bibitem [{\citenamefont {Israel}(1968)}]{Israel:1967za}%
  \BibitemOpen
  \bibfield  {author} {\bibinfo {author} {\bibfnamefont {W.}~\bibnamefont {Israel}},\ }\href {https://doi.org/10.1007/BF01645859} {\bibfield  {journal} {\bibinfo  {journal} {Commun. Math. Phys.}\ }\textbf {\bibinfo {volume} {8}},\ \bibinfo {pages} {245} (\bibinfo {year} {1968})}\BibitemShut {NoStop}%
\bibitem [{\citenamefont {Carter}(1971)}]{Carter:1971zc}%
  \BibitemOpen
  \bibfield  {author} {\bibinfo {author} {\bibfnamefont {B.}~\bibnamefont {Carter}},\ }\href {https://doi.org/10.1103/PhysRevLett.26.331} {\bibfield  {journal} {\bibinfo  {journal} {Phys. Rev. Lett.}\ }\textbf {\bibinfo {volume} {26}},\ \bibinfo {pages} {331} (\bibinfo {year} {1971})}\BibitemShut {NoStop}%
\bibitem [{\citenamefont {Wald}(1971)}]{Wald:1971iw}%
  \BibitemOpen
  \bibfield  {author} {\bibinfo {author} {\bibfnamefont {R.~M.}\ \bibnamefont {Wald}},\ }\href {https://doi.org/10.1103/PhysRevLett.26.1653} {\bibfield  {journal} {\bibinfo  {journal} {Phys. Rev. Lett.}\ }\textbf {\bibinfo {volume} {26}},\ \bibinfo {pages} {1653} (\bibinfo {year} {1971})}\BibitemShut {NoStop}%
\bibitem [{\citenamefont {Bekenstein}(1972{\natexlab{a}})}]{Bekenstein:1971hc}%
  \BibitemOpen
  \bibfield  {author} {\bibinfo {author} {\bibfnamefont {J.~D.}\ \bibnamefont {Bekenstein}},\ }\href {https://doi.org/10.1103/PhysRevD.5.1239} {\bibfield  {journal} {\bibinfo  {journal} {Phys. Rev. D}\ }\textbf {\bibinfo {volume} {5}},\ \bibinfo {pages} {1239} (\bibinfo {year} {1972}{\natexlab{a}})}\BibitemShut {NoStop}%
\bibitem [{\citenamefont {Bekenstein}(1972{\natexlab{b}})}]{Bekenstein:1972ky}%
  \BibitemOpen
  \bibfield  {author} {\bibinfo {author} {\bibfnamefont {J.~D.}\ \bibnamefont {Bekenstein}},\ }\href {https://doi.org/10.1103/PhysRevD.5.2403} {\bibfield  {journal} {\bibinfo  {journal} {Phys. Rev. D}\ }\textbf {\bibinfo {volume} {5}},\ \bibinfo {pages} {2403} (\bibinfo {year} {1972}{\natexlab{b}})}\BibitemShut {NoStop}%
\bibitem [{\citenamefont {Bekenstein}(1972{\natexlab{c}})}]{Bekenstein:1972ny}%
  \BibitemOpen
  \bibfield  {author} {\bibinfo {author} {\bibfnamefont {J.~D.}\ \bibnamefont {Bekenstein}},\ }\href {https://doi.org/10.1103/PhysRevLett.28.452} {\bibfield  {journal} {\bibinfo  {journal} {Phys. Rev. Lett.}\ }\textbf {\bibinfo {volume} {28}},\ \bibinfo {pages} {452} (\bibinfo {year} {1972}{\natexlab{c}})}\BibitemShut {NoStop}%
\bibitem [{\citenamefont {Teitelboim}(1972)}]{Teitelboim:1972qx}%
  \BibitemOpen
  \bibfield  {author} {\bibinfo {author} {\bibfnamefont {C.}~\bibnamefont {Teitelboim}},\ }\href {https://doi.org/10.1103/PhysRevD.5.2941} {\bibfield  {journal} {\bibinfo  {journal} {Phys. Rev. D}\ }\textbf {\bibinfo {volume} {5}},\ \bibinfo {pages} {2941} (\bibinfo {year} {1972})}\BibitemShut {NoStop}%
\bibitem [{\citenamefont {Adler}\ and\ \citenamefont {Pearson}(1978)}]{Adler:1978dp}%
  \BibitemOpen
  \bibfield  {author} {\bibinfo {author} {\bibfnamefont {S.~L.}\ \bibnamefont {Adler}}\ and\ \bibinfo {author} {\bibfnamefont {R.~B.}\ \bibnamefont {Pearson}},\ }\href {https://doi.org/10.1103/PhysRevD.18.2798} {\bibfield  {journal} {\bibinfo  {journal} {Phys. Rev. D}\ }\textbf {\bibinfo {volume} {18}},\ \bibinfo {pages} {2798} (\bibinfo {year} {1978})}\BibitemShut {NoStop}%
\bibitem [{\citenamefont {Richards}\ \emph {et~al.}(2025)\citenamefont {Richards}, \citenamefont {Dima}, \citenamefont {Ferguson},\ and\ \citenamefont {Witek}}]{Richards:2025ows}%
  \BibitemOpen
  \bibfield  {author} {\bibinfo {author} {\bibfnamefont {C.}~\bibnamefont {Richards}}, \bibinfo {author} {\bibfnamefont {A.}~\bibnamefont {Dima}}, \bibinfo {author} {\bibfnamefont {D.}~\bibnamefont {Ferguson}},\ and\ \bibinfo {author} {\bibfnamefont {H.}~\bibnamefont {Witek}},\ }\href@noop {} {\bibfield  {journal} {\bibinfo  {journal} {Phys. Rev. D, in press}\ } (\bibinfo {year} {2025})},\ \Eprint {https://arxiv.org/abs/2501.14034} {arXiv:2501.14034 [gr-qc]} \BibitemShut {NoStop}%
\bibitem [{\citenamefont {Herdeiro}\ and\ \citenamefont {Radu}(2014)}]{Herdeiro:2014goa}%
  \BibitemOpen
  \bibfield  {author} {\bibinfo {author} {\bibfnamefont {C.~A.~R.}\ \bibnamefont {Herdeiro}}\ and\ \bibinfo {author} {\bibfnamefont {E.}~\bibnamefont {Radu}},\ }\href {https://doi.org/10.1103/PhysRevLett.112.221101} {\bibfield  {journal} {\bibinfo  {journal} {Phys. Rev. Lett.}\ }\textbf {\bibinfo {volume} {112}},\ \bibinfo {pages} {221101} (\bibinfo {year} {2014})},\ \Eprint {https://arxiv.org/abs/1403.2757} {arXiv:1403.2757 [gr-qc]} \BibitemShut {NoStop}%
\bibitem [{\citenamefont {Chodosh}\ and\ \citenamefont {Shlapentokh-Rothman}(2017)}]{Chodosh:2015oma}%
  \BibitemOpen
  \bibfield  {author} {\bibinfo {author} {\bibfnamefont {O.}~\bibnamefont {Chodosh}}\ and\ \bibinfo {author} {\bibfnamefont {Y.}~\bibnamefont {Shlapentokh-Rothman}},\ }\href {https://doi.org/10.1007/s00220-017-2998-3} {\bibfield  {journal} {\bibinfo  {journal} {Commun. Math. Phys.}\ }\textbf {\bibinfo {volume} {356}},\ \bibinfo {pages} {1155} (\bibinfo {year} {2017})},\ \Eprint {https://arxiv.org/abs/1510.08025} {arXiv:1510.08025 [gr-qc]} \BibitemShut {NoStop}%
\bibitem [{\citenamefont {Siemonsen}\ \emph {et~al.}(2023)\citenamefont {Siemonsen}, \citenamefont {Mondino}, \citenamefont {Egana-Ugrinovic}, \citenamefont {Huang}, \citenamefont {Baryakhtar},\ and\ \citenamefont {East}}]{Siemonsen:2022ivj}%
  \BibitemOpen
  \bibfield  {author} {\bibinfo {author} {\bibfnamefont {N.}~\bibnamefont {Siemonsen}}, \bibinfo {author} {\bibfnamefont {C.}~\bibnamefont {Mondino}}, \bibinfo {author} {\bibfnamefont {D.}~\bibnamefont {Egana-Ugrinovic}}, \bibinfo {author} {\bibfnamefont {J.}~\bibnamefont {Huang}}, \bibinfo {author} {\bibfnamefont {M.}~\bibnamefont {Baryakhtar}},\ and\ \bibinfo {author} {\bibfnamefont {W.~E.}\ \bibnamefont {East}},\ }\href {https://doi.org/10.1103/PhysRevD.107.075025} {\bibfield  {journal} {\bibinfo  {journal} {Phys. Rev. D}\ }\textbf {\bibinfo {volume} {107}},\ \bibinfo {pages} {075025} (\bibinfo {year} {2023})},\ \Eprint {https://arxiv.org/abs/2212.09772} {arXiv:2212.09772 [astro-ph.HE]} \BibitemShut {NoStop}%
\bibitem [{\citenamefont {Xin}\ and\ \citenamefont {Most}(2025)}]{Xin:2024trp}%
  \BibitemOpen
  \bibfield  {author} {\bibinfo {author} {\bibfnamefont {S.}~\bibnamefont {Xin}}\ and\ \bibinfo {author} {\bibfnamefont {E.~R.}\ \bibnamefont {Most}},\ }\href {https://doi.org/10.1103/PhysRevD.111.063050} {\bibfield  {journal} {\bibinfo  {journal} {Phys. Rev. D}\ }\textbf {\bibinfo {volume} {111}},\ \bibinfo {pages} {063050} (\bibinfo {year} {2025})},\ \Eprint {https://arxiv.org/abs/2406.02992} {arXiv:2406.02992 [astro-ph.HE]} \BibitemShut {NoStop}%
\end{thebibliography}%

\end{document}